\documentclass[twocolumn,10pt]{asme2e}

\usepackage{caption}
\usepackage{subcaption}
\usepackage{graphicx}
\usepackage{tikz}
\usepackage{lipsum}
\usepackage{amssymb}
\confshortname{OMAE2022}
\conffullname{the ASME 2022 41th International \\
 Conferences on Ocean, Offshore and Arctic Engineering}

\confdate{June 5-10}
\confyear{2022}
\confcity{Hamburg}
\confcountry{Germany}

\papernum{Under consideration to be published in OMAE 2022}

\title{PROPULSIVE PERFORMANCE OF MORPHING AND HEAVING FOIL}

\author{Pragalbh Dev Singh, Ishan Neogi, Vardhan Niral Shah, Vaibhav Joshi\thanks{Address all correspondence to this author.}
    \affiliation{
	Department of Mechanical Engineering\\
	Birla Institute of Technology \& Science Pilani\\
	K K Birla Goa Campus\\
	Goa, India 403726\\
    Email: vaibhavj@goa.bits-pilani.ac.in
    }	
}

\graphicspath{ {Images/} }
\begin{document}

\maketitle    

\begin{abstract}
{\it Biological locomotion, observed in the flexible wings of birds and insects, bodies and fins of
aquatic mammals and fishes, consists of their ability to morph the wings/fins. The morphing
capability holds significance in the ability of fishes to swim upstream without spending too
much energy and that of birds to glide for extended periods of time. Simplifying the wing or
fins to a foil, morphing refers to the ability of the foil to change its camber smoothly, without
sharp bends on the foil surface. This allows precise control over flow separation and vortex
shedding. Compared to conventional trailing-edge extensions or flaps, used in rudders and
elevators in submarines and ships, morphing foils provide better control of thrust and lift
characteristics. This study aims at understanding the importance of the morphing of foil
combined with a sinusoidal heaving motion on thrust generation. A two-dimensional variational
stabilized Petrov-Galerkin moving mesh framework is utilized for modelling the incompressible
low Reynolds number flow across the flapping foil. The morphing motion is characterized by the
extent of morphing, measured as an angle of deviation from the initial camber, and the point of
initiation of morphing on the foil as a percentage of its chord length. The effect of the foil
morphing and the heaving motion on the propulsive performance are investigated. The extent
of morphing is varied from -30$^\circ$ and 30$^\circ$, and the point of initiation ranges from 15\% to
50\% of the chord. The Reynolds and Strouhal numbers for the study are 1100 and 0.2,
respectively. The results from the current work can pave the way for enhanced engineering
designs in bio-mimetics and give insights on design conditions for optimal thrust performance.}
\end{abstract}

\section*{INTRODUCTION}
Flexible flapping of wings or foil-defined structures like fins is a mechanism of lift and thrust generation commonly found in avian and aquatic life. Birds and fishes flap their wings and fins to propel themselves into the air or through the water. Thus, inheritance of the evolutionary designs present in biological sources can prove to be advantageous and more efficient in optimizing the lift and thrust generation \cite{CHEN2017112, BOCKMANN20168, TRIANTAFYLLOU1993205}. Such bio-inspired designs, like ornithopters, have attracted significant attention due to their non-reliance on traditional propulsion mechanisms like propellers and jets, resemblance to wildlife, high efficiency in some regimes \cite{doi:10.2514/5.9781600866654.0001.0010} and  potential in micro-to-nano scale robotic applications. A key feature of biologically-inspired designs is the flexibility of structures which has been of interest to researchers recently with the advent of soft robots and compliant mechanisms implemented in aircraft wing extensions \cite{10.1117/12.483869, kota_1999}.

A simplification of the motion of wings and fins may be ascribed to a combination of three independent motions, morphing, heaving, and pitching. The combination of heaving and pitching has been extensively studied, both numerically and experimentally, and in isolation, flexible foils have been studied as an analogue for morphing. Biologically inspired unmanned underwater vehicles have experimented with flapping motion as a source for thrust, however, the current applications are limited to rigid foils where the flapping motion is approximated by combined heaving and pitching motions\cite{Betz1912, Lighthill_book_1975, Knoller1909}. Extensive studies have been performed to understand the flapping dynamics of a single foil \cite{WU2020106712} and the effect of various parameters like chord ratio, phase difference, and gap ratio for the tandem foil configuration \cite{Akhtar2007, Rival_2011, Broering2012, Lin2019, Sampath2020, Kinsey2012, KARBASIAN2015816, Pan2016, muscutt_weymouth_ganapathisubramani_2017, Ma2019, OMAE_Joshi_2021, tandem2021}. 

While rigid foils are pragmatically sound with respect to manufacturing and structural constraints, wings and fins in nature are highly flexible, and the chord-wise deformation of the foil adds to the efficiency and performance of propulsion. The change in the camber of the foils (hereby referred to as morphing of foils) can allow for precise control over lift and drag characteristics, thus altering flow structures and enhancing thrust performance and/or efficiency. Uncontrolled flexibility (two-way coupling between foil and freestream flow) was found to have some benefits in thrust performance within some regimes \cite{Heathcote2004}. Studies focusing on morphing investigated the performance of morphing at high Reynolds number flows which is characterized by the formation of Kelvin–Helmholtz vortices \cite{LI201846, Alexander2015, Lyu2015, Scheller2015, Jodin2017, TO2019102595, Zhang2021, Simiriotis2019, HANG2020105612, Burdette2018DesignOA, eguea_fuel_2020}. Observations from these studies support the use of morphing for a favorable $C_L/C_D$ ratio. These results provide motivation to study the behavior of morphing in the current work.

The muscles within the wings of birds and bodies of fish allow them to bend their bodies in precise ways, and the intention of this study is to mimic that pre-determined control of morphing. If the flexibility and deformation of the foil is controlled and prescribed along side heaving and/or pitching, benefits of both worlds can be gained and the performance characteristics of the mechanism may be tuned to any application. The applications of such a fully controllable variable camber foil (i.e., morphing foil) are immense, and find uses in the development of micro air vehicles, Autonomous Underwater Vehicles (AUVs), nano-robotics in a myriad of fields and energy harvesting systems.

The geometric manipulation of the morphing represents two control mechanisms of flapping in aquatic and avian locomotion.
Let us take the example of fish to understand these mechanisms. The first is the ability of fish to only morph a certain section of their body. A fish may only flap its caudal fin or morph its whole body to move. The second control mechanism is the ability to control the amount of deflection. This study focuses on the effect of the morph position (i.e. the location of the onset of morphing along the chord of the foil) and the morph amplitude (deflection of the trailing edge from mean position) on the propulsive performance of a foil that is heaving and morphing. 


In order to truly understand and mimic the thrust generation capabilities of birds and fishes, it is necessary to combine heaving, pitching and morphing together in one study. Moreover, there is a necessity to conduct studies in low Reynolds number flows as a large number of applications correspond to this regime such as marine and offshore station-keeping using low-speed maneuvering robots and low-speed hovering flight of drones, where stability and agility become important. 
 As per the knowledge of the authors, the literature lacks in the studies dealing with a combination of morphing with heaving or pitching motion.
 The present study analyses the thrust performance of a morphing and heaving foil at low Reynolds number regime, and analyzes the effect of morphing position and amplitude on the propulsive performance of the foil.

A symmetrical NACA 0015 foil is considered at the Reynolds number $Re = 1100$. The foil heaves with an amplitude of one chord length. The phase difference between heaving and morphing is taken to be \(\pi/2\) ahead in favor of heaving, i.e., the morph of the foil will be greatest when the foil is at its central heave position. The onset of morphing varies between 0\% and 50\% of the chord length, and the morph amplitude ranges from 10$^\circ$ to 60$^\circ$. The numerical framework solves the incompressible Navier-Stokes equations in two-dimensions, discretized by variational stabilized Petrov-Galerkin finite element formulation in a moving-mesh Arbitrary Lagrangian-Eulerian (ALE) framework. The formulation has been implemented as an in-house solver for computing two-dimensional fluid-structure interactions.

The outline of the article is as follows. The numerical framework of the study is discussed first, where the flow modelling equations and discretization, mesh characteristics and the mechanism of morphing are laid out in detail. Key results and trends are presented thereafter, with emphasis on the difference between pitching and morphing, and the individual impacts of morph amplitude and position. The study is then concluded with key findings and opportunities for future work are highlighted.
\section*{NUMERICAL FRAMEWORK}
In this section, we describe the numerical framework utilized for simulating the combined heaving and morphing of the foil. The prescribed motion on the foil is described along with the flow modeling in the moving mesh framework.

\subsection*{Prescribed Motion of the Structure}
The structure (a foil in the present case) is prescribed a heaving and morphing motion, both of which are varying sinusoidally with time. The heave position is given by $h(t)$ and the amount of trailing edge deflection (i.e. morph amount) is denoted by the variable $\theta(t)$,
\begin{align}
    h(t) &= h_0\mathrm{sin}(2\pi ft + \phi_h),\label{heaveeq} \\
    \theta(t) &= \theta_0\mathrm{sin}(2\pi ft), \label{morpheeq}
\end{align}
where \(h_0\) and \(\theta_0\) are the amplitudes of heaving and morphing respectively, $f$ denotes the frequency of motion and \(\phi_h\) is the phase difference between morphing and heaving. 
The onset position of morphing is denoted by $x_0 \%$ of the of the chord length. There is no morphing effect from the leading edge of the foil till the morphing onset position.

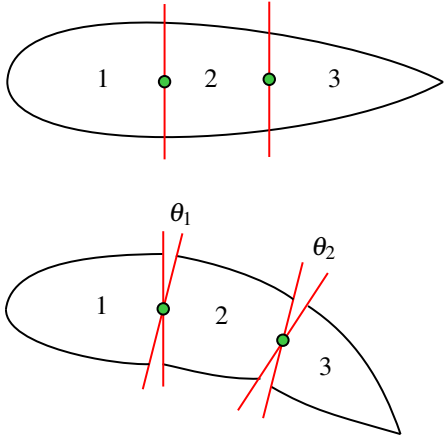
\begin{figure}[!htbp]
    \centering
\tikzset{every picture/.style={line width=0.75pt}} 

\begin{tikzpicture}[x=0.75pt,y=0.75pt,yscale=-1,xscale=1]

\draw    (290,228.73) .. controls (310.67,240.73) and (322.67,243.4) .. (355.33,252.73) ;
\draw    (235.33,217.4) .. controls (248,220.07) and (265.33,225.4) .. (284.67,224.73) ;
\draw [color={rgb, 255:red, 255; green, 0; blue, 0 }  ,draw opacity=1 ]   (306,166) -- (286,244.67) ;
\draw [color={rgb, 255:red, 255; green, 0; blue, 0 }  ,draw opacity=1 ]   (318.67,171.4) -- (273.33,240.73) ;
\draw    (242,162.73) .. controls (260.67,166.07) and (286,173.4) .. (301.33,184.73) ;
\draw [color={rgb, 255:red, 255; green, 0; blue, 0 }  ,draw opacity=1 ]   (245.33,151.4) -- (225.33,230.07) ;
\draw    (157,75) .. controls (158.8,23) and (328.8,49) .. (376.8,75) ;
\draw    (157,75) .. controls (157.8,116) and (316.8,108) .. (376.8,75) ;
\draw [color={rgb, 255:red, 255; green, 0; blue, 0 }  ,draw opacity=1 ]   (236.32,35.64) -- (236.32,114.04) ;
\draw [color={rgb, 255:red, 255; green, 0; blue, 0 }  ,draw opacity=1 ]   (289.12,34.44) -- (289.12,112.84) ;
\draw  [fill={rgb, 255:red, 64; green, 199; blue, 67 }  ,fill opacity=1 ] (233.44,74.84) .. controls (233.44,73.25) and (234.73,71.96) .. (236.32,71.96) .. controls (237.91,71.96) and (239.2,73.25) .. (239.2,74.84) .. controls (239.2,76.43) and (237.91,77.72) .. (236.32,77.72) .. controls (234.73,77.72) and (233.44,76.43) .. (233.44,74.84) -- cycle ;
\draw  [fill={rgb, 255:red, 64; green, 199; blue, 67 }  ,fill opacity=1 ] (286.24,73.64) .. controls (286.24,72.05) and (287.53,70.76) .. (289.12,70.76) .. controls (290.71,70.76) and (292,72.05) .. (292,73.64) .. controls (292,75.23) and (290.71,76.52) .. (289.12,76.52) .. controls (287.53,76.52) and (286.24,75.23) .. (286.24,73.64) -- cycle ;
\draw    (156.33,189.67) .. controls (159.33,164.47) and (206.67,162.07) .. (235.33,161.8) ;
\draw [color={rgb, 255:red, 255; green, 0; blue, 0 }  ,draw opacity=1 ]   (235.65,150.31) -- (235.65,228.71) ;
\draw  [fill={rgb, 255:red, 64; green, 199; blue, 67 }  ,fill opacity=1 ] (232.77,189.51) .. controls (232.77,187.92) and (234.06,186.63) .. (235.65,186.63) .. controls (237.24,186.63) and (238.53,187.92) .. (238.53,189.51) .. controls (238.53,191.1) and (237.24,192.39) .. (235.65,192.39) .. controls (234.06,192.39) and (232.77,191.1) .. (232.77,189.51) -- cycle ;
\draw  [fill={rgb, 255:red, 64; green, 199; blue, 67 }  ,fill opacity=1 ] (293.12,205.33) .. controls (293.12,203.74) and (294.41,202.45) .. (296,202.45) .. controls (297.59,202.45) and (298.88,203.74) .. (298.88,205.33) .. controls (298.88,206.92) and (297.59,208.21) .. (296,208.21) .. controls (294.41,208.21) and (293.12,206.92) .. (293.12,205.33) -- cycle ;
\draw    (308.67,188.07) .. controls (332,203.4) and (346,225.4) .. (355.33,252.73) ;
\draw    (156.33,189.67) .. controls (156,208.73) and (211.33,218.07) .. (228.67,217.4) ;

\draw (200.52,66.8) node [anchor=north west][inner sep=0.75pt]   [align=left] {1};
\draw (254.92,66.8) node [anchor=north west][inner sep=0.75pt]   [align=left] {2};
\draw (317.32,66.93) node [anchor=north west][inner sep=0.75pt]   [align=left] {3};
\draw (199.85,181.47) node [anchor=north west][inner sep=0.75pt]   [align=left] {1};
\draw (260.39,186.8) node [anchor=north west][inner sep=0.75pt]   [align=left] {2};
\draw (312.72,212.33) node [anchor=north west][inner sep=0.75pt]   [align=left] {3};
\draw (237,134.4) node [anchor=north west][inner sep=0.75pt]    {$\theta _{1}$};
\draw (309,151.73) node [anchor=north west][inner sep=0.75pt]    {$\theta _{2}$};
\end{tikzpicture}
\caption{MORPHING MECHANISM}
\label{Fig:1}
\end{figure}

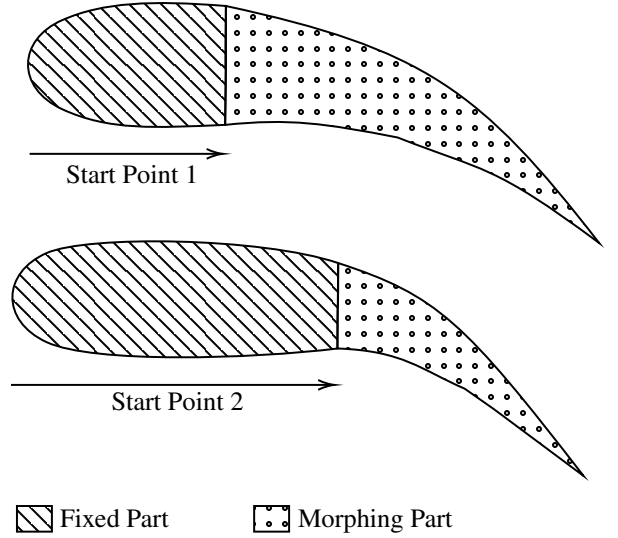
\begin{figure}
    \centering
\tikzset{
pattern size/.store in=\mcSize, 
pattern size = 5pt,
pattern thickness/.store in=\mcThickness, 
pattern thickness = 0.3pt,
pattern radius/.store in=\mcRadius, 
pattern radius = 1pt}
\makeatletter
\pgfutil@ifundefined{pgf@pattern@name@_g2ihfc2bn}{
\pgfdeclarepatternformonly[\mcThickness,\mcSize]{_g2ihfc2bn}
{\pgfqpoint{0pt}{-\mcThickness}}
{\pgfpoint{\mcSize}{\mcSize}}
{\pgfpoint{\mcSize}{\mcSize}}
{
\pgfsetcolor{\tikz@pattern@color}
\pgfsetlinewidth{\mcThickness}
\pgfpathmoveto{\pgfqpoint{0pt}{\mcSize}}
\pgfpathlineto{\pgfpoint{\mcSize+\mcThickness}{-\mcThickness}}
\pgfusepath{stroke}
}}
\makeatother

 
\tikzset{
pattern size/.store in=\mcSize, 
pattern size = 5pt,
pattern thickness/.store in=\mcThickness, 
pattern thickness = 0.3pt,
pattern radius/.store in=\mcRadius, 
pattern radius = 1pt}
\makeatletter
\pgfutil@ifundefined{pgf@pattern@name@_laspikef5}{
\makeatletter
\pgfdeclarepatternformonly[\mcRadius,\mcThickness,\mcSize]{_laspikef5}
{\pgfpoint{-0.5*\mcSize}{-0.5*\mcSize}}
{\pgfpoint{0.5*\mcSize}{0.5*\mcSize}}
{\pgfpoint{\mcSize}{\mcSize}}
{
\pgfsetcolor{\tikz@pattern@color}
\pgfsetlinewidth{\mcThickness}
\pgfpathcircle\pgfpointorigin{\mcRadius}
\pgfusepath{stroke}
}}
\makeatother

 
\tikzset{
pattern size/.store in=\mcSize, 
pattern size = 5pt,
pattern thickness/.store in=\mcThickness, 
pattern thickness = 0.3pt,
pattern radius/.store in=\mcRadius, 
pattern radius = 1pt}
\makeatletter
\pgfutil@ifundefined{pgf@pattern@name@_v7trz2ue9}{
\pgfdeclarepatternformonly[\mcThickness,\mcSize]{_v7trz2ue9}
{\pgfqpoint{0pt}{-\mcThickness}}
{\pgfpoint{\mcSize}{\mcSize}}
{\pgfpoint{\mcSize}{\mcSize}}
{
\pgfsetcolor{\tikz@pattern@color}
\pgfsetlinewidth{\mcThickness}
\pgfpathmoveto{\pgfqpoint{0pt}{\mcSize}}
\pgfpathlineto{\pgfpoint{\mcSize+\mcThickness}{-\mcThickness}}
\pgfusepath{stroke}
}}
\makeatother

 
\tikzset{
pattern size/.store in=\mcSize, 
pattern size = 5pt,
pattern thickness/.store in=\mcThickness, 
pattern thickness = 0.3pt,
pattern radius/.store in=\mcRadius, 
pattern radius = 1pt}
\makeatletter
\pgfutil@ifundefined{pgf@pattern@name@_ho60y3cek}{
\makeatletter
\pgfdeclarepatternformonly[\mcRadius,\mcThickness,\mcSize]{_ho60y3cek}
{\pgfpoint{-0.5*\mcSize}{-0.5*\mcSize}}
{\pgfpoint{0.5*\mcSize}{0.5*\mcSize}}
{\pgfpoint{\mcSize}{\mcSize}}
{
\pgfsetcolor{\tikz@pattern@color}
\pgfsetlinewidth{\mcThickness}
\pgfpathcircle\pgfpointorigin{\mcRadius}
\pgfusepath{stroke}
}}
\makeatother

 
\tikzset{
pattern size/.store in=\mcSize, 
pattern size = 5pt,
pattern thickness/.store in=\mcThickness, 
pattern thickness = 0.3pt,
pattern radius/.store in=\mcRadius, 
pattern radius = 1pt}
\makeatletter
\pgfutil@ifundefined{pgf@pattern@name@_npcrdnkar}{
\pgfdeclarepatternformonly[\mcThickness,\mcSize]{_npcrdnkar}
{\pgfqpoint{0pt}{-\mcThickness}}
{\pgfpoint{\mcSize}{\mcSize}}
{\pgfpoint{\mcSize}{\mcSize}}
{
\pgfsetcolor{\tikz@pattern@color}
\pgfsetlinewidth{\mcThickness}
\pgfpathmoveto{\pgfqpoint{0pt}{\mcSize}}
\pgfpathlineto{\pgfpoint{\mcSize+\mcThickness}{-\mcThickness}}
\pgfusepath{stroke}
}}
\makeatother

 
\tikzset{
pattern size/.store in=\mcSize, 
pattern size = 5pt,
pattern thickness/.store in=\mcThickness, 
pattern thickness = 0.3pt,
pattern radius/.store in=\mcRadius, 
pattern radius = 1pt}
\makeatletter
\pgfutil@ifundefined{pgf@pattern@name@_x3au099r3}{
\makeatletter
\pgfdeclarepatternformonly[\mcRadius,\mcThickness,\mcSize]{_x3au099r3}
{\pgfpoint{-0.5*\mcSize}{-0.5*\mcSize}}
{\pgfpoint{0.5*\mcSize}{0.5*\mcSize}}
{\pgfpoint{\mcSize}{\mcSize}}
{
\pgfsetcolor{\tikz@pattern@color}
\pgfsetlinewidth{\mcThickness}
\pgfpathcircle\pgfpointorigin{\mcRadius}
\pgfusepath{stroke}
}}
\makeatother
\tikzset{every picture/.style={line width=0.75pt}} 

\begin{tikzpicture}[x=0.75pt,y=0.75pt,yscale=-0.75,xscale=0.75]

\draw  [pattern=_g2ihfc2bn,pattern size=6pt,pattern thickness=0.75pt,pattern radius=0pt, pattern color={rgb, 255:red, 0; green, 0; blue, 0}] (150.7,99.85) .. controls (177.7,87.85) and (210.7,86.35) .. (260.7,90.35) .. controls (260.7,169.35) and (261.2,90.35) .. (260.2,170.35) .. controls (212.7,172.35) and (182.2,174.35) .. (150.7,159.85) .. controls (119.2,145.35) and (120.7,112.85) .. (150.7,99.85) -- cycle ;
\draw  [pattern=_laspikef5,pattern size=6pt,pattern thickness=0.75pt,pattern radius=0.75pt, pattern color={rgb, 255:red, 0; green, 0; blue, 0}] (260.7,90.35) .. controls (392.7,118.85) and (431.7,146.35) .. (511.7,249.85) .. controls (439.7,196.35) and (434.7,201.85) .. (376.2,178.85) .. controls (327.2,169.35) and (303.7,165.85) .. (260.2,170.35) .. controls (261.2,90.35) and (259.7,170.35) .. (260.7,90.35) -- cycle ;
\draw  [pattern=_v7trz2ue9,pattern size=6pt,pattern thickness=0.75pt,pattern radius=0pt, pattern color={rgb, 255:red, 0; green, 0; blue, 0}] (140.2,256.6) .. controls (167.2,244.6) and (284.2,245.85) .. (336.2,263.35) .. controls (335.7,320.85) and (336.2,263.85) .. (335.7,320.85) .. controls (278.2,327.85) and (171.7,331.1) .. (140.2,316.6) .. controls (108.7,302.1) and (110.2,269.6) .. (140.2,256.6) -- cycle ;
\draw  [pattern=_ho60y3cek,pattern size=6pt,pattern thickness=0.75pt,pattern radius=0.75pt, pattern color={rgb, 255:red, 0; green, 0; blue, 0}] (336.2,263.35) .. controls (402.7,284.85) and (421.2,303.1) .. (501.2,406.6) .. controls (429.2,353.1) and (458.2,373.85) .. (421.2,347.85) .. controls (392.2,335.35) and (380.2,322.35) .. (335.7,320.85) .. controls (335.7,263.35) and (336.2,320.35) .. (336.2,263.35) -- cycle ;
\draw    (128.7,189.85) -- (256.7,189.85) ;
\draw [shift={(258.7,189.85)}, rotate = 180] [color={rgb, 255:red, 0; green, 0; blue, 0 }  ][line width=0.75]    (10.93,-3.29) .. controls (6.95,-1.4) and (3.31,-0.3) .. (0,0) .. controls (3.31,0.3) and (6.95,1.4) .. (10.93,3.29)   ;
\draw    (116.2,345.35) -- (331.7,345.35) ;
\draw [shift={(333.7,345.35)}, rotate = 180] [color={rgb, 255:red, 0; green, 0; blue, 0 }  ][line width=0.75]    (10.93,-3.29) .. controls (6.95,-1.4) and (3.31,-0.3) .. (0,0) .. controls (3.31,0.3) and (6.95,1.4) .. (10.93,3.29)   ;
\draw  [pattern=_npcrdnkar,pattern size=6pt,pattern thickness=0.75pt,pattern radius=0pt, pattern color={rgb, 255:red, 0; green, 0; blue, 0}] (120.2,426.85) -- (143.7,426.85) -- (143.7,444.1) -- (120.2,444.1) -- cycle ;
\draw  [pattern=_x3au099r3,pattern size=6pt,pattern thickness=0.75pt,pattern radius=0.75pt, pattern color={rgb, 255:red, 0; green, 0; blue, 0}] (279.7,426.85) -- (303.2,426.85) -- (303.2,444.1) -- (279.7,444.1) -- cycle ;

\draw (151.2,195.1) node [anchor=north west][inner sep=0.75pt]   [align=left] {Start Point 1};
\draw (181.7,348.1) node [anchor=north west][inner sep=0.75pt]   [align=left] {Start Point 2};
\draw (147.2,427.1) node [anchor=north west][inner sep=0.75pt]   [align=left] {Fixed Part};
\draw (307.2,426.6) node [anchor=north west][inner sep=0.75pt]   [align=left] {Morphing Part};
\end{tikzpicture}
\caption{VISUALIZATION OF THE START POINT LOCATION FOR MORPHING}
\label{Fig:2}
\end{figure}

The mechanics of morphing is designed in such a way that a smooth surface of the morphed foil can be obtained. 
To acoomplish this, the morphing has been approximated as the rotation of several minute rigid body sections, as represented in Fig. \ref{Fig:1}. The required morphable part of the foil is divided into 100 sections. The number chosen here is arbitrary, the smoothness of the foil is well defined at 100 and any larger number will not significantly increase the smoothness of the surface. Each  section is then rotated by an angle of \(\theta(t)/100\) about the central point in the foil, relative to the section just preceding it to obtain the net morph state. In Fig. \ref{Fig:1}, the foil is divided into 3 sections, the latter two of which rotate by an angle of \(\theta_1\) and \(\theta_2\) respectively, thus giving total morph angle = \(\theta_1 + \theta_2\). For this case, \(\theta_1\) = \(\theta_2\) = Total Morph angle / 2. The circles represent the points of rotation and each section rotates about the circle just preceding it. The visualization of the starting morphing location is depicted in Fig. \ref{Fig:2}. 

\subsection*{Flow Modeling}
Two-dimensional incompressible Navier-Stokes equations are solved in a moving mesh framework in the current formulation. The momentum and the continuity equations, respectively, are given as:
\begin{align}
\label{Navstokes}
\left.\rho^\mathrm{f}\frac{\partial \boldsymbol{v}^\mathrm{f}}{\partial t}\right\vert_{\chi}+ \rho^\mathrm{f}(\boldsymbol{v}^\mathrm{f}-\boldsymbol{w})\cdot\nabla \boldsymbol{v}^\mathrm{f} &= \nabla\cdot\boldsymbol{\sigma}^\mathrm{f} + \rho^\mathrm{f} \boldsymbol{b}^\mathrm{f},\ \text{on}\ \Omega^\mathrm{f}(t) \\
\nabla \cdot \boldsymbol{v}^\mathrm{f} = 0. \label{continuity}
\end{align}
Here, $\boldsymbol{w}$ is the mesh velocity, \( \boldsymbol{v}^\mathrm{f}\) is the fluid velocity with \( v^\mathrm{f}_x\), and \( v^\mathrm{f}_y\) its two components in x and y directions, respectively. The Cauchy stress tensor \(\boldsymbol{\sigma}^\mathrm{f}\) for a Newtonian fluid is written as $\boldsymbol{\sigma}^\mathrm{f} = -p\boldsymbol{I} + \mu^\mathrm{f}(\nabla\boldsymbol{v}^\mathrm{f} + (\nabla\boldsymbol{v}^\mathrm{f})^T)$,
where \(\mu^\mathrm{f}\) is the dynamic viscosity of the fluid, p is the fluid pressure, and \(\boldsymbol{I}\) is the identity matrix. The body force and fluid density are represented by \(\boldsymbol{b}^\mathrm{f}\) and \(\rho^\mathrm{f}\) respectively. \(\boldsymbol{\chi}\) denotes the ALE referential coordinates in Eq. (\ref{Navstokes}).

Temporal  discretization for the governing equations is carried out using the Generalized-\(\alpha\) method \cite{Gen_alpha}, and finite element variational stabilized method is employed for the spatial discretization. The variational formulation for the flow equations can be found in \cite{tandem2021, FSI_Book}. 

\subsection*{Fluid Structure Interface}

The fluid-structure interaction is one-way in the current scenario, as the structural displacements are known \textit{a-priori}. By equating the fluid and structure velocities, the kinematic continuity condition is satisfied at the fluid-structure interface as 
\begin{equation}
    \label{kinematic continuity}
    \boldsymbol{v}^\mathrm{f}(\boldsymbol{\varphi}(\boldsymbol{X},t),t) = \boldsymbol{v}^\mathrm{s}(\boldsymbol{X},t), \forall\ \boldsymbol{X}\ \in \Gamma^\mathrm{fs}.
\end{equation}

Here, $\boldsymbol{\varphi}$ is a bijection mapping from the structural initial position $\boldsymbol{X}$ at $t = 0$ to that at an arbitrary time $t$, i.e. $\boldsymbol{\varphi}(\boldsymbol{X},t) = \boldsymbol{X} + \boldsymbol{u}^\mathrm{s}(\boldsymbol{X},t)$, where $\Gamma^\mathrm{fs}$ denotes fluid structure interface and $\boldsymbol{u}^\mathrm{s}$ is the structural displacement. 

\begin{figure}[!htbp]
    \centering
    \includegraphics[width = 0.45\textwidth]{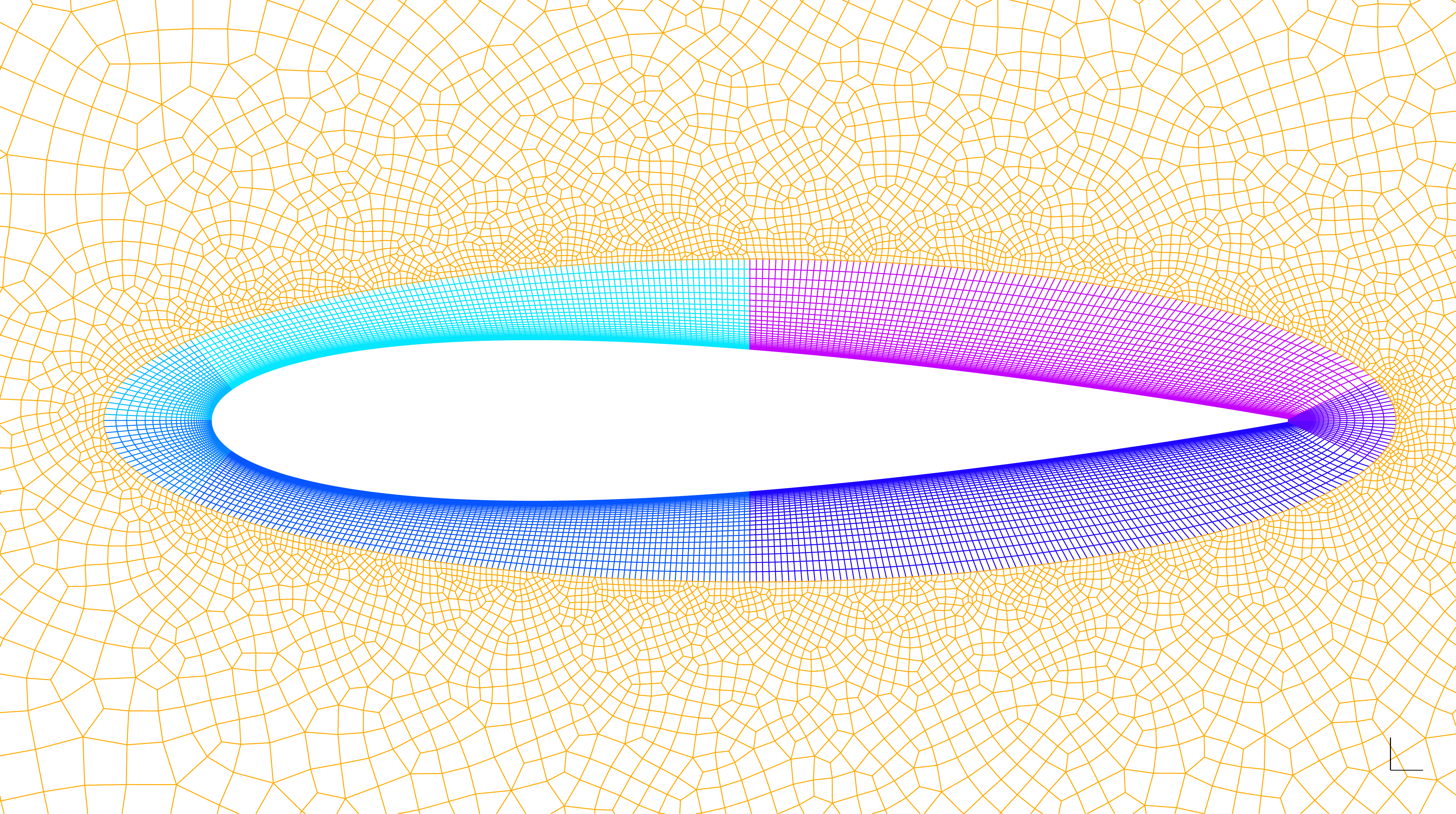}
    \caption{THE COMPUTATIONAL MESH FOR THE FOIL ALONG WITH THE REFINED BOUNDARY LAYER MESH}
    \label{fig:mesh}
\end{figure}

The verification of the formulation along with mesh convergence and validation has been carried out in the previous works \cite{OMAE_Joshi_2021, tandem2021} and will not be dealt with in the current work. The mesh convergence study performed is applicable for the mesh utilized for the current study. This ensures that the mesh is refined enough to capture the flow physics. A representative close-up view of the mesh is shown in Fig. \ref{fig:mesh}.

\subsection*{Parameters of Interest}
The performance of the foil is determined by evaluating the integrated values of the forces on the foil surface. The instantaneous coefficients for the foil can be written as:
\begin{align}
    C_D &= \frac{F_x}{\frac{1}{2}\rho^\mathrm{f} U_{\infty}^2 cl} = \frac{1}{\frac{1}{2}\rho^\mathrm{f} U_{\infty}^2 cl} \int_{\Gamma^{\mathrm{fs}}(t)} (\boldsymbol{\sigma}^\mathrm{f}\cdot \boldsymbol{n}) \cdot \boldsymbol{n}_x d\Gamma, \\
    C_T &= \frac{-F_x}{\frac{1}{2}\rho^\mathrm{f} U_{\infty}^2 cl} = -C_D,\\
    C_L &= \frac{F_y}{\frac{1}{2}\rho^\mathrm{f} U_{\infty}^2 cl} = \frac{1}{\frac{1}{2}\rho^\mathrm{f} U_{\infty}^2 cl} \int_{\Gamma^{\mathrm{fs}}(t)} (\boldsymbol{\sigma}^\mathrm{f}\cdot \boldsymbol{n}) \cdot \boldsymbol{n}_y d\Gamma,\\
    C_P &= \frac{P}{\frac{1}{2}\rho^\mathrm{f}U_{\infty}^3 cl} = \frac{-(\boldsymbol{F}\cdot\boldsymbol{v})}{\frac{1}{2}\rho^\mathrm{f}U_{\infty}^3 cl},
\end{align}
where $C_D$, $C_T$, $C_L$ and $C_P$ are the drag, thrust, lift and power coefficients, respectively. The chord length and the span of the foil are denoted by $c$ and $l=1$, respectively. The freestream velocity is represented by $U_{\infty}$. The integrated force on the surface of the foil is given by $\boldsymbol{F} = (F_x, F_y)$ with its components as $F_x$ and $F_y$ in $X$ and $Y$ directions, respectively. The velocity at the points on the surface of the foil is represented by $\boldsymbol{v}$. The propulsive efficiency can be written as $\eta = C_{T,\mathrm{mean}}/C_{P,\mathrm{mean}}$, where $X_{,\mathrm{mean}}$ indicates the time-averaged value of $X$ over a time period $T$.



\section*{RESULTS AND DISCUSSION}
In this section, we analyse the effects of the morphing on the lift and thrust of the foil, and investigate the flow dynamics using vorticity plots. We aim to computationally study the effects of morphing position and amplitude on the propulsive performance of the foil. For all the cases, the non-dimensional heave amplitude $h_0/c = 1$, reduced frequency $fc/U_{\infty} = 0.2$, Reynolds number $Re = \rho^\mathrm{f} U_{\infty}c/\mu^\mathrm{f} = 1100$ and $\phi_h = \pi/2$, where $U_{\infty}$ is the freestream velocity.

\subsection*{Statistics of Flow Coefficients}
\begin{figure}
    \centering
    \begin{subfigure}[b]{0.4\textwidth}
        \centering
        \includegraphics[trim = {10cm 0 11cm 0}, clip, width = \textwidth]{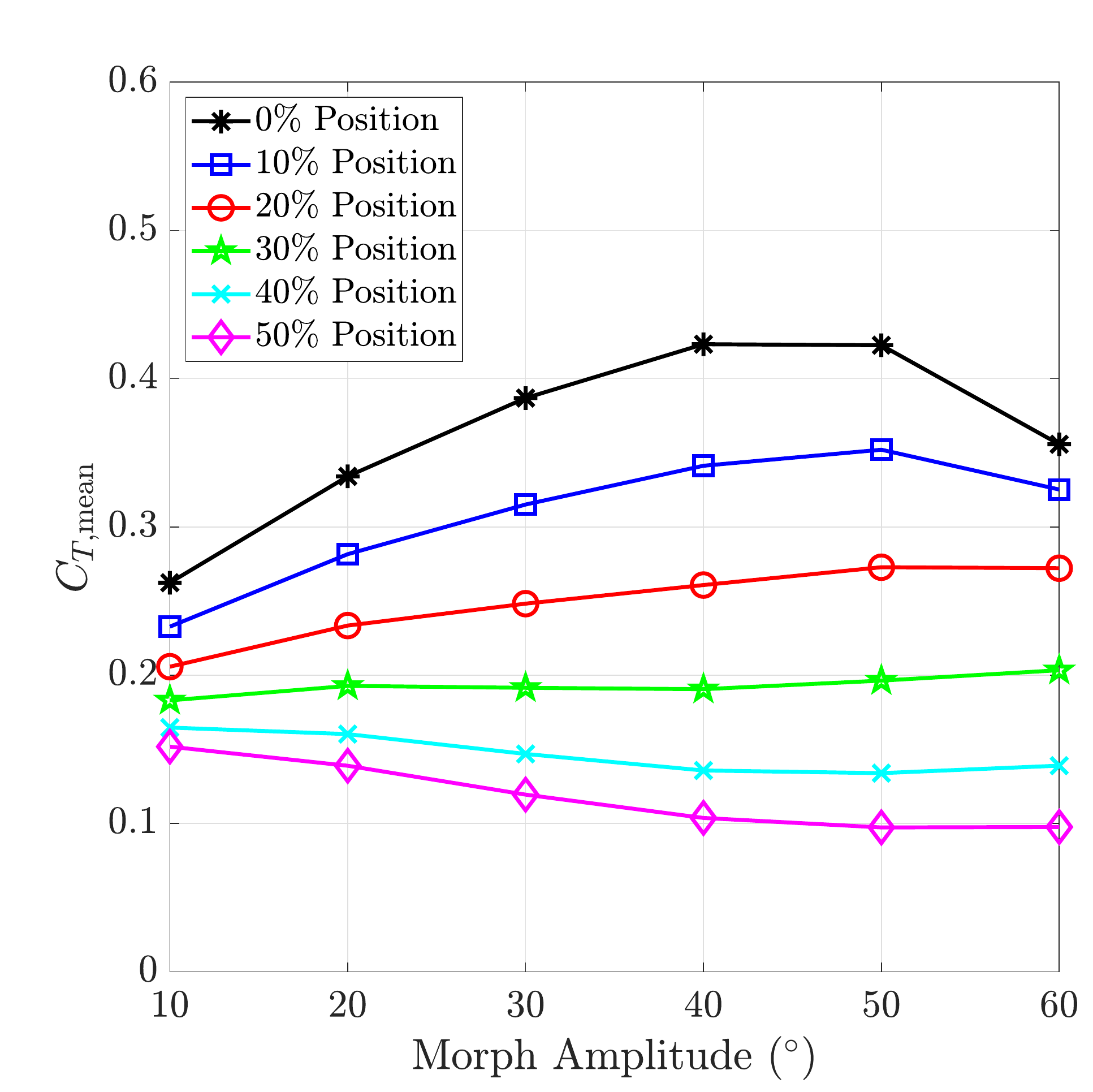}
        \caption{}
    \end{subfigure}
    
    \begin{subfigure}[b]{0.4\textwidth}
        \centering
        \includegraphics[trim = {10cm 0 11cm 0}, clip, width = \textwidth]{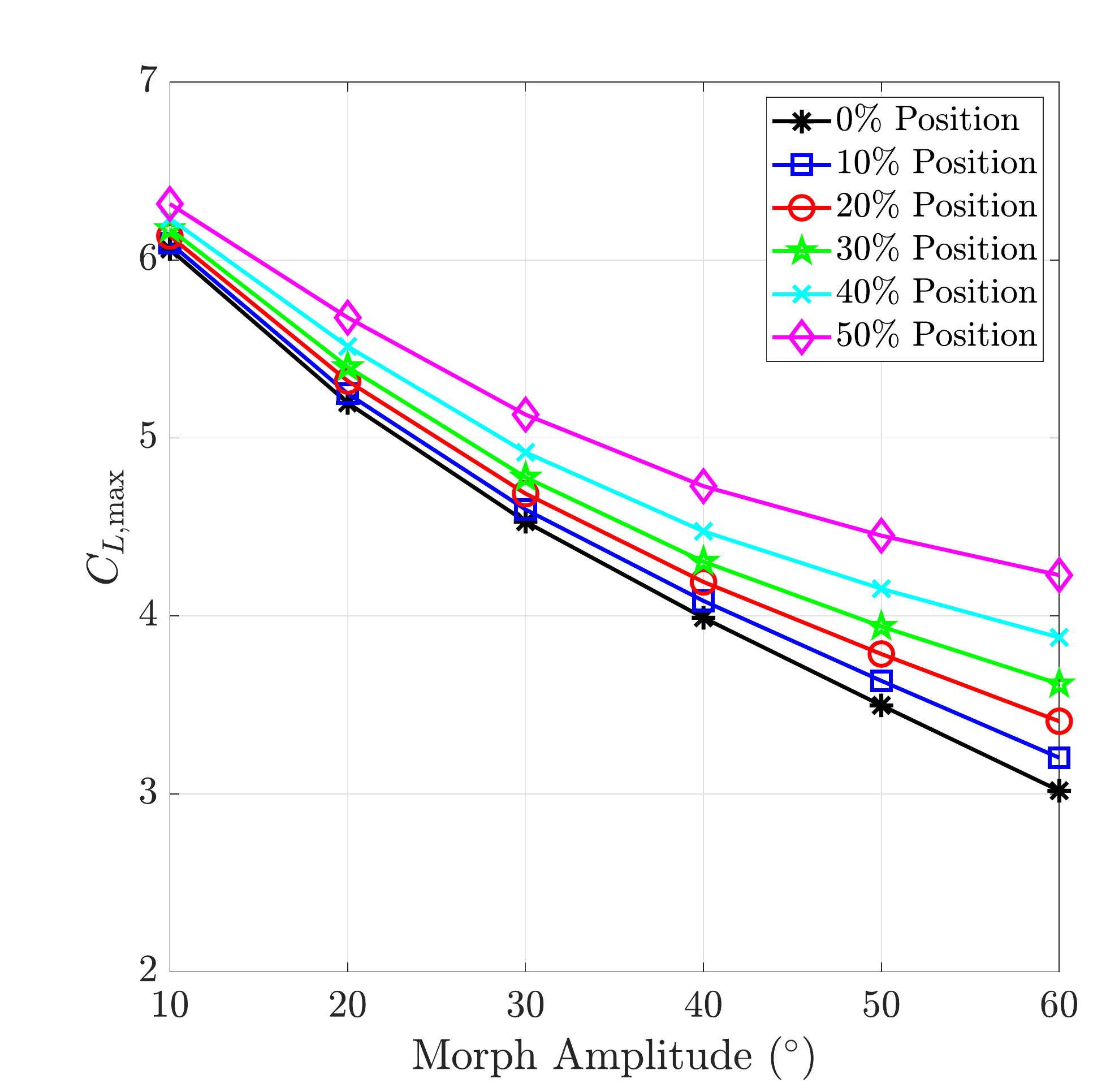}
        \caption{}
    \end{subfigure}
    
    \caption{VARIATION IN THE FORCE COEFFICIENTS WITH MORPH AMPLITUDE AND POSITION: (a) \(C_{T,\mathrm{mean}}\), AND (b) \(C_{L,\mathrm{max}}\)}
    \label{CT_CL_Charts}
\end{figure}

The thrust and lift coefficients characterize the flow dynamics of the morphing and heaving foil. The time-averaged thrust coefficient $C_{T,\mathrm{mean}}$ over a morphing cycle is shown in Fig. \ref{CT_CL_Charts}(a). For a fixed morph amplitude, $C_{T,\mathrm{mean}}$ is observed to decrease as the morph location tends towards the trailing edge of the foil. On the other hand, if the morph position is fixed, the variation in the mean thrust with the morph amplitude is quite interesting. For the positions in proximity to the leading edge of the foil (0\% - 20\%), the thrust is noticed to increase with the morph amplitude till $\theta_0 = 50^{\circ}$. The thrust then slightly decreases at $\theta_0 = 60^{\circ}$. As the morph location is increased towards the trailing edge, the averaged thrust coefficient is observed to decrease with the morph amplitude.

The maximum lift coefficient over a time period is depicted in Fig. \ref{CT_CL_Charts}(b). As the morph position increases from the leading edge towards mid-chord length, the maximum lift is noted to increase for a fixed morph amplitude. Moreover, $C_{L,\mathrm{max}}$ decreases monotonically with the morph amplitude for a fixed morph position. However, the amount of decrease in $C_{L,\mathrm{max}}$ reduces as the morph location tends towards the trailing edge.
\begin{figure}
    \centering
    \begin{subfigure}[b]{0.4\textwidth}
        \centering
        \includegraphics[trim = {10cm 0 11cm 0}, clip, width = \textwidth]{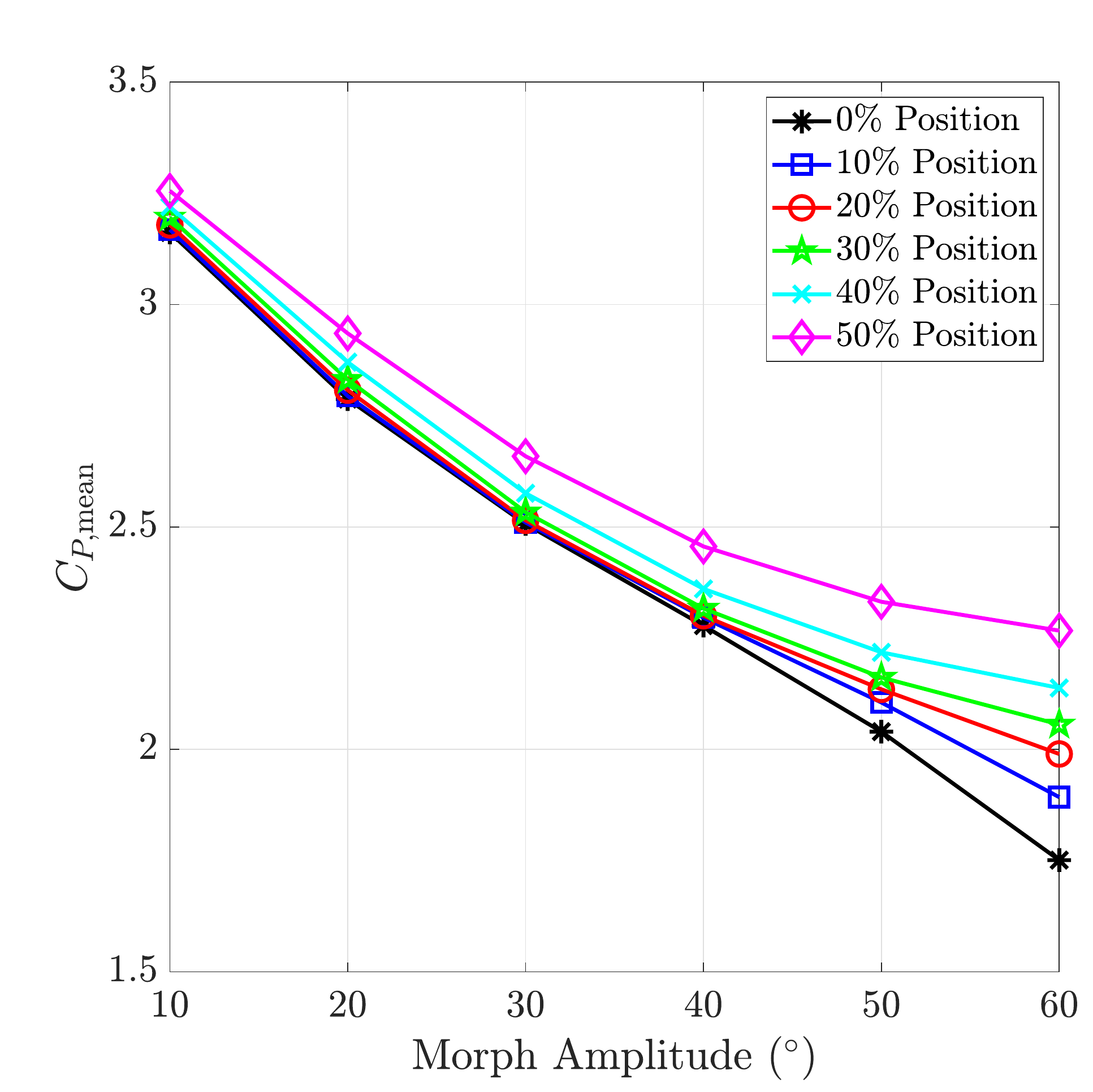}
        \caption{}
    \end{subfigure}
    
    \begin{subfigure}[b]{0.4\textwidth}
        \centering
        \includegraphics[trim = {10cm 0 11cm 1cm}, clip, width = \textwidth]{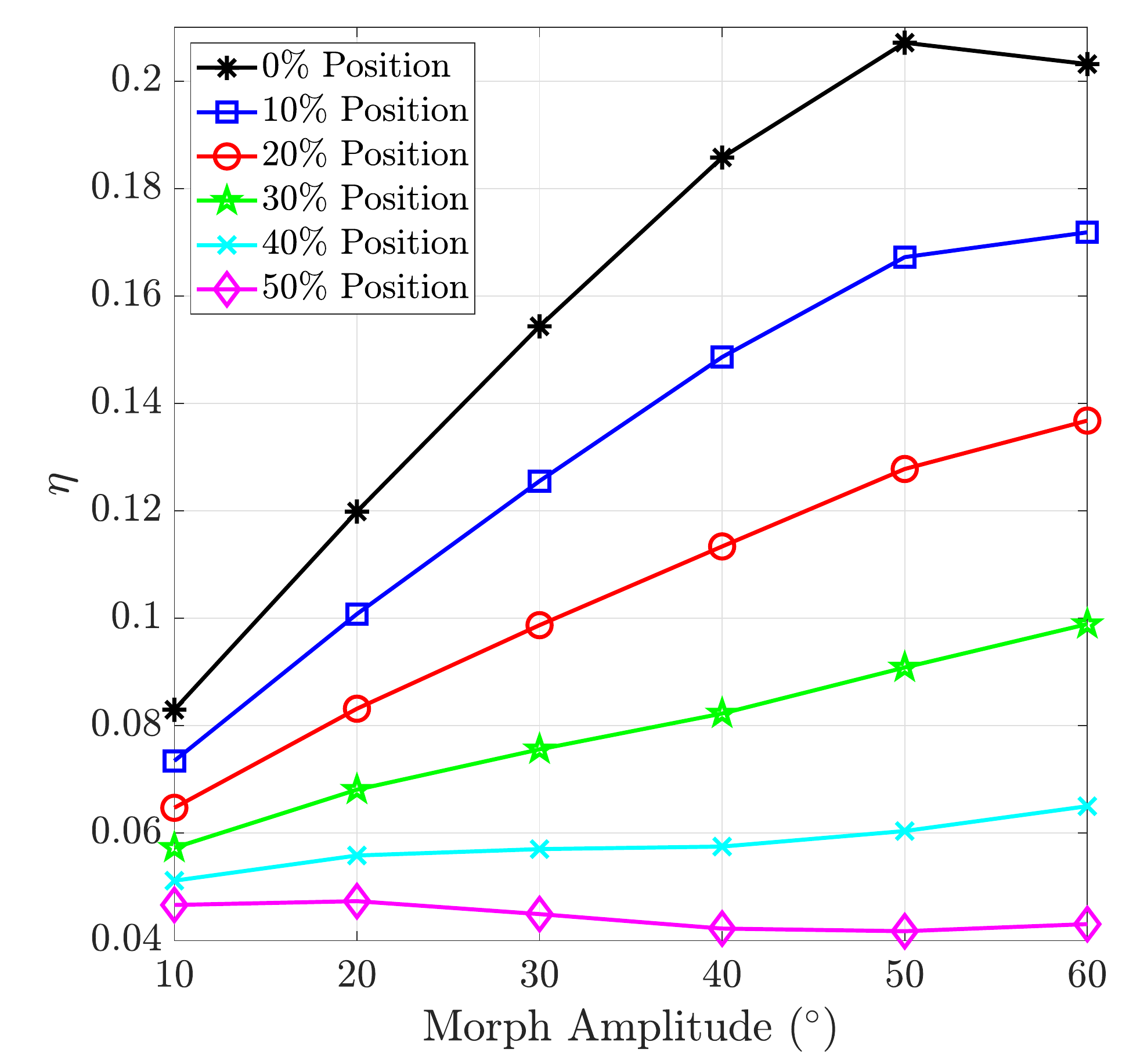}
        \caption{}
    \end{subfigure}
    \caption{VARIATION OF (a) MEAN POWER COEFFICIENT $C_{P,\mathrm{mean}}$ AND (b) PROPULSIVE EFFICIENCY $\eta$, WITH THE MORPH AMPLITUDE AND POSITION}
    \label{CP_eta_Charts}
\end{figure}
The mean power coefficient which represents the input power to the morphing foil is shown in Fig. \ref{CP_eta_Charts}(a). It is observed that the coefficient decreases smoothly with increase in morph amplitude for a fixed morph position. The combined trends of the mean thrust and power coefficients lead to the propulsive efficiency, which is plotted in Fig. \ref{CP_eta_Charts}(b). For positions near the leading edge (0\% - 20\%), the efficiency increases with morph amplitude. This is a consequence of the fact that the averaged thrust also increases and the mean power decreases. At the position of 30\%, the mean thrust remains the same with morph amplitude, but power coefficient decreases, resulting in an increasing efficiency. For the starting morphing locations of 40\% and 50\%, the change in efficiency is not that profound.

\begin{figure*}[!htbp]
        \centering
        \begin{subfigure}[b]{0.32\textwidth}
            \centering
            \includegraphics[trim = {10cm 0 11cm 1cm}, clip, width = \textwidth]{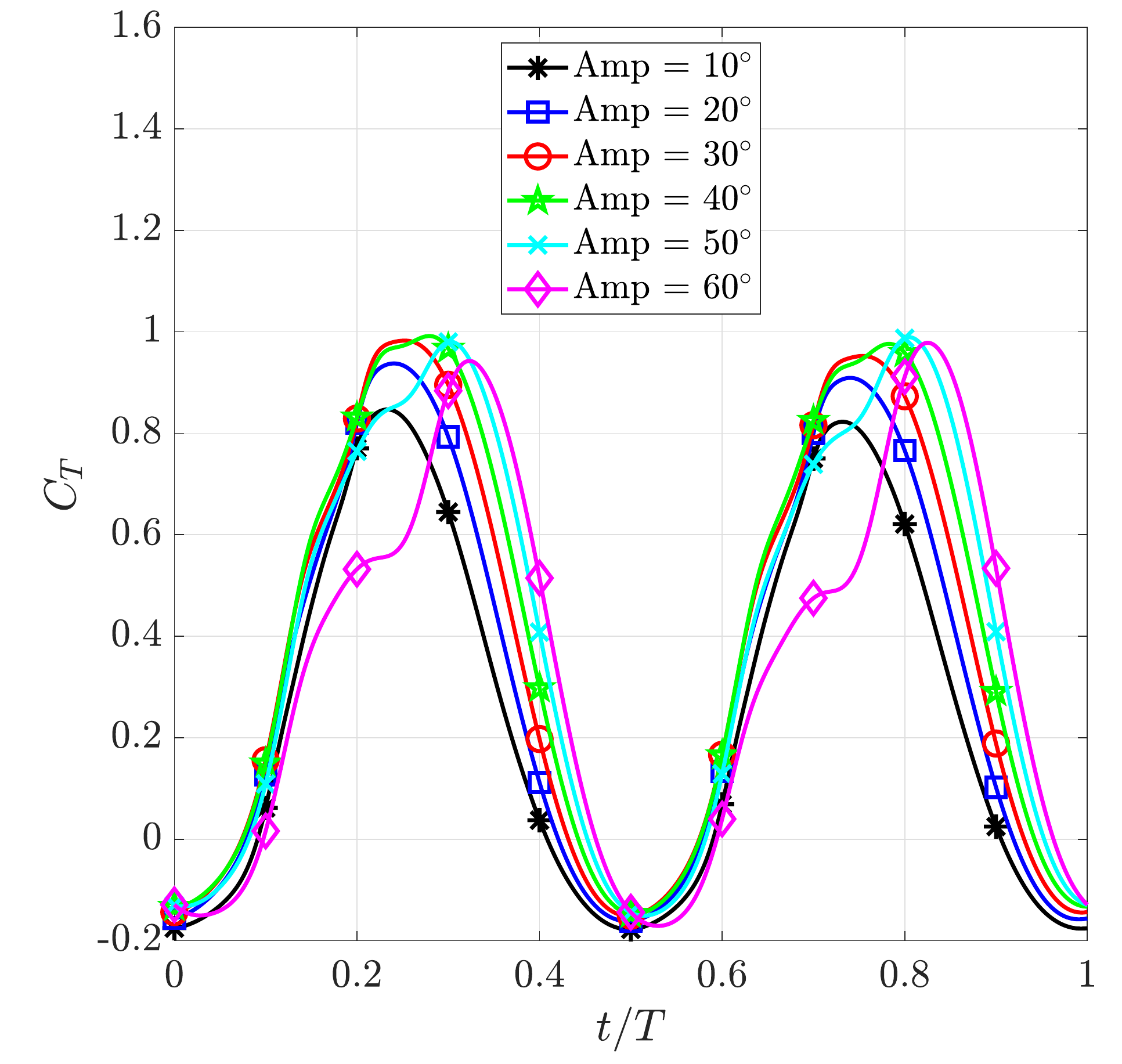}
            \caption{}
        \end{subfigure}%
        \begin{subfigure}[b]{0.32\textwidth}
            \centering
            \includegraphics[trim = {10cm 0 11cm 1cm}, clip, width = \textwidth]{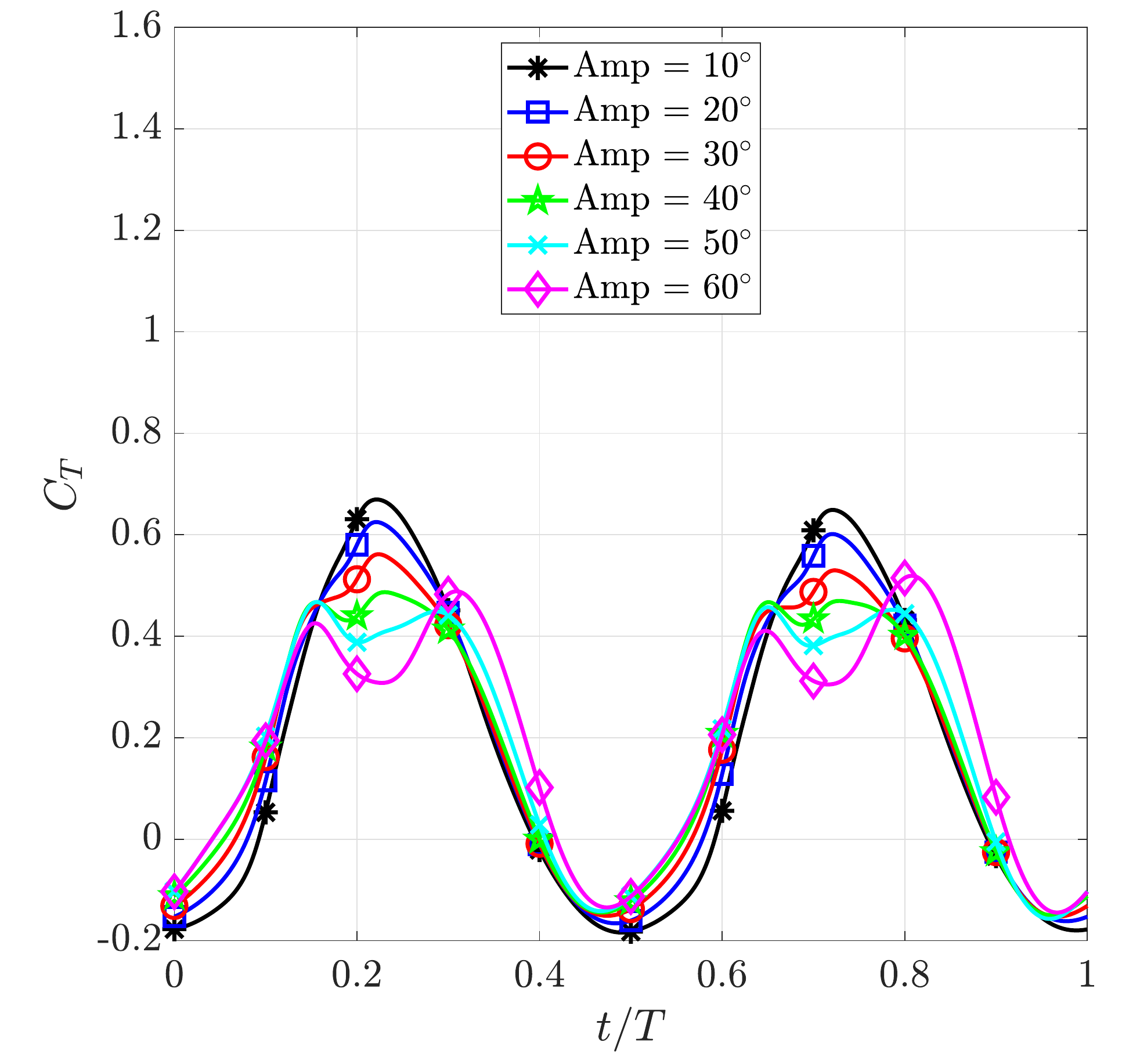}
            \caption{}
        \end{subfigure}%
        \begin{subfigure}[b]{0.32\textwidth}
            \centering
            \includegraphics[trim = {10cm 0 11cm 1cm}, clip, width = \textwidth]{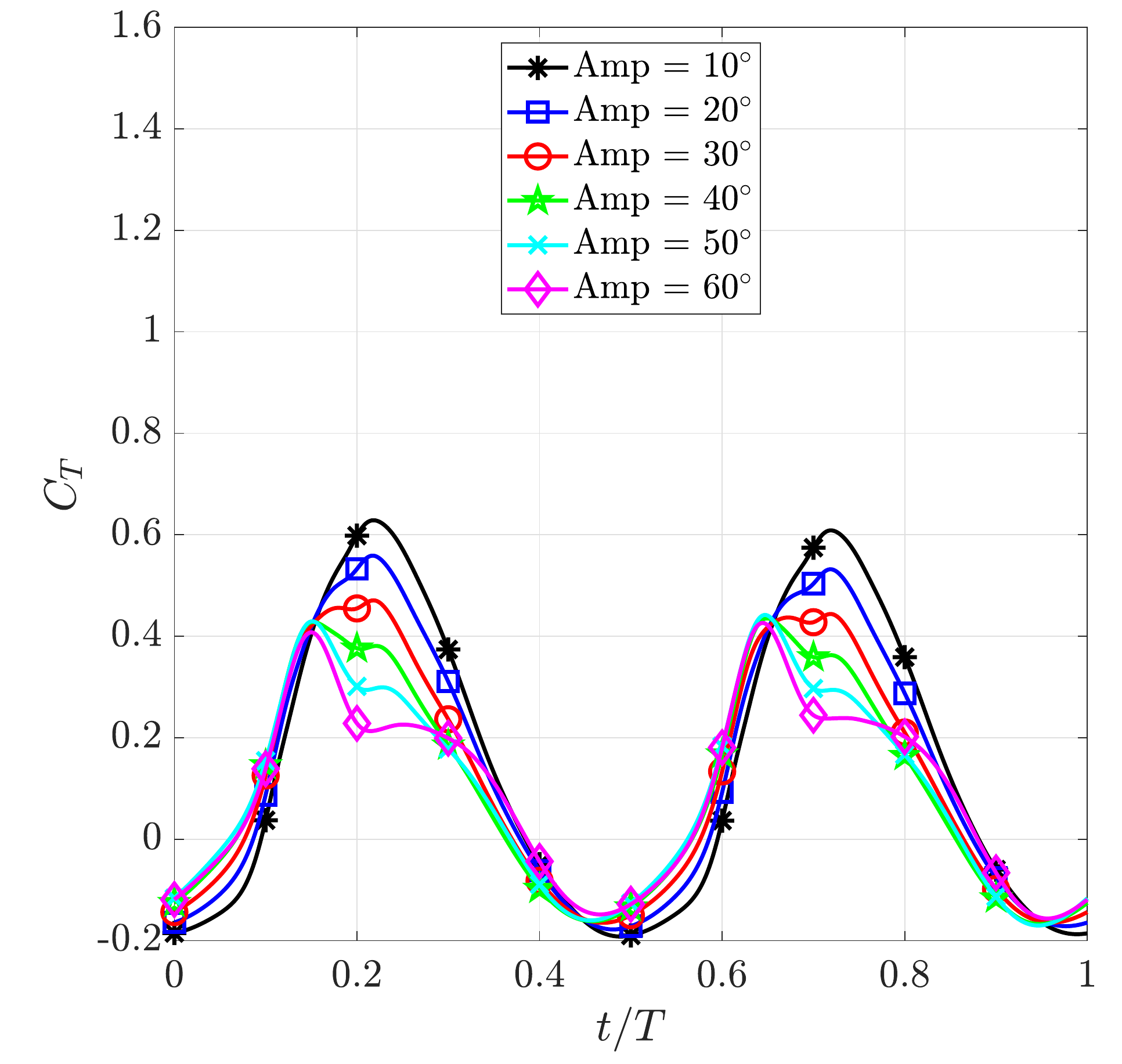}
            \caption{}
        \end{subfigure}
        \caption{VARIATION IN THE THRUST COEFFICIENT $C_T$ WITH TIME FOR MORPH POSITION OF (a) 0\%, (b) 30\% AND (c) 50\%}
        \label{CT_vs_time_amp}
 \end{figure*}
\begin{figure*}[!htbp]
        \centering
        \begin{subfigure}[b]{0.32\textwidth}
            \centering
            \includegraphics[trim = {10cm 0 11cm 1cm}, clip, width = \textwidth]{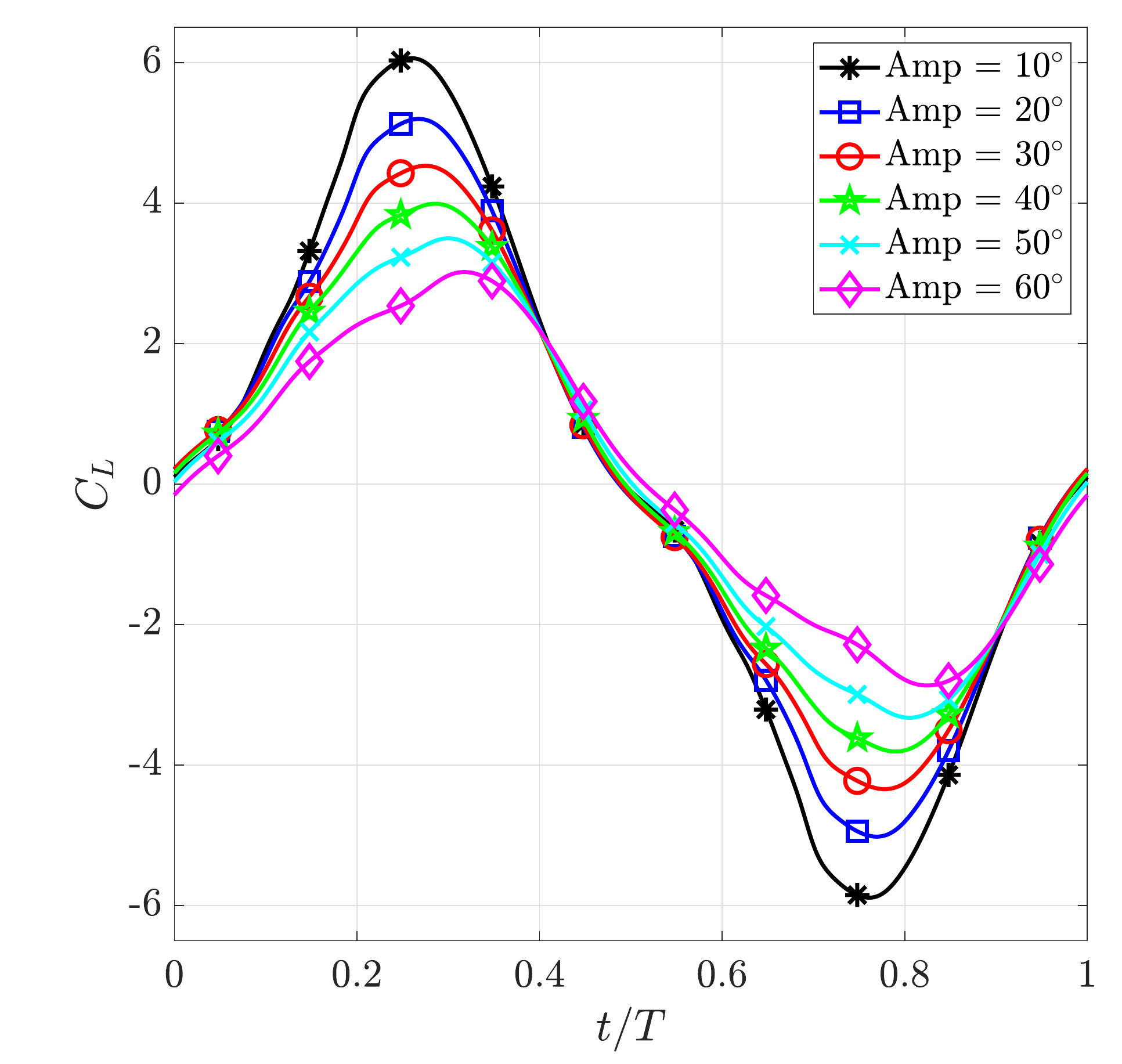}
            \caption{}
        \end{subfigure}%
        \begin{subfigure}[b]{0.32\textwidth}
            \centering
            \includegraphics[trim = {10cm 0 11cm 1cm}, clip, width = \textwidth]{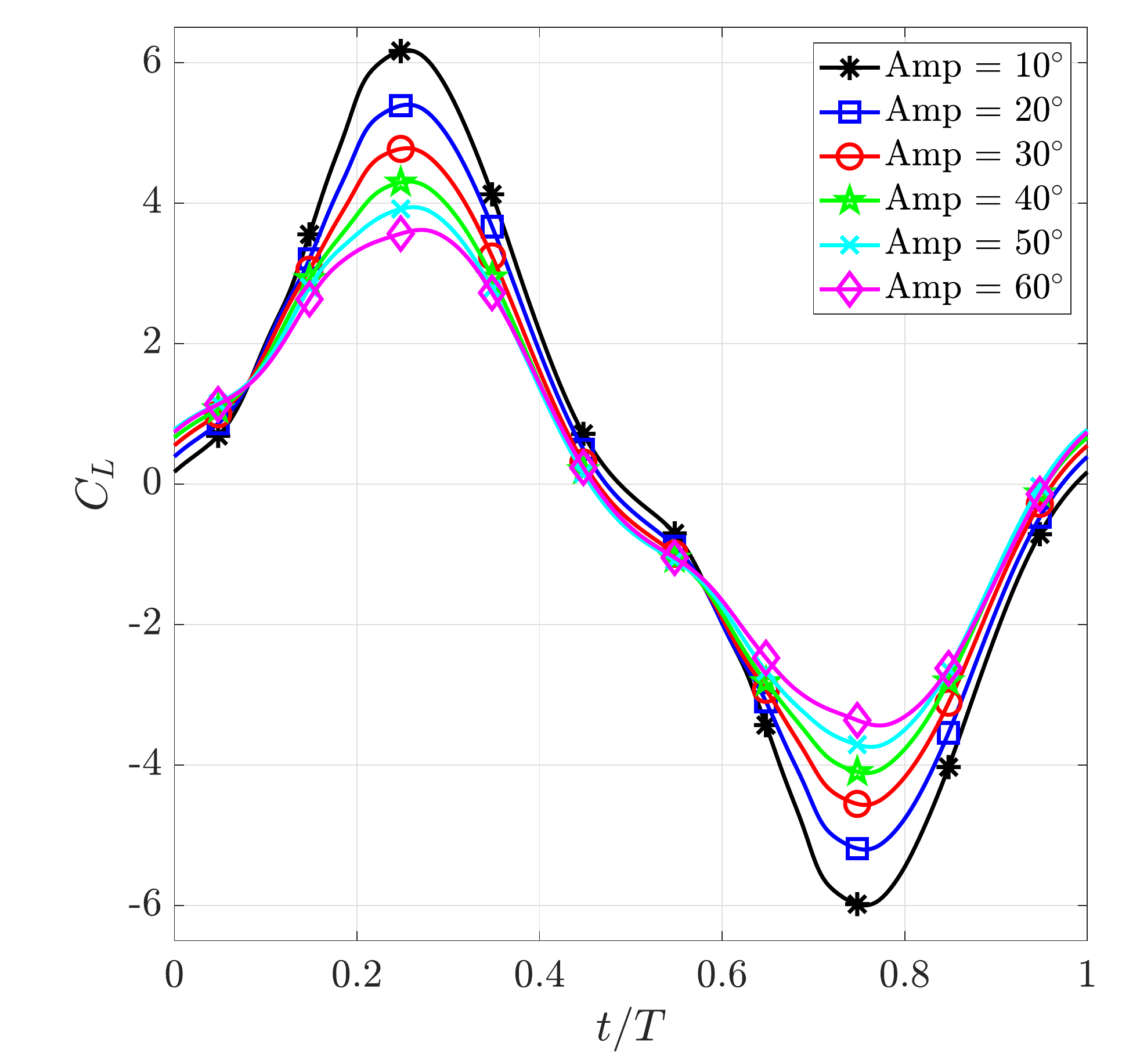}
            \caption{}
        \end{subfigure}%
        \begin{subfigure}[b]{0.32\textwidth}
            \centering
            \includegraphics[trim = {10cm 0 11cm 1cm}, clip, width = \textwidth]{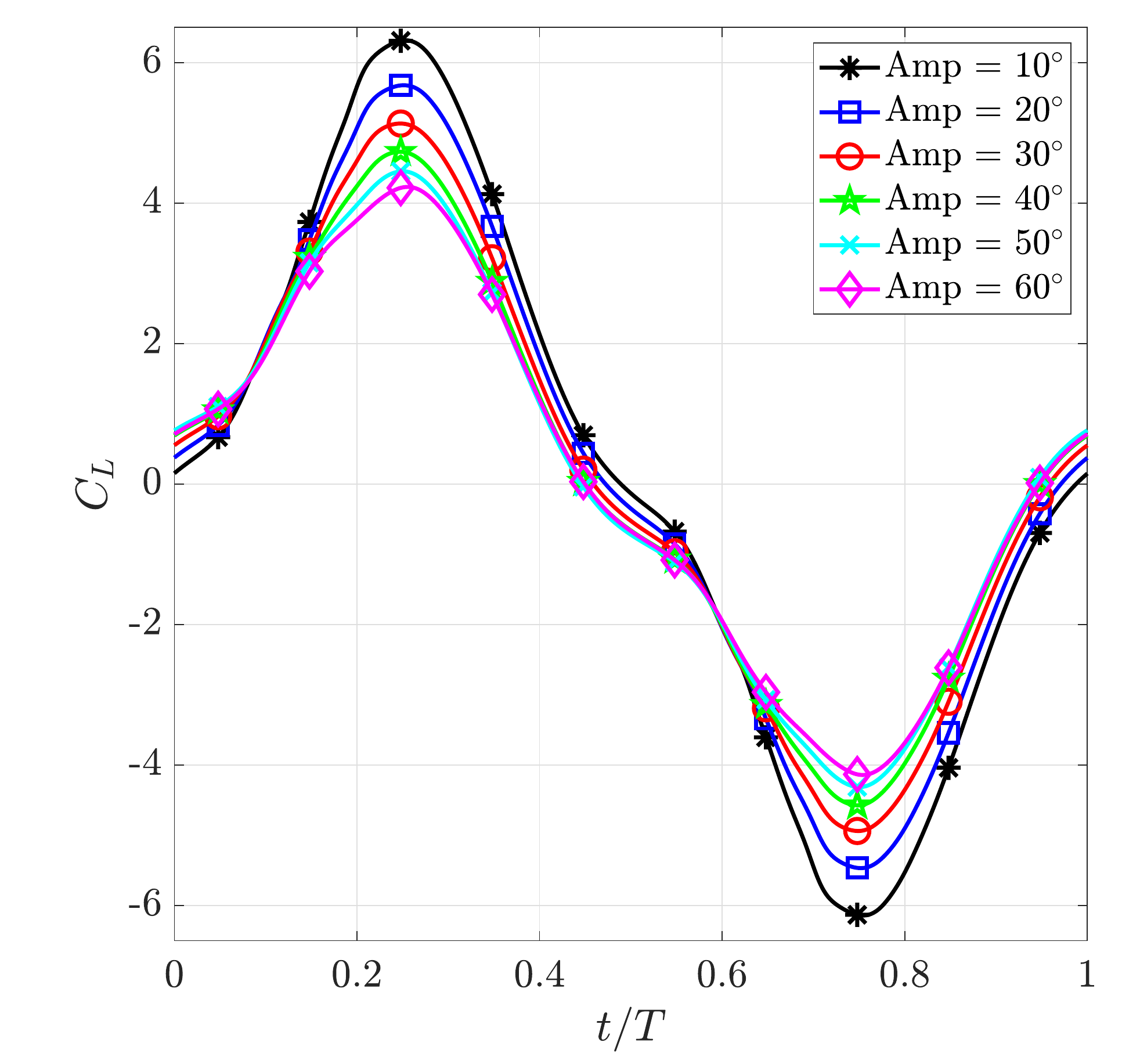}
            \caption{}
        \end{subfigure}
        \caption{VARIATION IN THE LIFT COEFFICIENT $C_L$ WITH TIME FOR MORPH POSITION OF (a) 0\%, (b) 30\% AND (c) 50\%}
        \label{CL_vs_time_amp}
 \end{figure*}

\underline{\textbf{Effect of Morph Amplitude}}:
To understand the dynamically changing force coefficients of the morphing and heaving foil, we plot the temporal variation in the thrust and lift coefficients at various representative morph positions in Figs. \ref{CT_vs_time_amp} and \ref{CL_vs_time_amp}, respectively. We will just focus on the downstroke ($t/T = [0, 0.5]$) of a cycle as the characteristics repeat during the upstroke. At 0\% morph position (Fig. \ref{CT_vs_time_amp}(a)), an increase in the maximum thrust coefficient is observed around quarter time period $t/T \approx 0.25$ with the morph amplitude till $50^{\circ}$. 
It is observed that the peak of the thrust gets delayed as the amplitude increases, with a prominent peak observed around $t/T \approx 0.35$ for $60^{\circ}$ morph amplitude. Furthermore, this peak spans for a lesser time compared to the other amplitude values. This leads to the decrease in $C_{T,\mathrm{mean}}$ at $60^{\circ}$ in Fig. \ref{CT_CL_Charts}(a). For the morph position of 30\% (Fig. \ref{CT_vs_time_amp}(b)), the peak of the thrust is noted to be divided into two peaks surrounding the quarter time period instant with increase in morph amplitude. Therefore, the averaged thrust coefficient is almost the same with change in the amplitude. With further increase in the morph position to 50\% shown in Fig. \ref{CT_vs_time_amp}(c), an increase in the morph amplitude shifts the peak before the quarter time period and the maximum value reduces, resulting in lower mean thrust values, as noticed in Fig. \ref{CT_CL_Charts}(a). 
 
The temporal variation of the lift coefficient does not indicate a significant difference across the various morph positions, depicted in Fig. \ref{CL_vs_time_amp}. Although, for a fixed morph position, the maximum lift coefficient decreases with the increase in amplitude. There is found to be a delay in the peak of the lift coefficient with increasing amplitude for morph positions near the leading edge (0\% in Fig. \ref{CL_vs_time_amp}(a)). To investigate the lift coefficient variation further, we evaluate the effective angle of attack for the morphing and heaving foil. It is the angle of the incoming flow as observed from the moving foil. It can be written as
\begin{align}
    \alpha_e(t) = \mathrm{tan}^{-1}\bigg( - \frac{v_{y,\mathrm{heave}}(t)}{U_{\infty}} \bigg) - \alpha(t),
\end{align}
where $v_{y,\mathrm{heave}}(t) = 2\pi f h_0 \mathrm{cos}(2\pi f t + \phi_h)$ is the heave translational velocity of the foil and $\alpha(t)$ is the angle of inclination of the trailing edge with respect to the leading edge, which can be computed as
\begin{align}
    \alpha(t) &= \mathrm{tan}^{-1}\bigg( \frac{y_{TE} - y_{LE}}{x_{TE} - x_{LE}} \bigg).
\end{align}
Here, $(x_{LE}, y_{LE})$ and $(x_{TE}, y_{TE})$ denote the coordinates of the leading and trailing edge of the moving foil, respectively. 
\begin{figure*}[!htbp]
    \centering
    \begin{subfigure}[b]{0.32\textwidth}
        \centering
        \includegraphics[trim = {10cm 0 10.5cm 1cm}, clip, width = \textwidth]{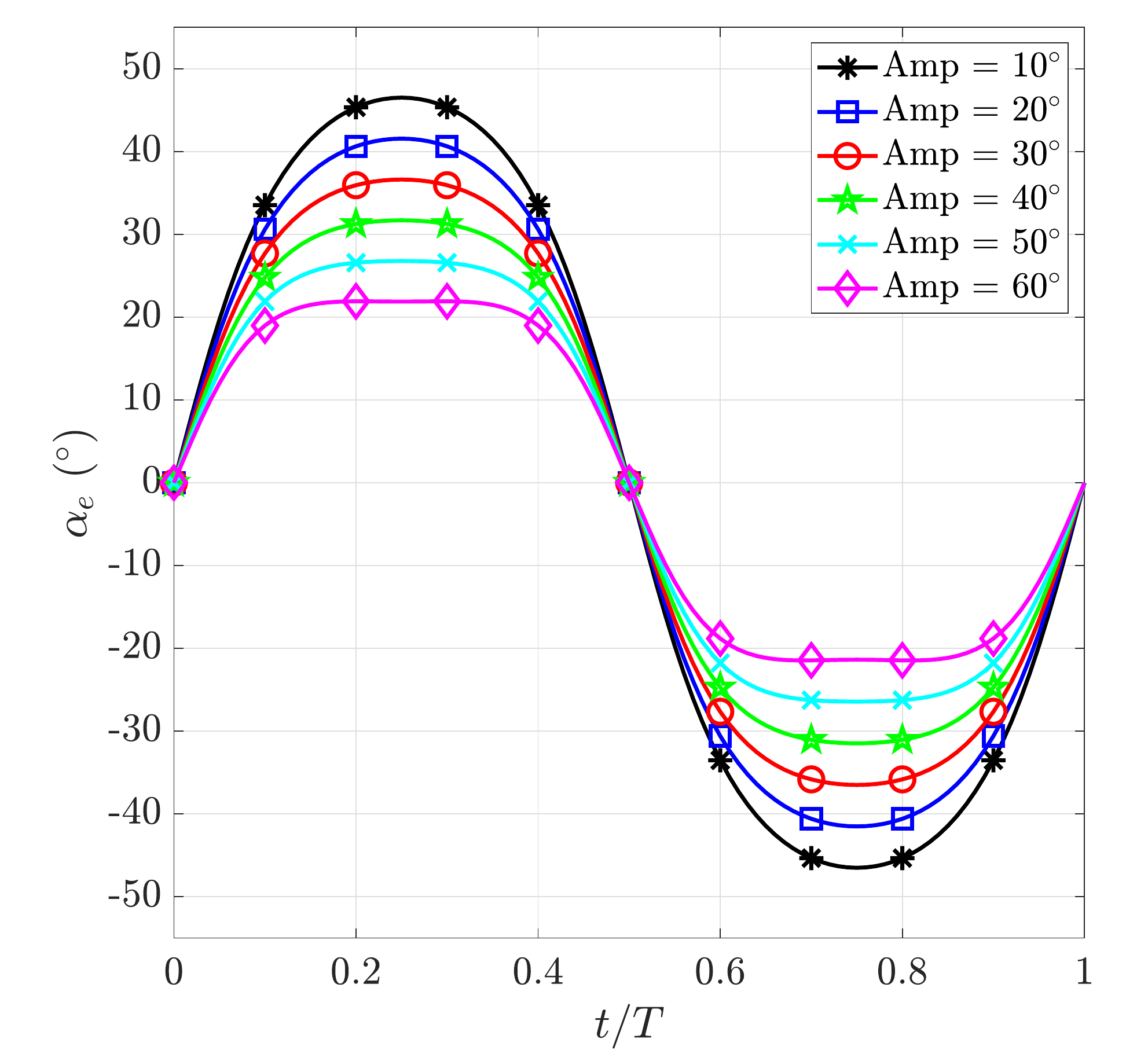}
        \caption{}
    \end{subfigure}%
    \begin{subfigure}[b]{0.32\textwidth}
        \centering
        \includegraphics[trim = {10cm 0 10.5cm 1cm}, clip, width = \textwidth]{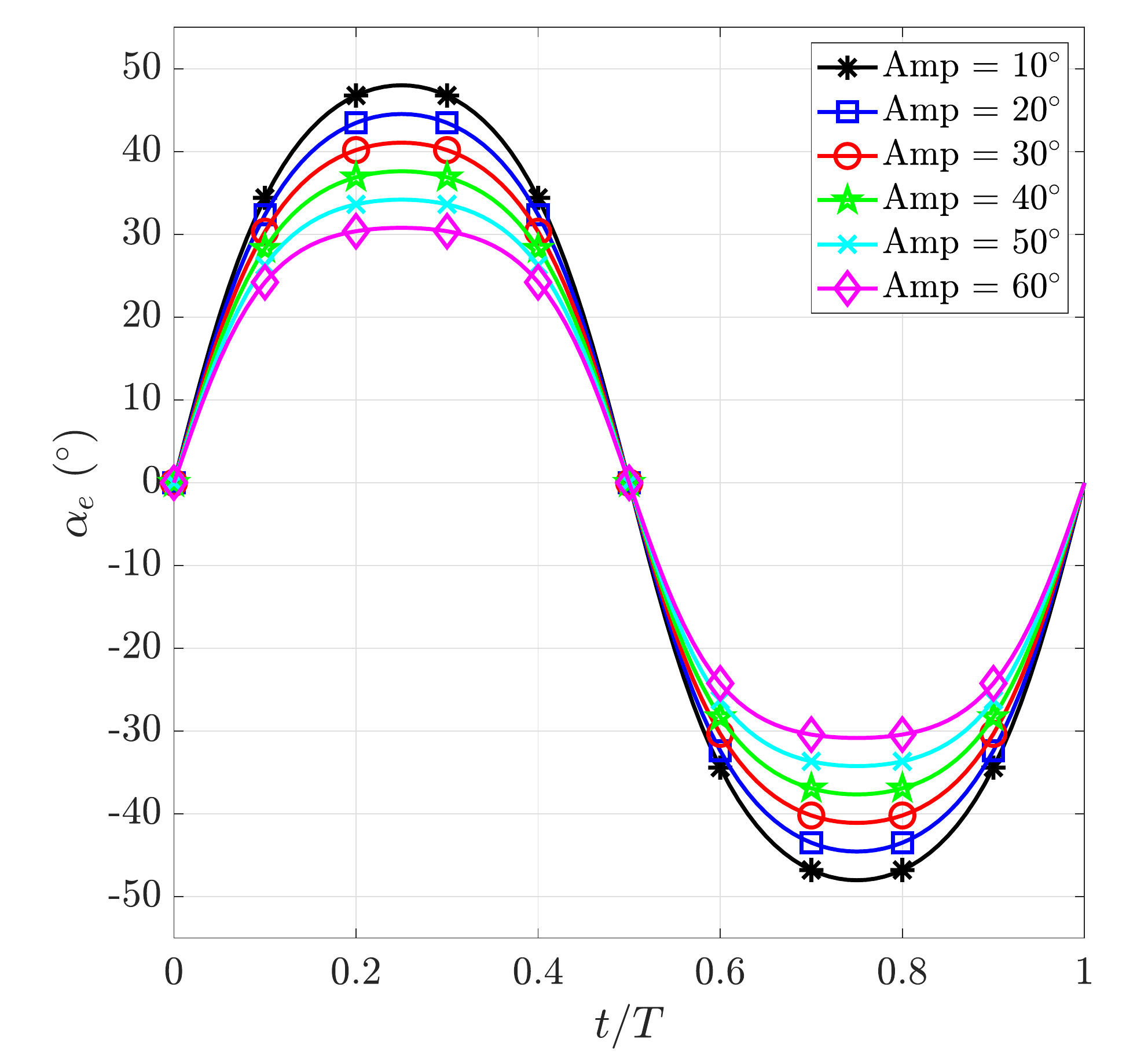}
        \caption{}
    \end{subfigure}
    \begin{subfigure}[b]{0.32\textwidth}
        \centering
        \includegraphics[trim = {10cm 0 10.5cm 1cm}, clip, width = \textwidth]{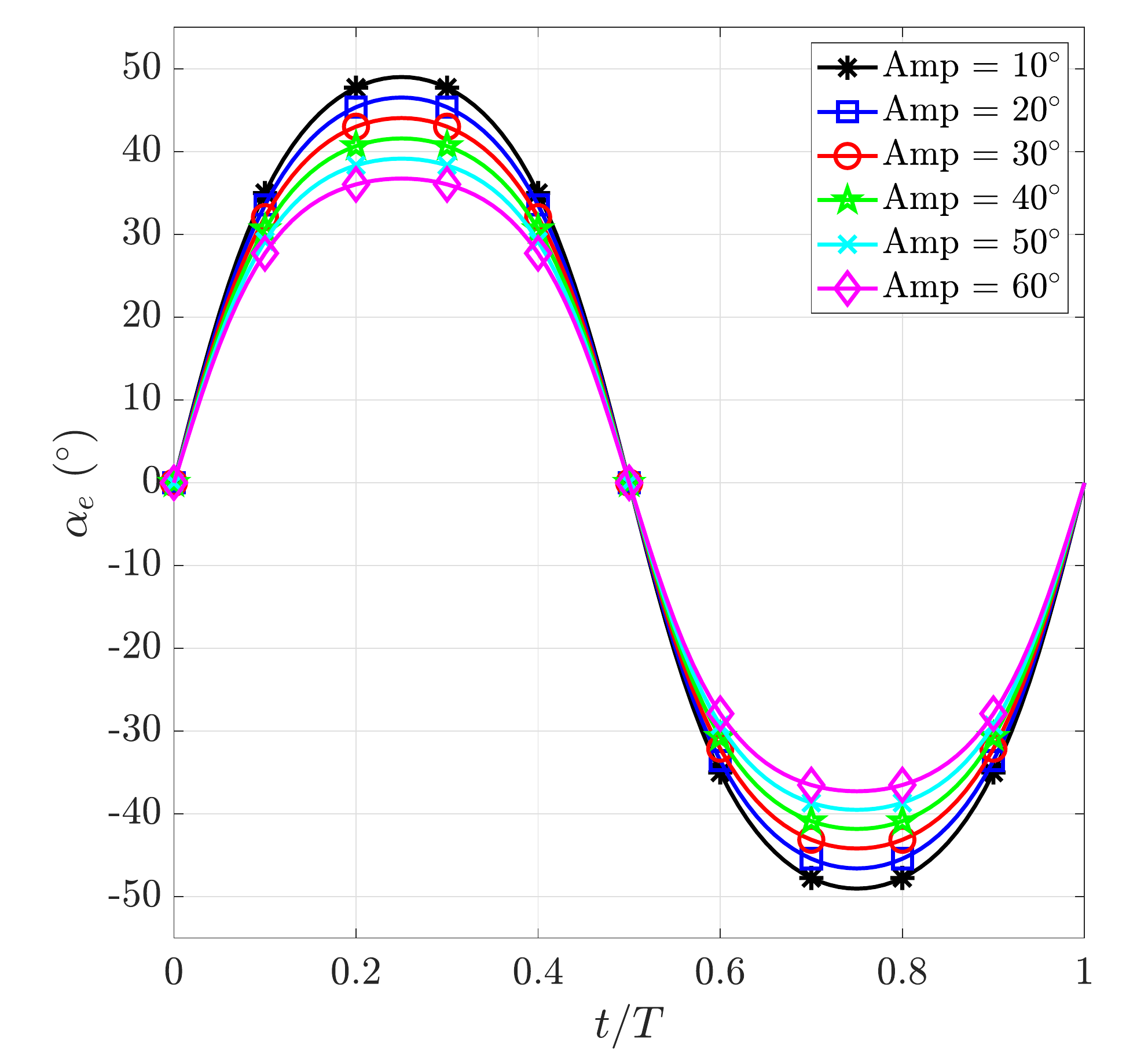}
        \caption{}
    \end{subfigure}
    \caption{VARIATION OF \(\alpha_e\) IN A CYCLE FOR A CONSTANT MORPH POSITION OF: (a) 0\%, (b) 30\%, AND (c) 50\%}
    \label{eAoA_plots_amp}
\end{figure*}

The temporal variation of the effective angle of attack in a time period for various morph positions is shown in Fig. \ref{eAoA_plots_amp}. It is observed that the effective angle of attack increases during the downstroke of the foil, reaches maximum around the quarter time period and then decreases to negative values as the upstroke progresses. We can see that the maximum $\alpha_e$ decreases with increasing amplitude for all the morph positions. This corroborates the findings in the variation of lift coefficient. 

\begin{figure*}[!htbp]
        \centering
        \begin{subfigure}[b]{0.32\textwidth}
            \centering
            \includegraphics[trim = {10cm 0 11cm 1cm}, clip, width = \textwidth]{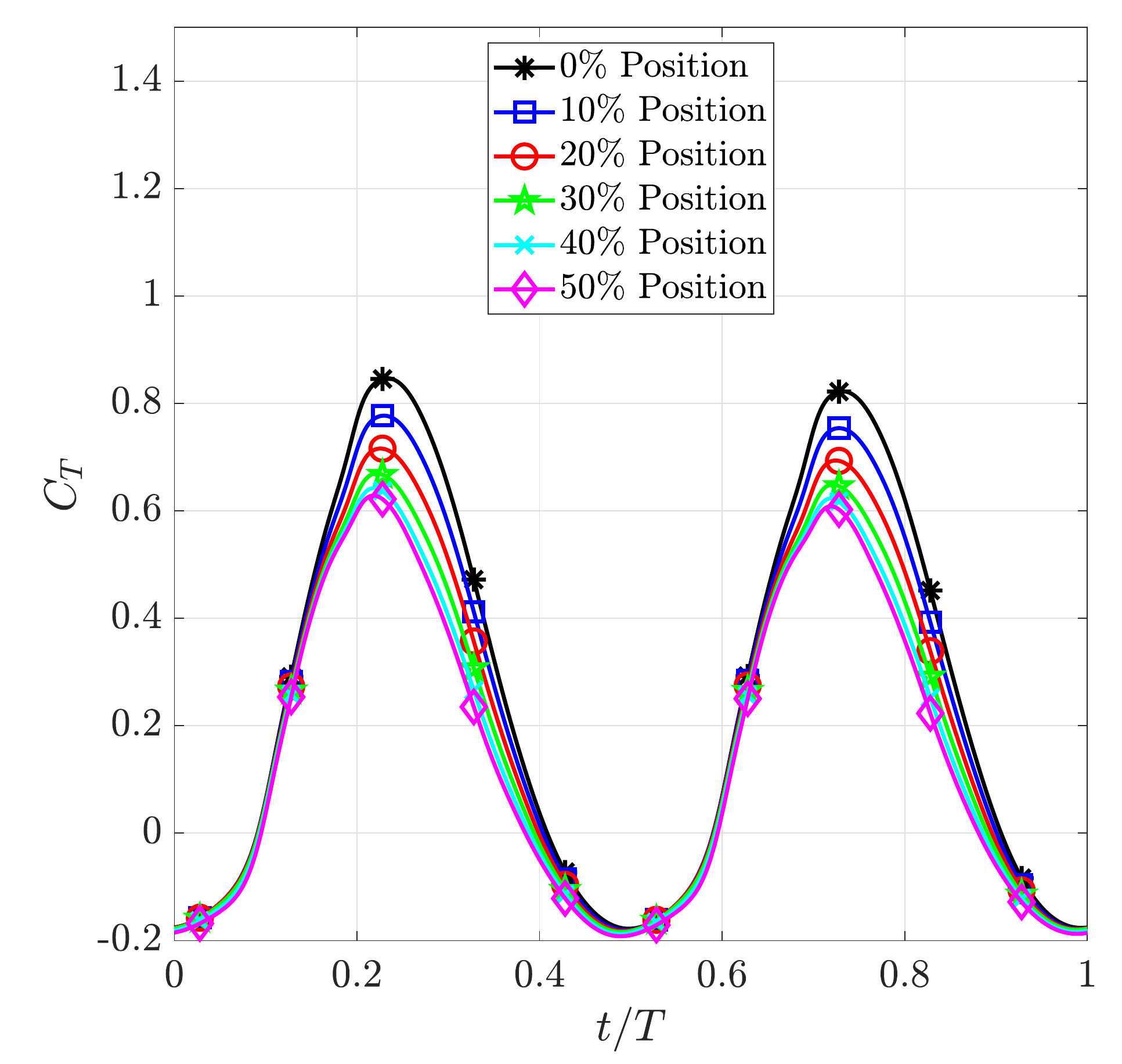}
            \caption{}
        \end{subfigure}%
        \begin{subfigure}[b]{0.32\textwidth}
            \centering
            \includegraphics[trim = {10cm 0 11cm 1cm}, clip, width = \textwidth]{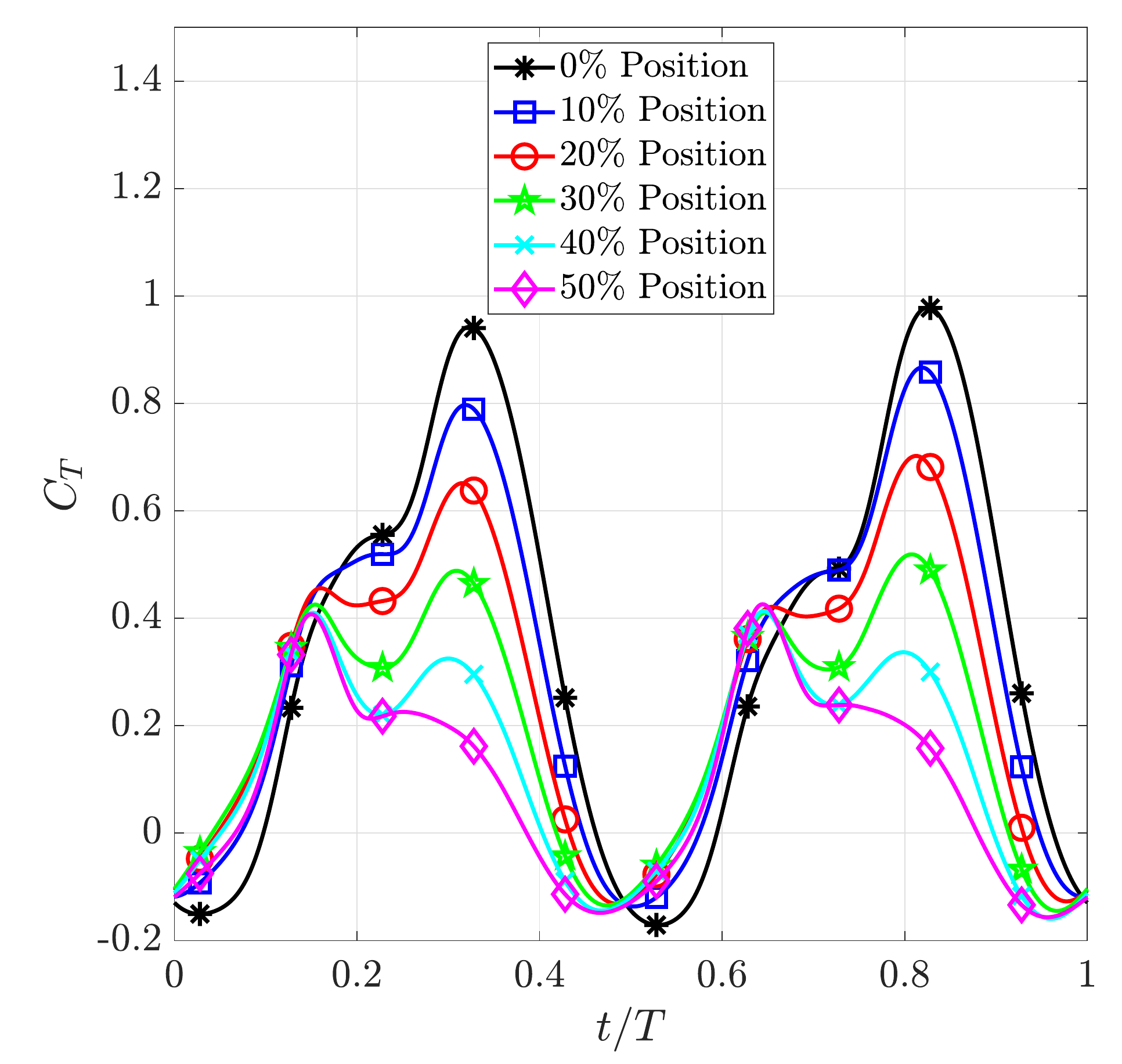}
            \caption{}
        \end{subfigure}%
        \caption{VARIATION IN THE THRUST COEFFICIENT $C_T$ WITH TIME FOR MORPH AMPLITUDE OF (a) $10^{\circ}$, AND (b) $60^{\circ}$}
        \label{CT_vs_time_pos}
 \end{figure*}

\begin{figure*}[!htbp]
        \centering
        \begin{subfigure}[b]{0.32\textwidth}
            \centering
            \includegraphics[trim = {10cm 0 11cm 1cm}, clip, width = \textwidth]{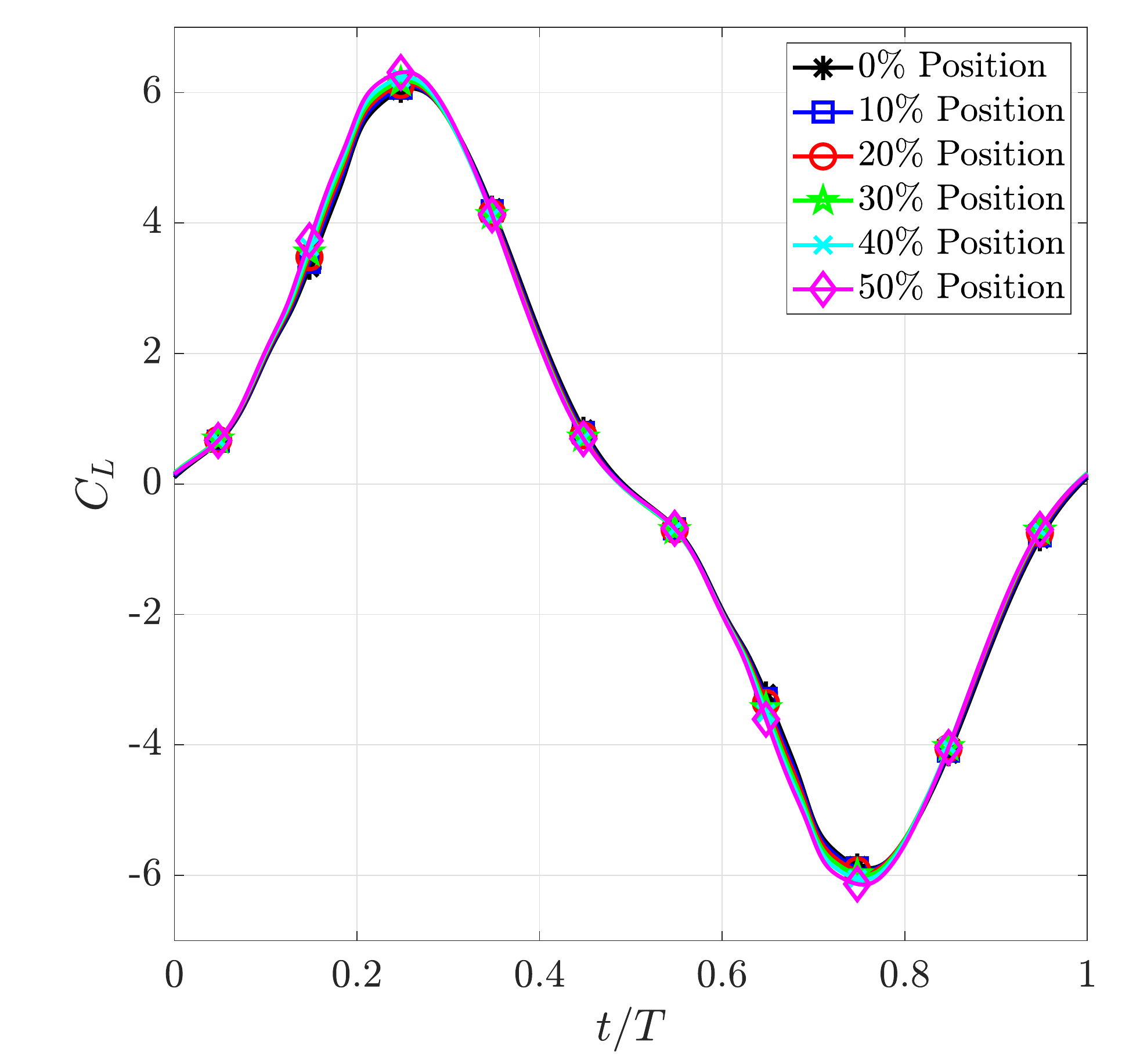}
            \caption{}
        \end{subfigure}%
        \begin{subfigure}[b]{0.32\textwidth}
            \centering
            \includegraphics[trim = {10cm 0 11cm 1cm}, clip, width = \textwidth]{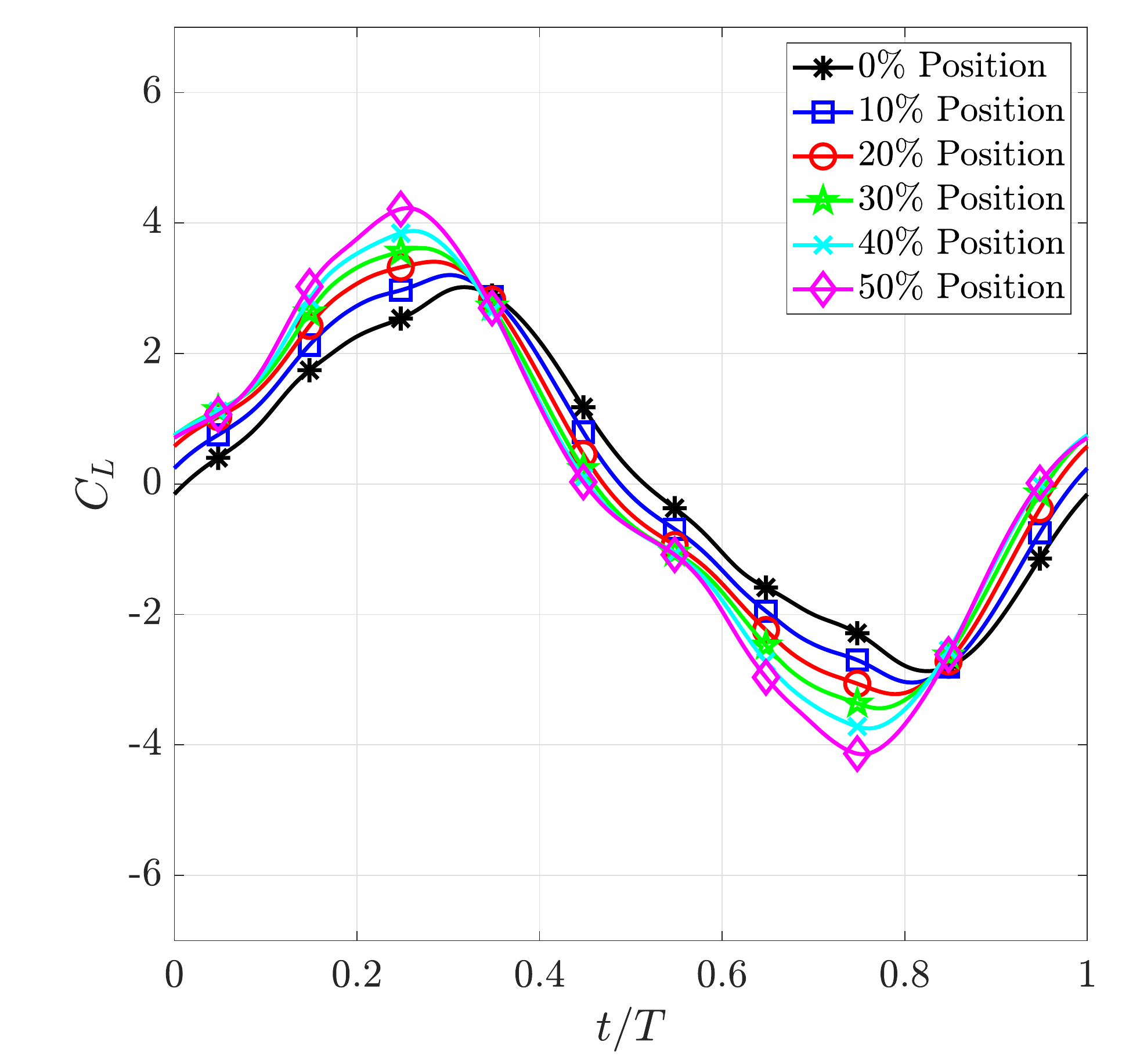}
            \caption{}
        \end{subfigure}%
        \caption{VARIATION IN THE LIFT COEFFICIENT $C_L$ WITH TIME FOR MORPH AMPLITUDE OF (a) $10^{\circ}$, AND (b) $60^{\circ}$}
        \label{CL_vs_time_pos}
 \end{figure*}

\underline{\textbf{Effect of Morph Position}}:
The temporal variation of thrust and lift coefficients in a time period for two representative amplitudes of $10^{\circ}$ and $60^{\circ}$ for various morph positions is shown in Figs. \ref{CT_vs_time_pos} and \ref{CL_vs_time_pos}, respectively. At the lower morph amplitude of $10^{\circ}$, the maximum thrust coefficient in a cycle decreases with an increase in the morph position (towards the trailing edge). The peak of the thrust coefficient is near the quarter time period for all the morph positions. This decrease in the peak of thrust gives lower averaged propulsion as the morph position is increased (Fig. \ref{CT_CL_Charts}(a)). For the higher value of morph amplitude of $60^{\circ}$, a similar observation can be made. With the increase in the morph position, the peak values of the thrust coefficient decrease, resulting in lower mean thrust.

\begin{figure*}[h]
    \centering
    \begin{subfigure}[b]{0.32\textwidth}
        \centering
        \includegraphics[trim = {10cm 0 11cm 1cm}, clip, width = \textwidth]{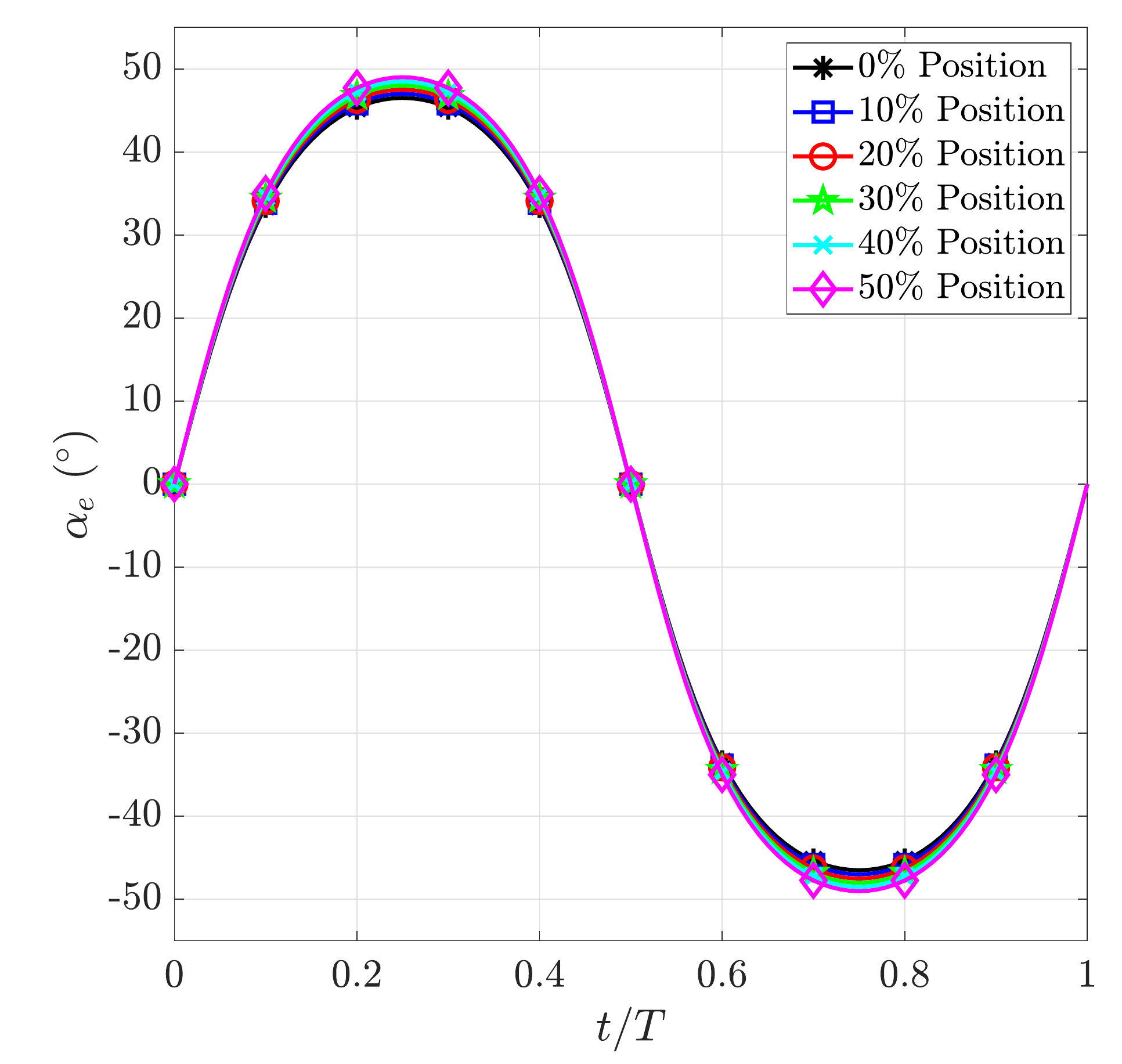}
        \caption{}
    \end{subfigure}%
    \begin{subfigure}[b]{0.32\textwidth}
        \centering
        \includegraphics[trim = {10cm 0 11cm 1cm}, clip, width = \textwidth]{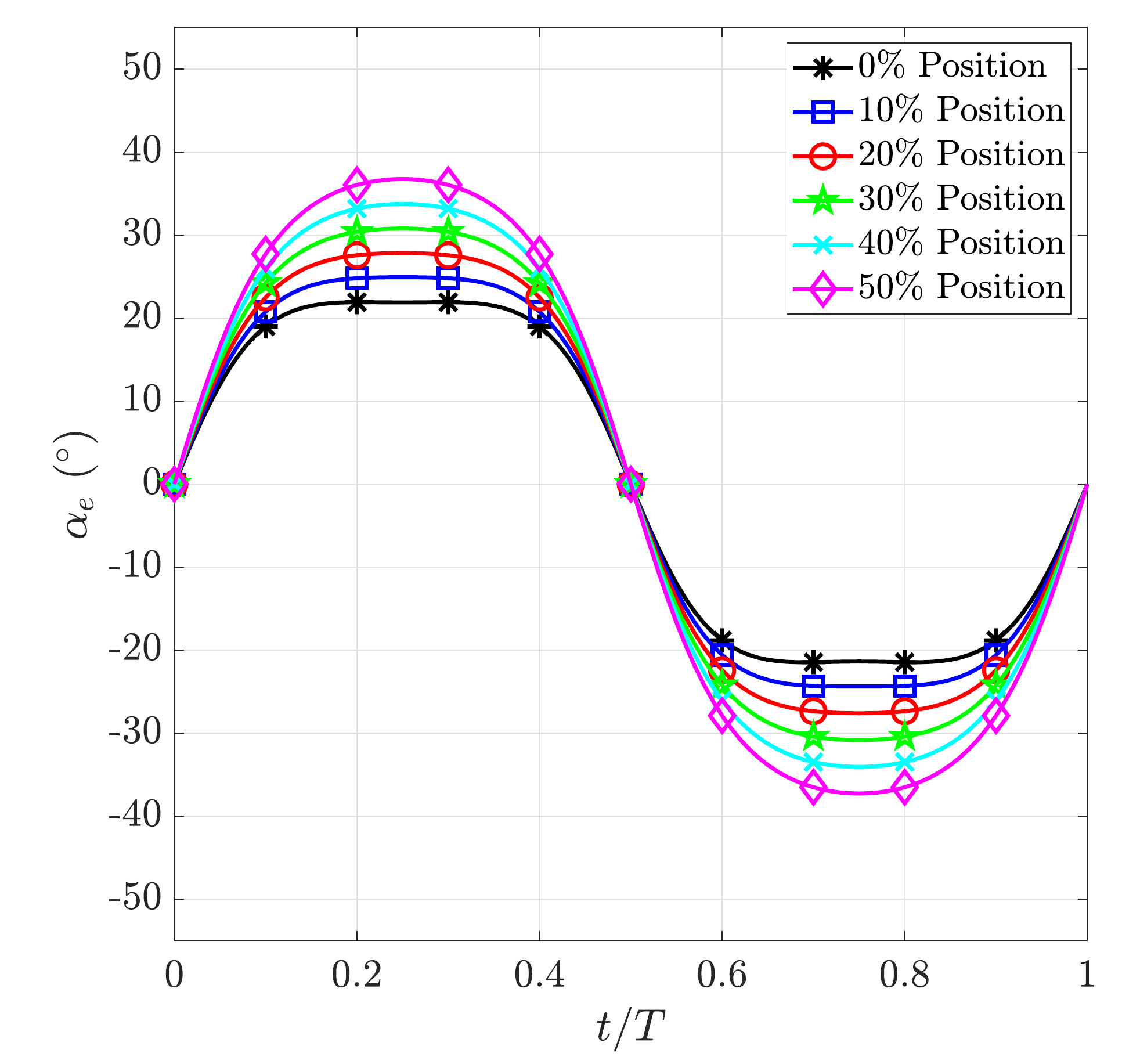}
        \caption{}
    \end{subfigure}
    \caption{VARIATION OF $\alpha_e$ IN A CYCLE FOR A CONSTANT MORPH AMPLITUDE OF: (a) $10^{\circ}$, AND (b) $60^{\circ}$}
    \label{eAoA_plots_pos}
\end{figure*}

The lift coefficient at $10^{\circ}$ morph amplitude (Fig. \ref{CL_vs_time_pos}(a)) shows negligible variation among different morph positions, which is corroborated by the effective angle of attack shown in Fig. \ref{eAoA_plots_pos}(a). As the amplitude is increased to $60^{\circ}$, maximum lift coefficient is observed for 50\% morph position, as a result of maximum effective angle of attack (Figs. \ref{CL_vs_time_pos}(b) and \ref{eAoA_plots_pos}(b)).

 
\subsection*{Flow Visualization}
\begin{figure}[!htbp]
    \centering
    \begin{subfigure}[h]{0.45\textwidth}
    \centering
        \includegraphics[width = 0.48\textwidth]{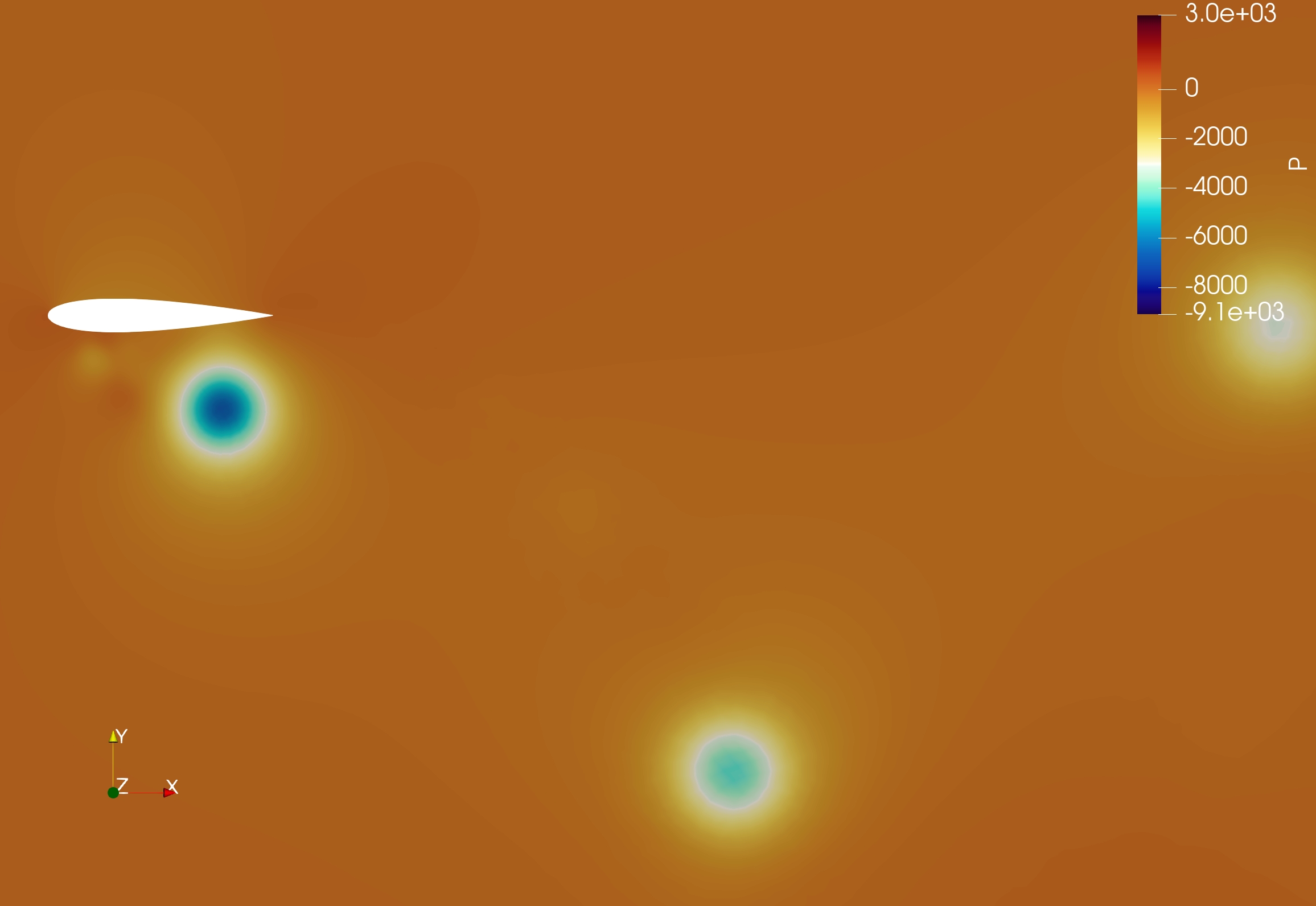}
        \includegraphics[width = 0.48\textwidth]{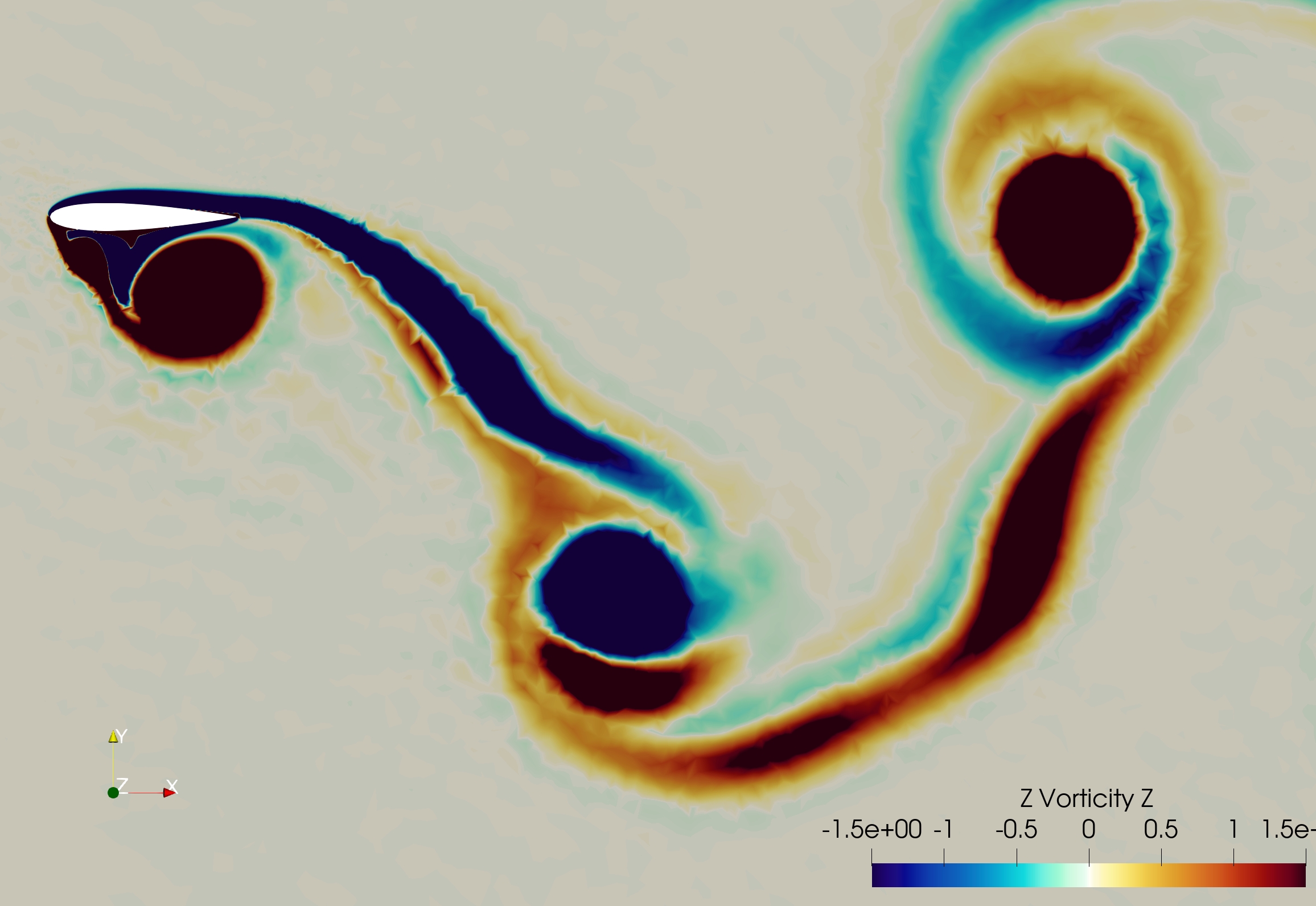}
        \caption{\(t/T = 0\)}
        \label{pres0}
    \end{subfigure}
 
    \begin{subfigure}[h]{0.45\textwidth}
    \centering
        \includegraphics[width = 0.48\textwidth]{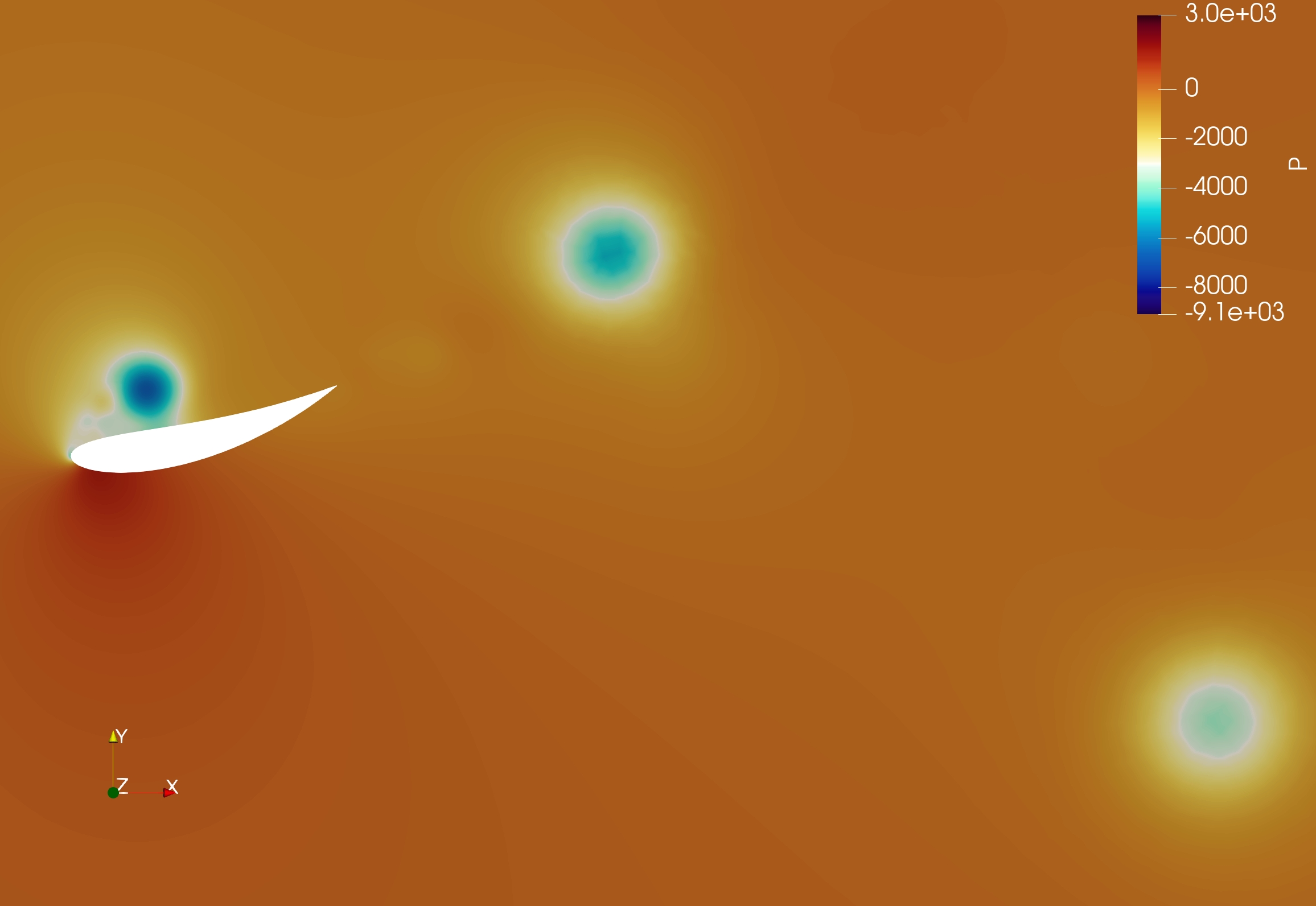}
        \includegraphics[width = 0.48\textwidth]{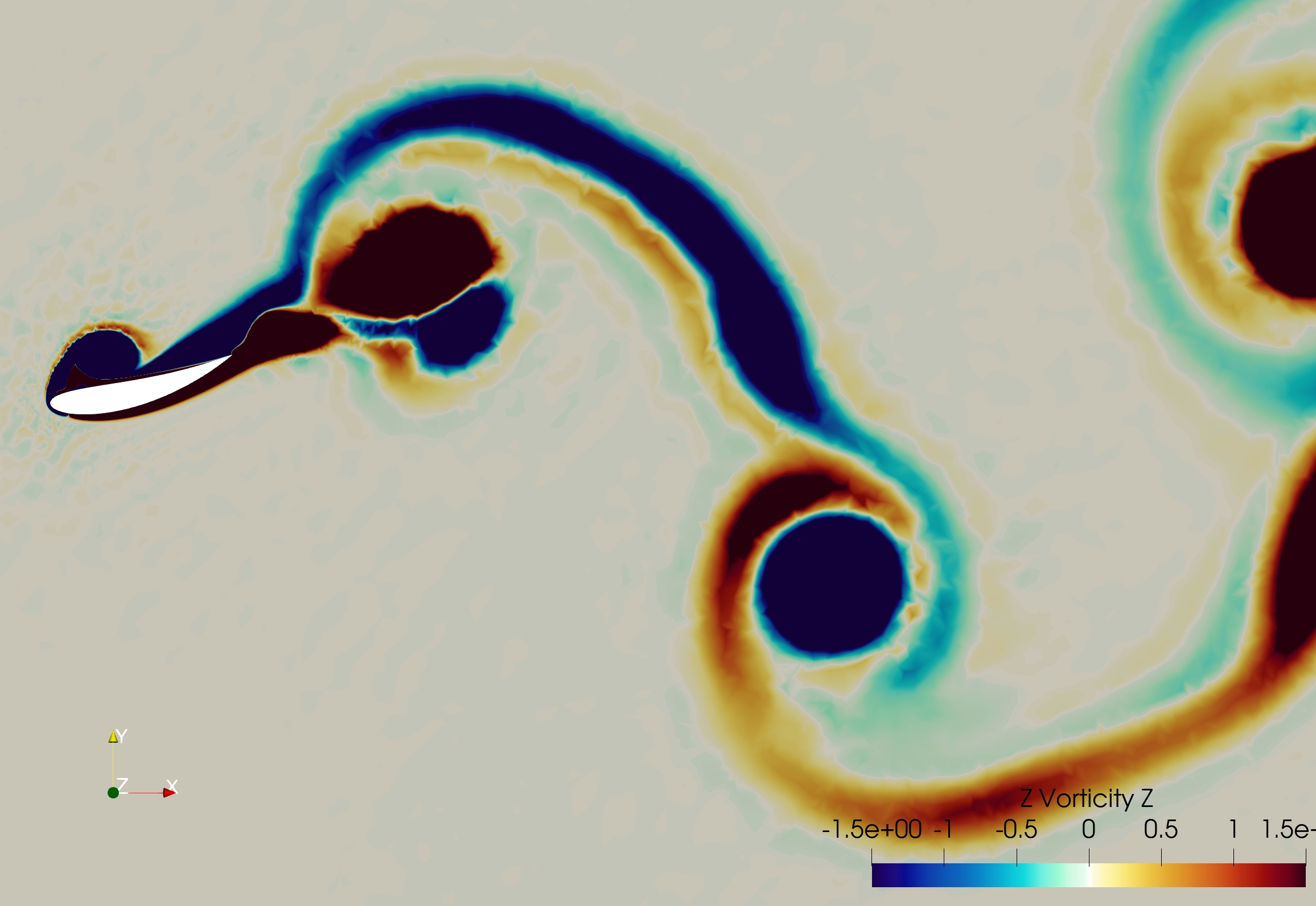}
        \caption{\(t/T = 0.25\)}
        \label{pres25}
    \end{subfigure}
    
    \begin{subfigure}[h]{0.45\textwidth}
    \centering
        \includegraphics[width = 0.48\textwidth]{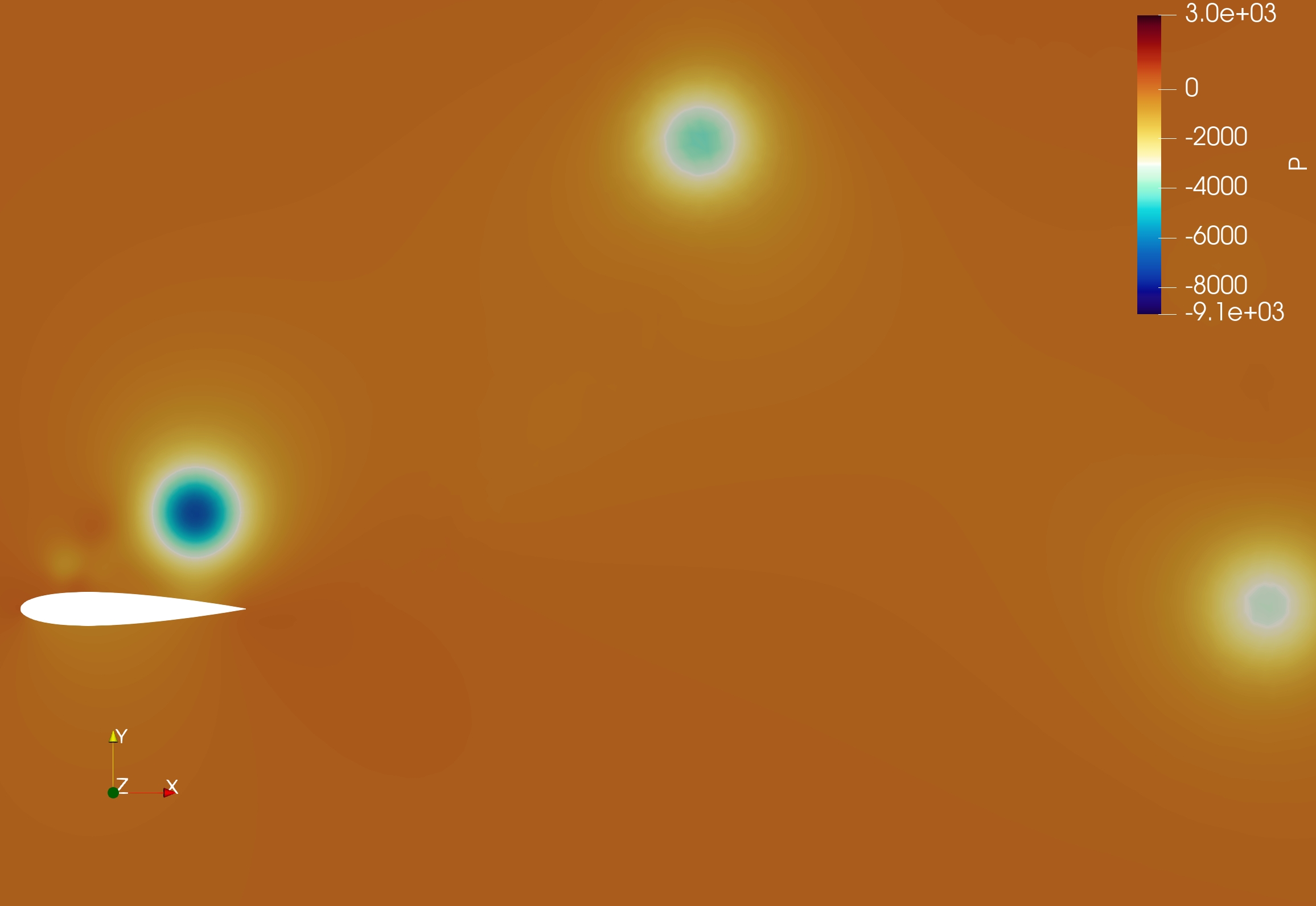}
        \includegraphics[width = 0.48\textwidth]{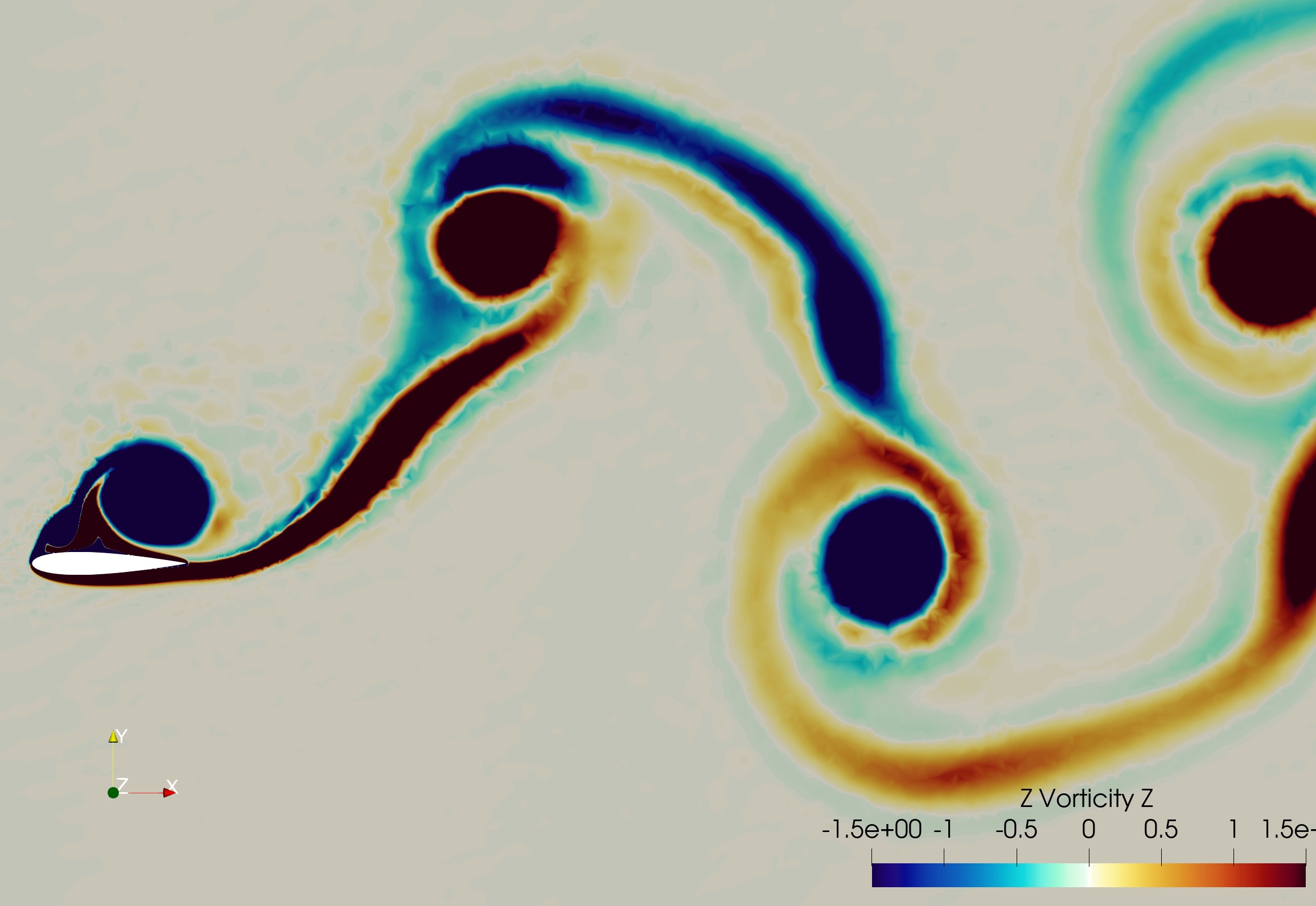}
        \caption{\(t/T = 0.50\)}
        \label{pres50}
    \end{subfigure}
    
    \begin{subfigure}[h]{0.45\textwidth}
    \centering
        \includegraphics[width = 0.48\textwidth]{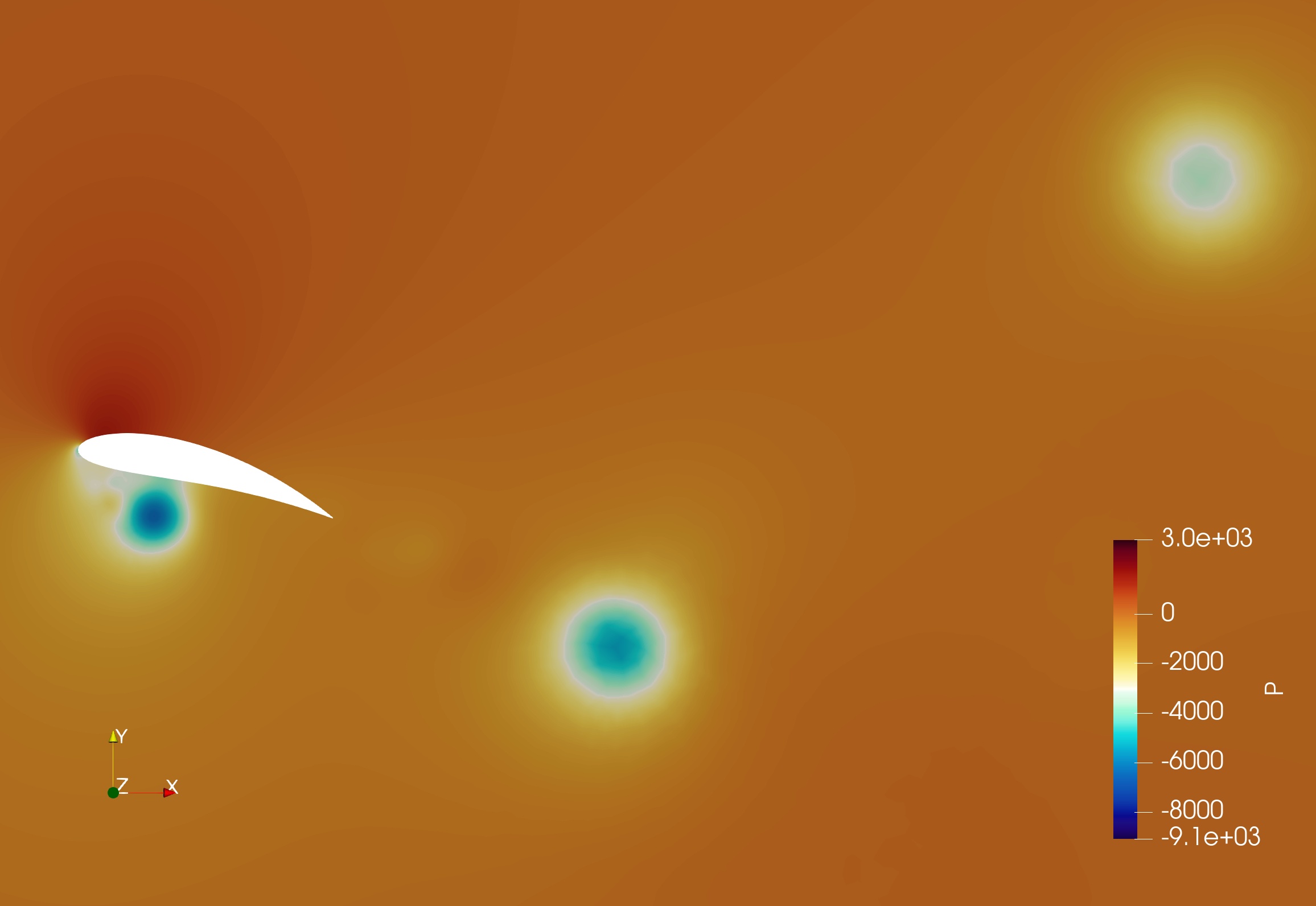}
        \includegraphics[width = 0.48\textwidth]{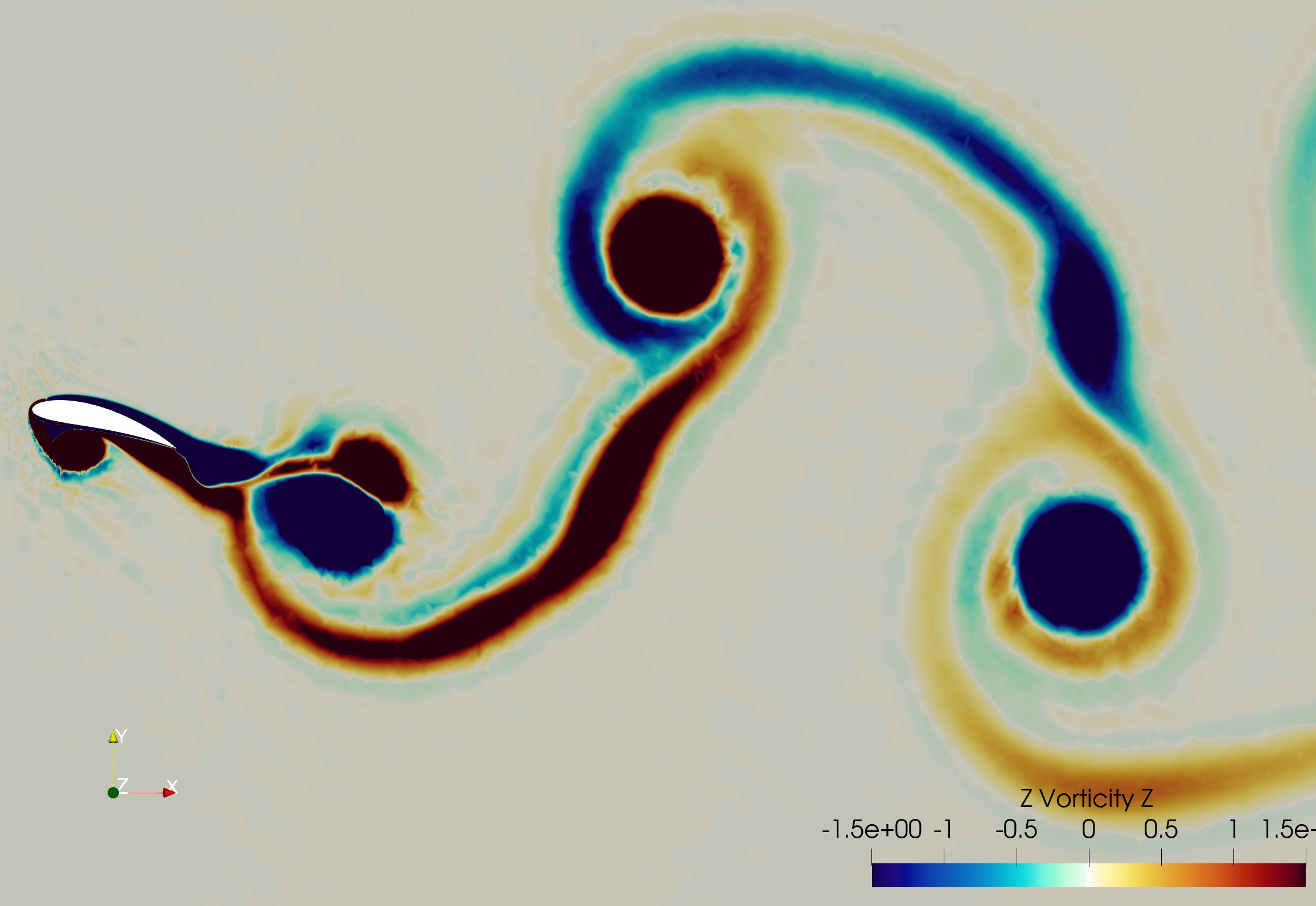}
        \caption{\(t/T = 0.75\)}
        \label{pres75}
    \end{subfigure}
    \caption{PRESSURE (LEFT) AND VORTEX CONTOURS (RIGHT) FOR A MORPHING FOIL WITH MORPH POSITION $0\%$ AND AMPLITUDE $30^{\circ}$ AT: (a) $t/T = 0$, (b) $t/T = 0.25$, (c) $t/T = 0.5$ AND (d) $t/T = 0.75$}
    \label{Gen Vorticity/Pressure Map}
\end{figure}

In this subsection, we further analyze the flow characteristics of the morphing and heaving foil with the help of the wake signature throughout a time period. To begin with, we select the case of morph amplitude $30^{\circ}$ and morph position 0\%. The wake of the morphing foil is visualized with the help of contours of Z-vorticity in Fig. \ref{Gen Vorticity/Pressure Map} along with the pressure contours. 
The parameters of study have been selected such that the morphing foil falls in the propulsive regime, indicated by the inverted von-K$\mathrm{\acute{a}}$rm$\mathrm{\acute{a}}$n vortex street, which leads to generation of thrust.

The motion of the foil consists of a downstroke (denoted by ``ds") (Fig. \ref{Gen Vorticity/Pressure Map}(a) to \ref{Gen Vorticity/Pressure Map}(c)) followed by an upstroke (``us") from Fig. \ref{Gen Vorticity/Pressure Map}(c) to \ref{Gen Vorticity/Pressure Map}(a) of next cycle). Two kinds of vortices, leading edge vortex (LEV) (circular) and trailing edge vortex (TEV) (elongated) are formed during the two strokes.
LEV plays a major role in thrust generation by creating a suction region on the upper surface of the foil during downstroke and on the lower surface during upstroke, as was discussed in \cite{tandem2021}. During downstroke, LEV generated is clockwise and negative (blue in color) on the upper surface of the foil (Fig. \ref{Gen Vorticity/Pressure Map}(c)). During the downstroke, the morph amplitude is maximum at the quarter time period, resulting in maximum projected area of the foil to the incoming flow. Flow conditions developing over the foil surface during this time will dictate its propulsive behavior \cite{tandem2021}. Vortex interactions which generate negative pressure on the upper surface and positive pressure on the lower surface of the foil during this period will lead to higher thrust generating conditions.

The vorticity and pressure contours of all the cases are analyzed to give an insight on the effect of morph position and amplitude on the thrust generation by the foil. We make an attempt to give some insight on the following questions:
\begin{enumerate}
\item How is morphing different from the previously studied flapping (combined pitching and heaving motion)?
\item How does the amplitude of morphing affect thrust performance?
\item How does the position of morphing affect thrust performance?
\end{enumerate}

\underline{\textbf{Comparison Between Morphing and Flapping Foils}}:
As can be observed from Fig. \ref{Gen Vorticity/Pressure Map}(a) at $t/T=0$, the leading edge vortex from the upstroke of the previous cycle [LEV-us(n - 1)] has fully developed on the lower surface of the morphing foil (referred here as foil-2 for the combined morphing and heaving motion). Compared with the flapping (combined pitching and heaving) motion in \cite{tandem2021} (referred here as foil-1) where the foil was given a pitching motion of amplitude $30^{\circ}$, there are subtle differences in the flow patterns albeit the fact that both the cases produce thrust-generating inverted von-K$\mathrm{\acute{a}}$rm$\mathrm{\acute{a}}$n vortex street.

It is observed that for a particular amplitude when the entire foil is morphed, i.e., morph position is 0\%, $C_{T,\mathrm{mean}} \approx 0.4$ is less than that of the flapping foil with similar amplitude of pitching motion ($C_{T,\mathrm{mean}} \approx 0.8$). The propulsive performance depends on the projected area of the foil to the incoming flow, duration for which favorable thrust-generating conditions exist, the amount of pressure differential across the foil and the shear forces acting on the surface. In contrast to the pitching motion where each point on the foil has identical angular deformation due to pitching, the morphing motion results in varying deformation along the chord of the foil. The points near the leading edge have the least amount of deformation compared to the trailing edge points. Consequently, the LEV is more prominent in the morphing motion compared to the flapping motion, which leads to suction pressure on the upper side of the foil during downstroke. However, a larger projected area to the incoming flow observed in the flapping motion results in higher thrust (force in the negative freestream direction), in comparison to the morphing foil. To summarize, the flapping motion (combined heaving and pitching) of the foil gives a more favorable condition for thrust generation compared to the present case of combined heaving and morphing due to enhanced component of the fluid force in the negative freestream (negative X) direction as a consequence of larger projected area of the foil to the incoming flow.

\underline{\textbf{Effect of Morph Amplitude}}:
With increase in the morph amplitude, the projected area to the incoming flow increases leading to increasing net force in the negative X direction. This is reflected as increase in the mean thrust (shown in Fig. \ref{CT_CL_Charts}(a)) for morph positions of 0\%-20\%. However, as the morph position of 50\% is reached, the thrust starts to decrease with the morph amplitude.

\begin{figure}[!htbp]
    \centering
    \begin{subfigure}[b]{0.45\textwidth}
            \centering
            \includegraphics[width = 0.48\textwidth]{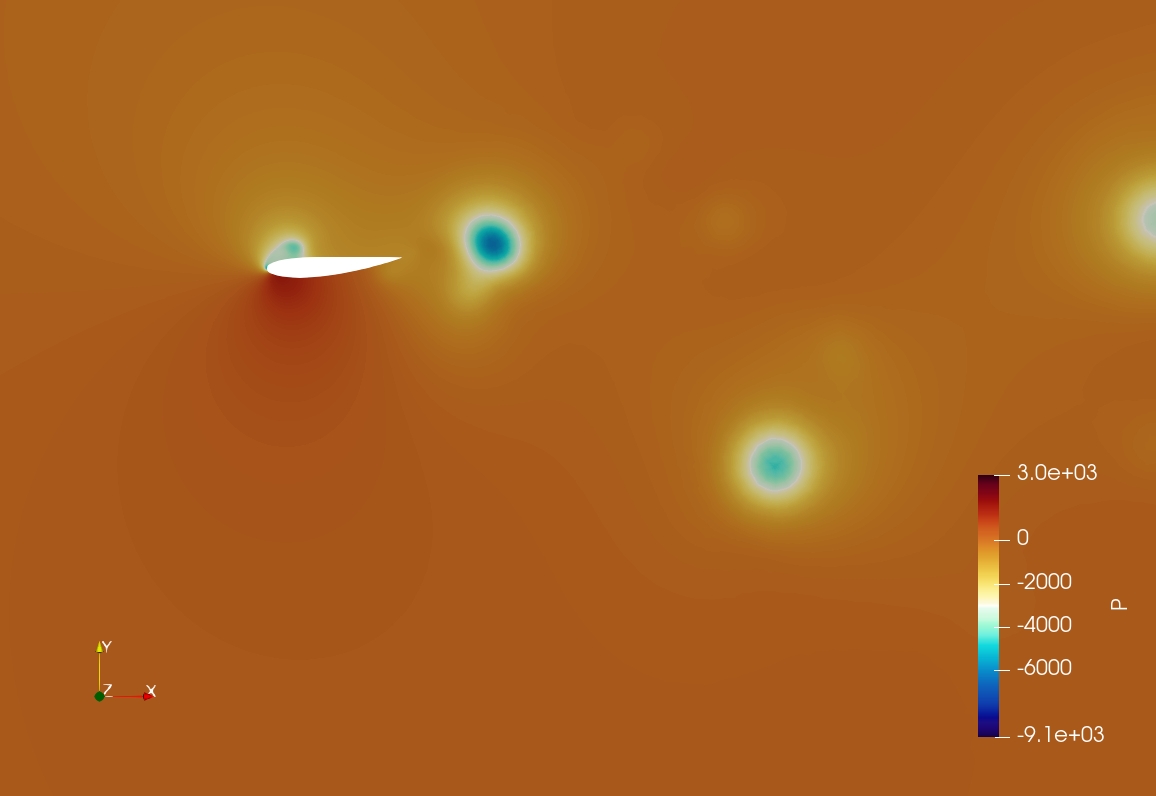}
            \hfill
            \includegraphics[width = 0.48\textwidth]{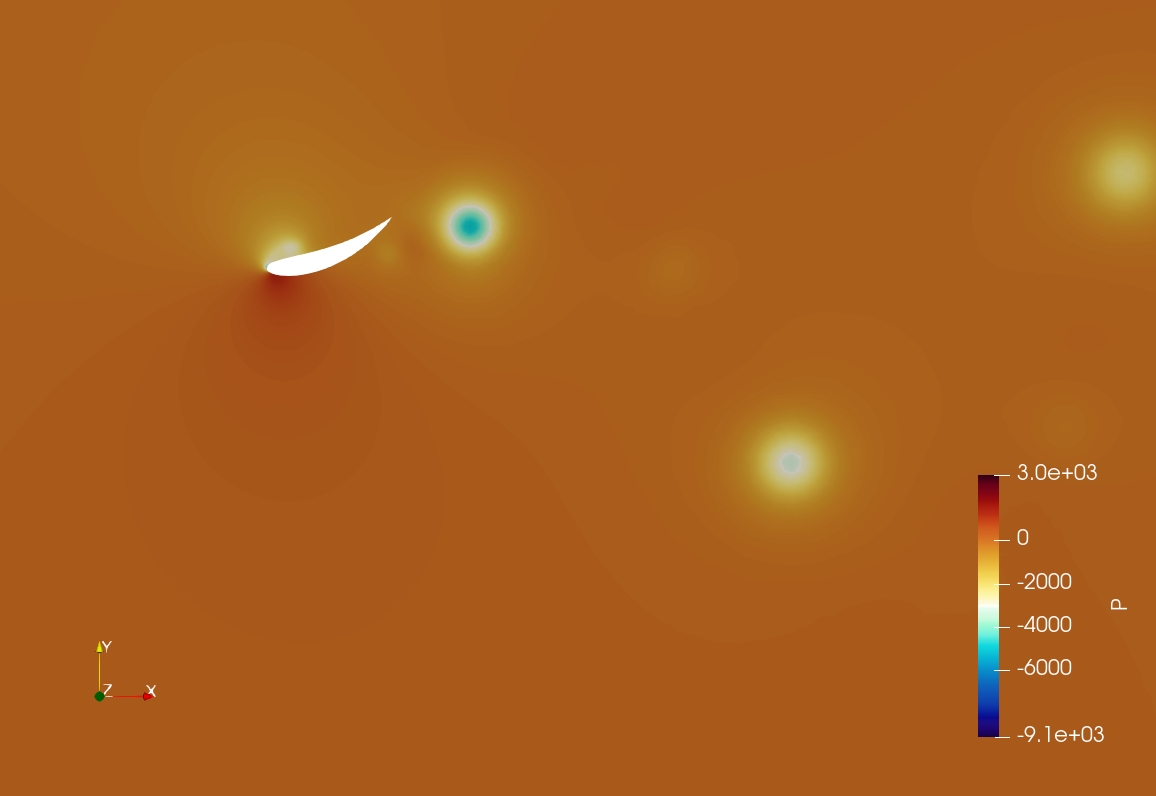}
            
            \includegraphics[width = 0.48\textwidth]{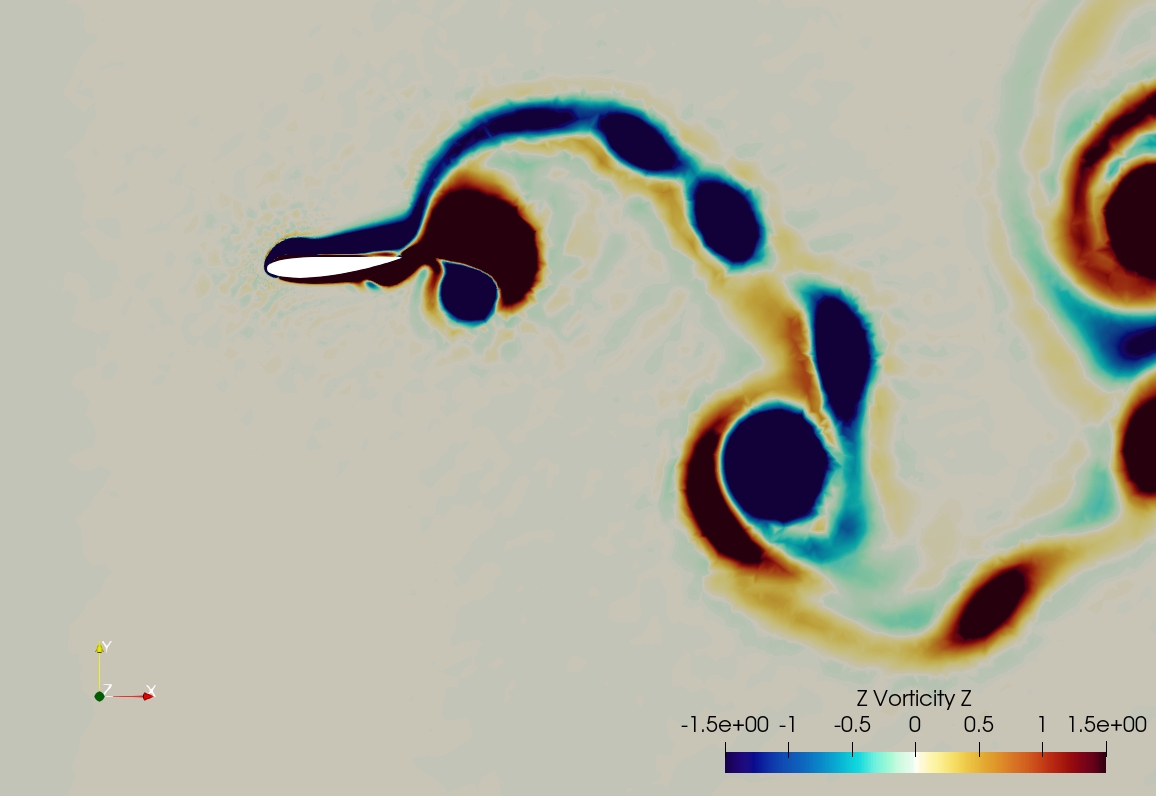}
            \hfill
            \includegraphics[width = 0.48\textwidth]{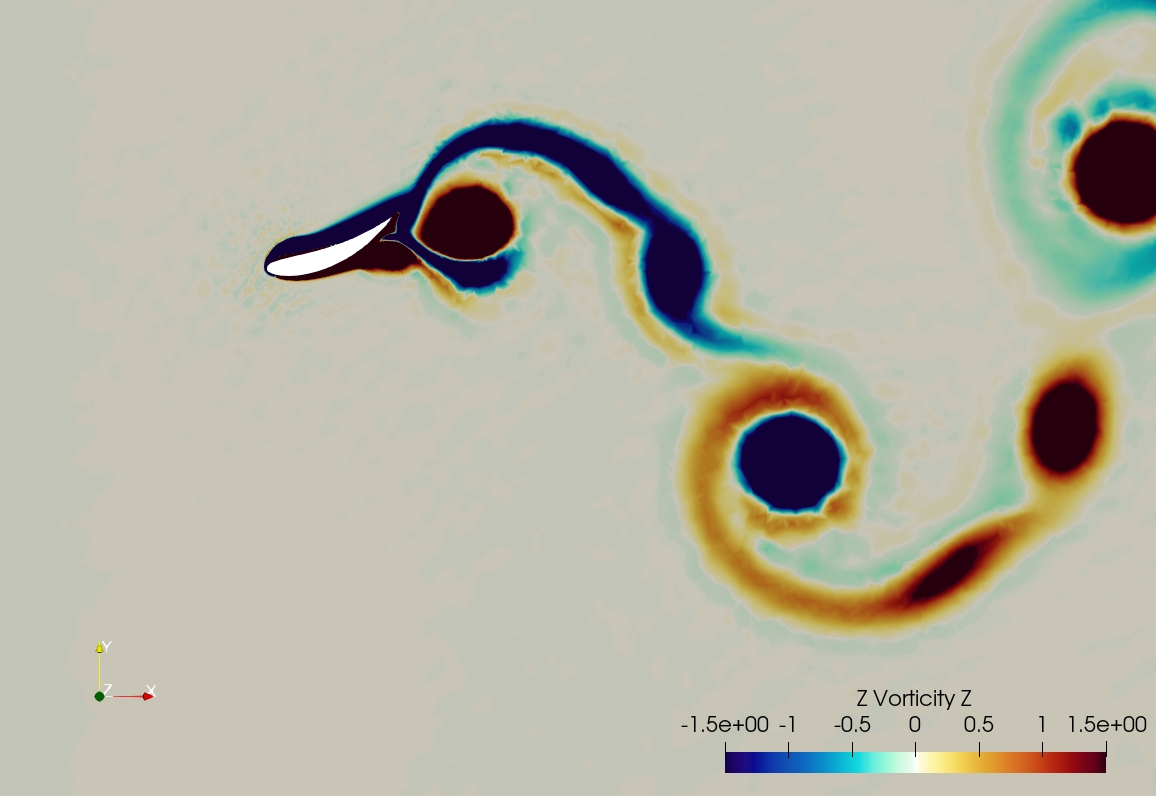}
        \caption{$t/T$ = 0.18}
    \end{subfigure}
    
    \begin{subfigure}[b]{0.45\textwidth}
            \centering
            \includegraphics[width = 0.48\textwidth]{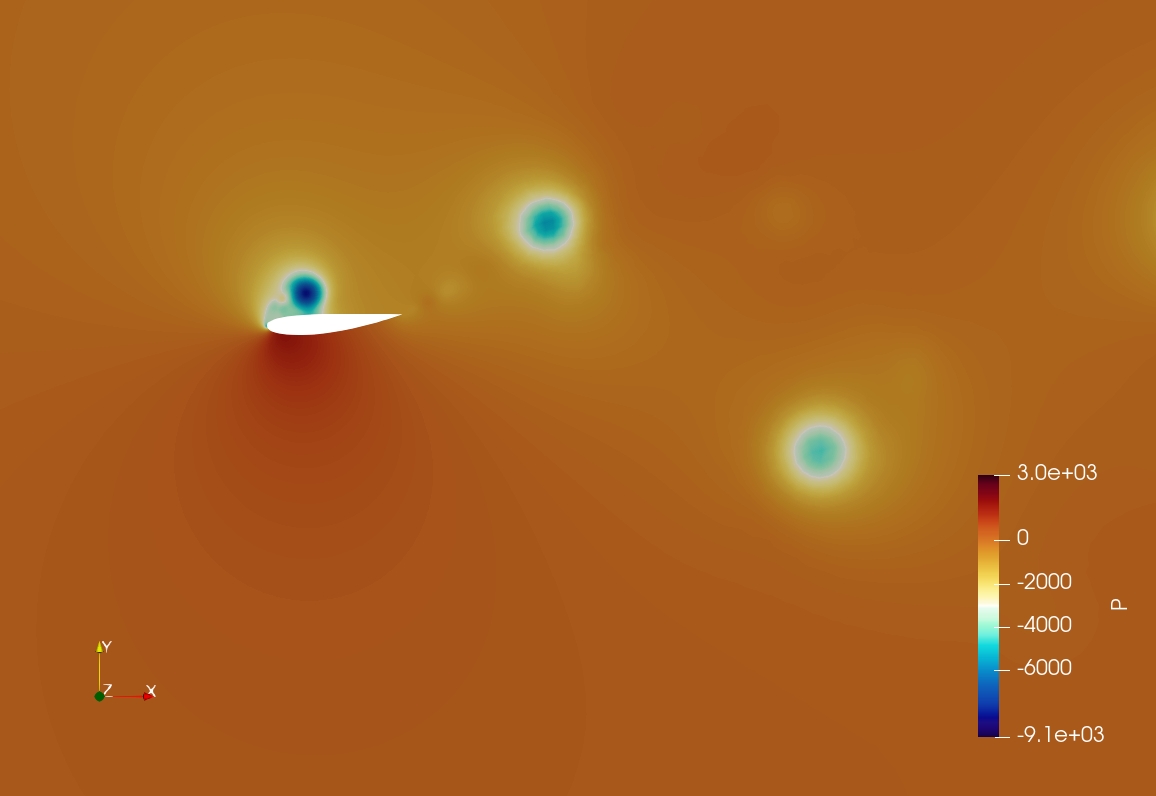}
            \hfill
            \includegraphics[width = 0.48\textwidth]{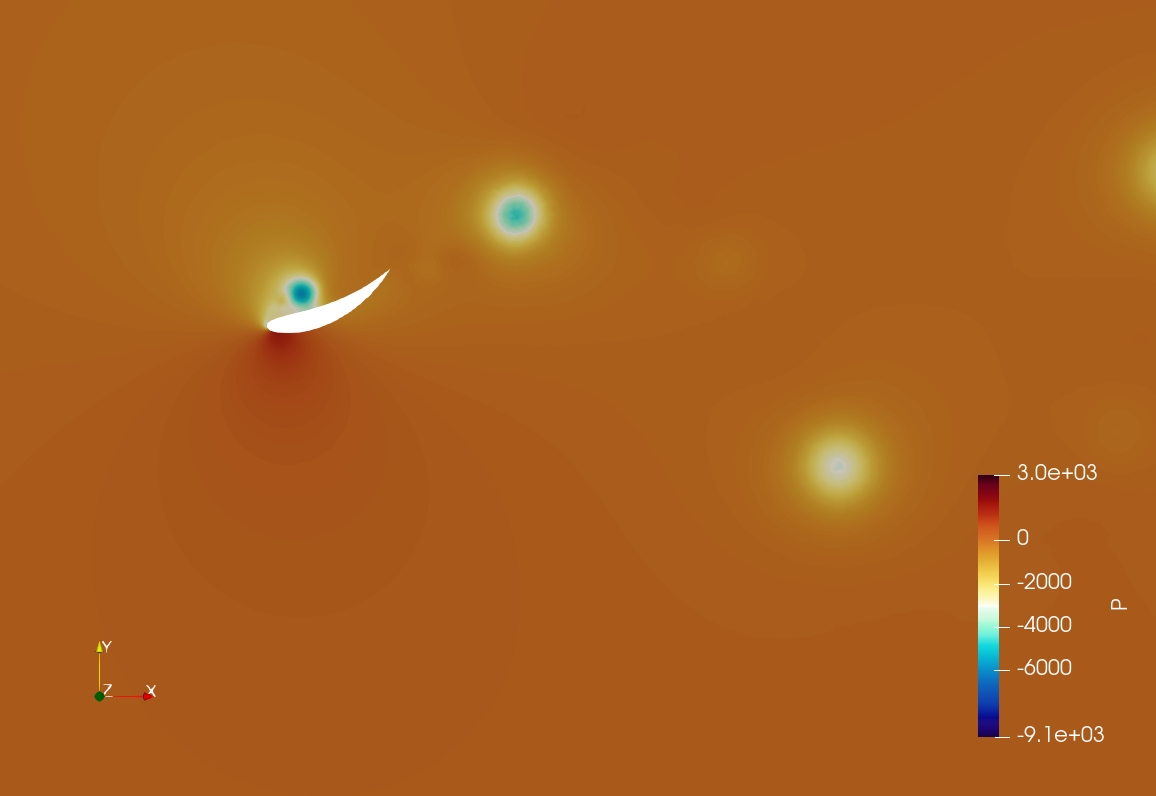}
            
            \includegraphics[width = 0.48\textwidth]{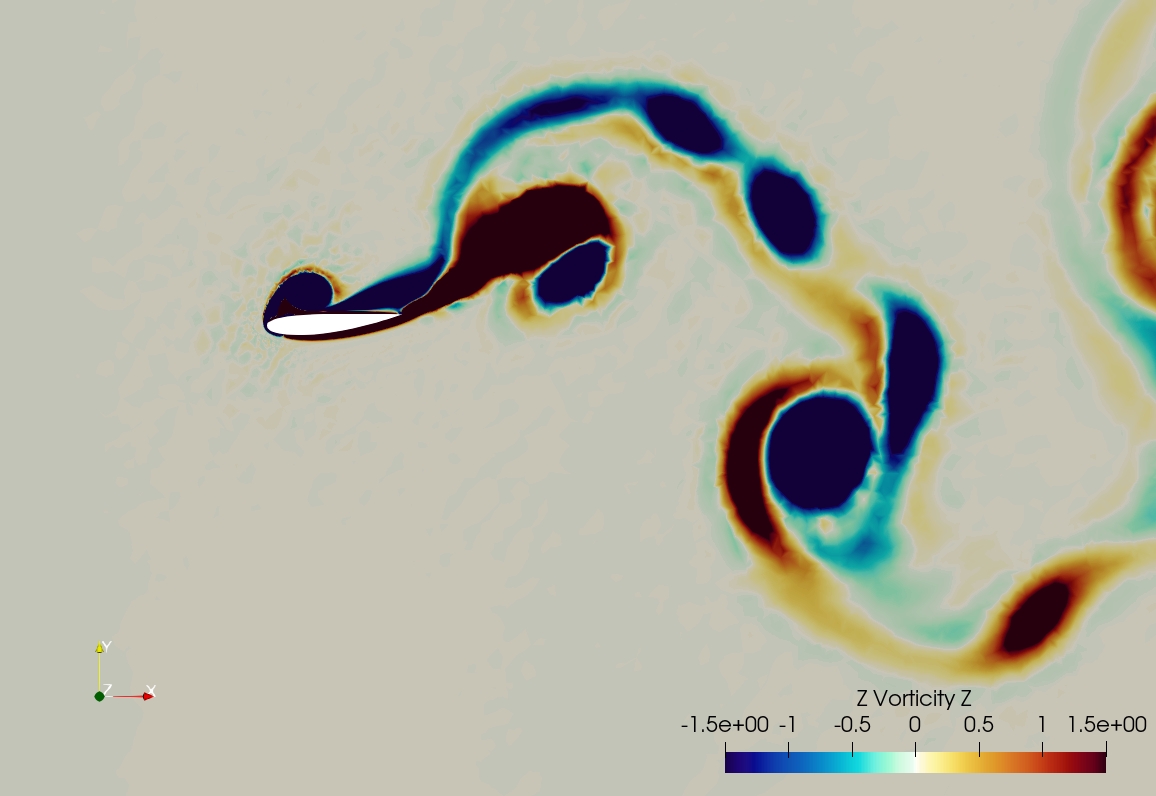}
            \hfill
            \includegraphics[width = 0.48\textwidth]{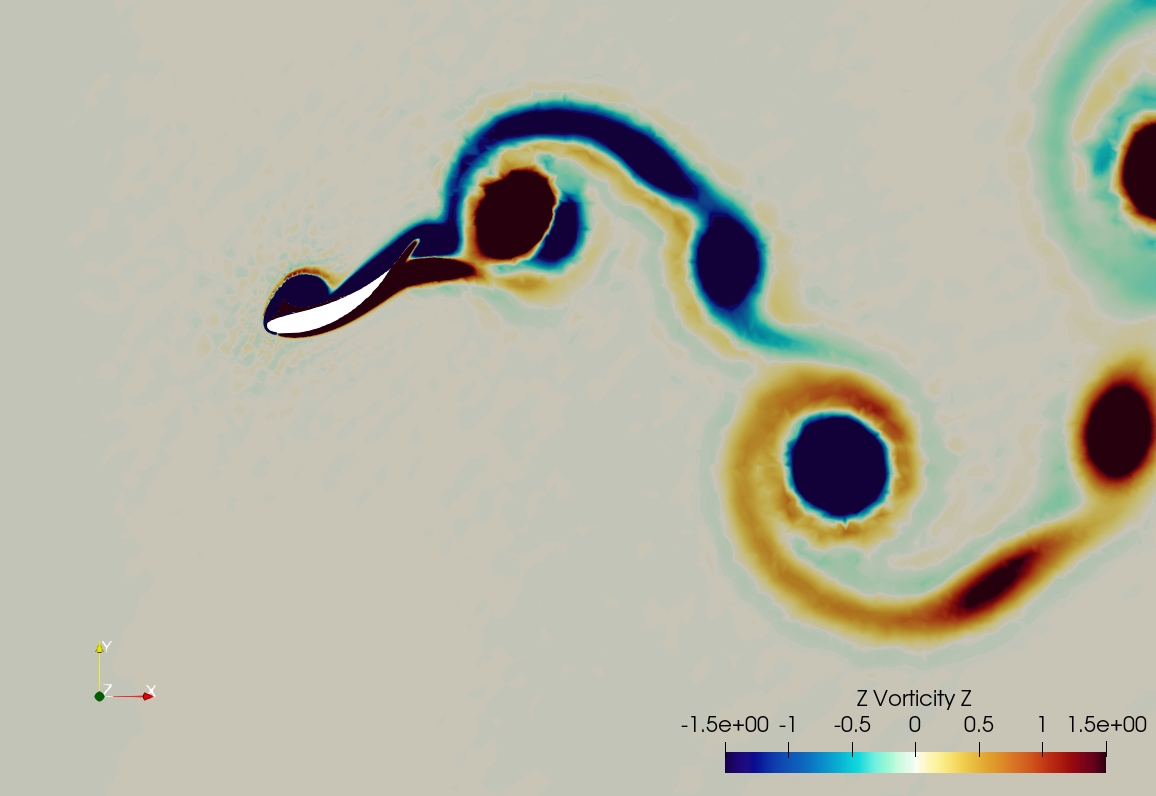}
        \caption{$t/T$ = 0.25}
    \end{subfigure}
    
    \begin{subfigure}[b]{0.45\textwidth}
            \centering
            \includegraphics[width = 0.48\textwidth]{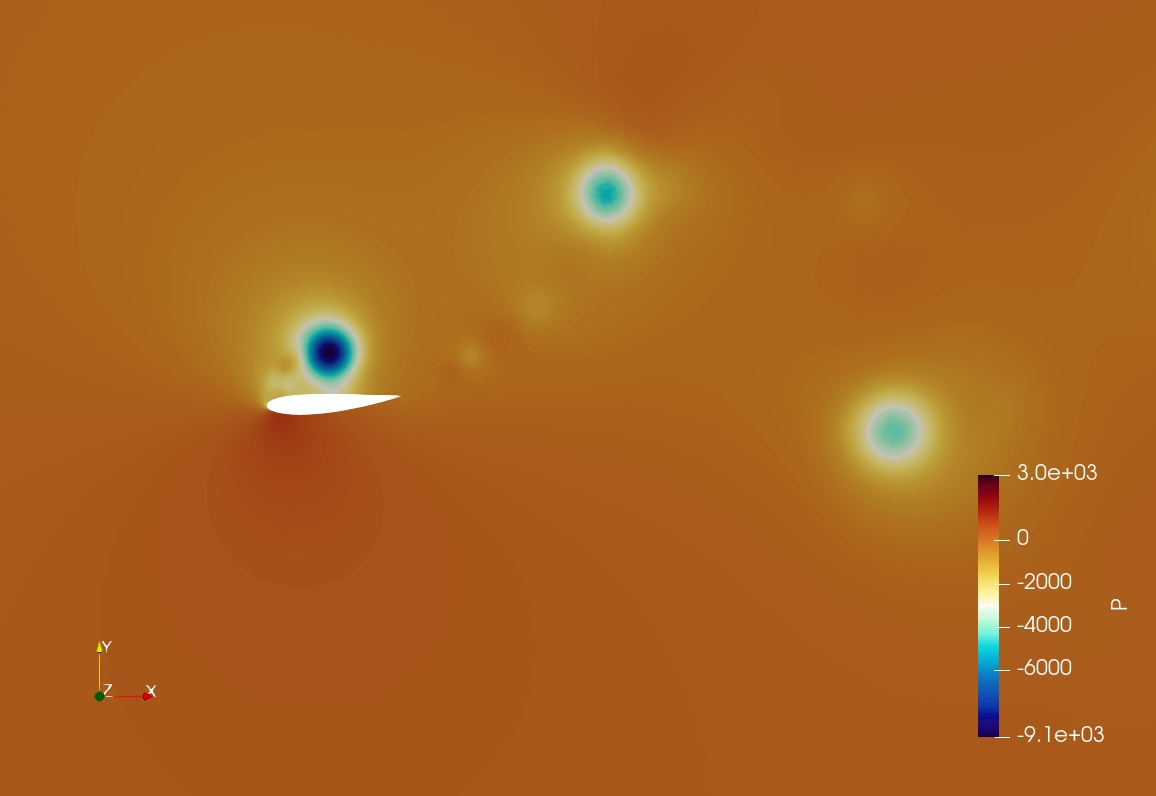}
            \hfill
            \includegraphics[width = 0.48\textwidth]{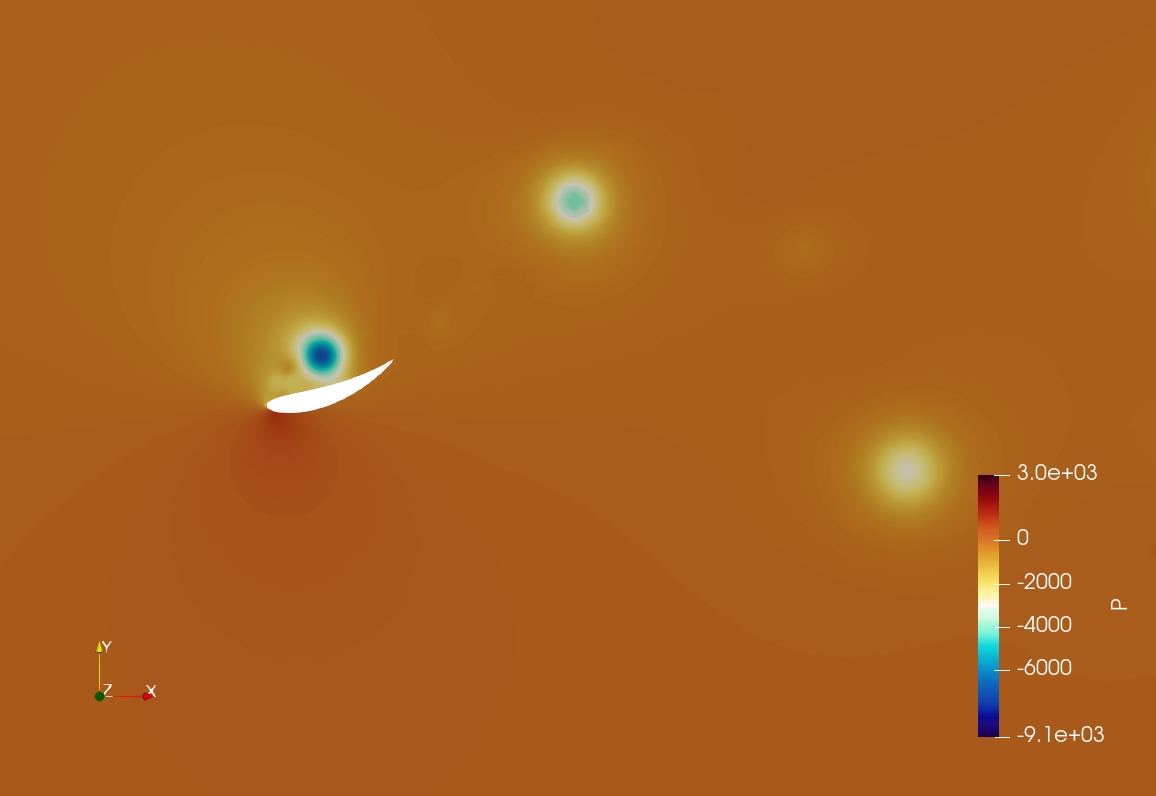}
            
            \includegraphics[width = 0.48\textwidth]{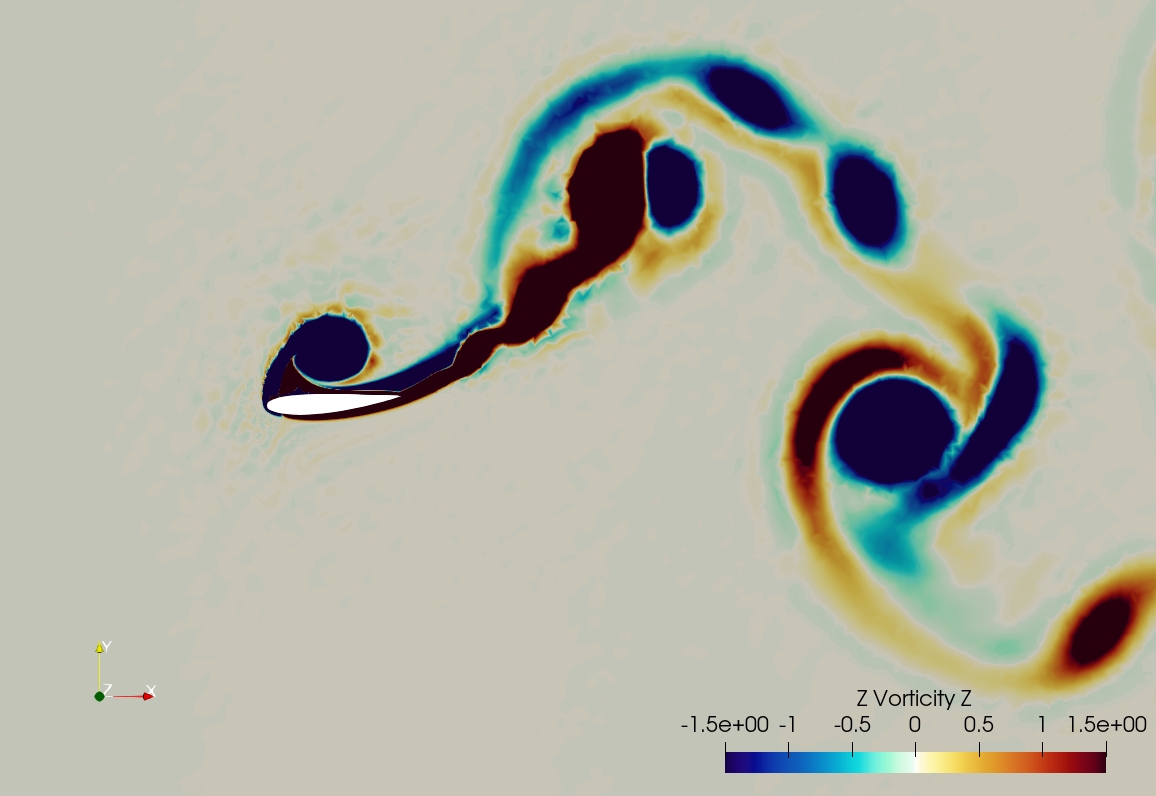}
            \hfill
            \includegraphics[width = 0.48\textwidth]{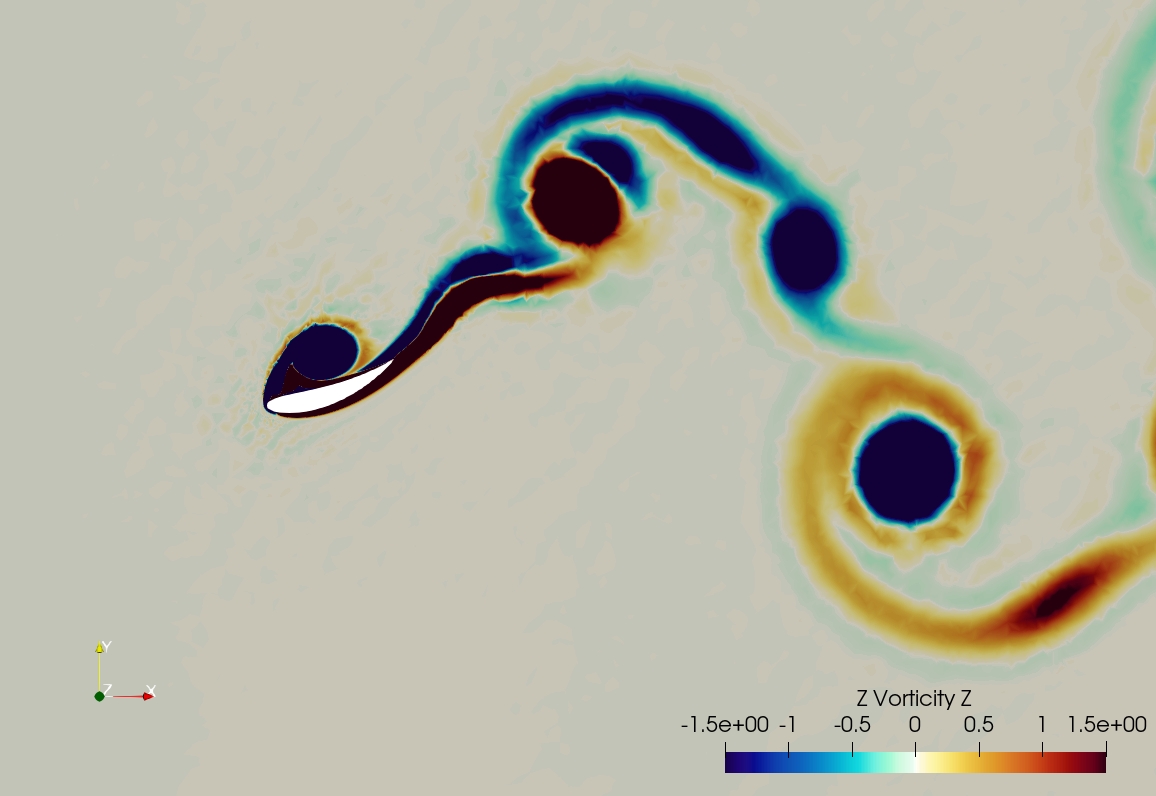}
        \caption{$t/T$ = 0.35}
    \end{subfigure}
    \caption{PRESSURE AND VORTICITY CONTOURS FOR THE MORPHING FOIL WITH MORPH POSITION OF 0\% AND MORPH AMPLITUDE OF 10\(^\circ\) (LEFT COLUMN) AND 50\(^\circ\) (RIGHT COLUMN) AT TIME INSTANCES (a) $t/T = 0.18$, (b) $t/T = 0.25$ AND (c) $t/T = 0.35$}
    \label{Amplitude Comparison with time}
\end{figure}

We next visualize the flow patterns for the morphing and heaving motion of the foil considering the constant morph position of 0\% and amplitudes of $10^{\circ}$ and $50^{\circ}$. The pressure and vorticity plots for these cases are shown in Fig. \ref{Amplitude Comparison with time} at different time instances in a morphing cycle. 

At time $t/T = 0.18$, the LEV-ds(n) starts to develop on the upper surface of the foil during the downstroke and is fully developed around $t/T = 0.35$. Comparing the two cases, we notice that the trailing edge of the foil morphs to a larger extent for $50^{\circ}$ morph amplitude, delaying the shedding of the LEV generated during downstroke. Thus, a suction pressure is maintained over the upper surface for a larger duration during downstroke compared to the $10^{\circ}$ case (Fig. \ref{Amplitude Comparison with time}(c)). Consequently, the above pressure differential leads to higher thrust for $50^{\circ}$ case compared to $10^{\circ}$ morph amplitude over extended duration in the downstroke (Fig. \ref{CT_vs_time_amp}(a)).

Therefore, the influence of morph amplitude on the propulsive performance of the morphing foil is an interplay between the increase in the projected area of the foil to the incoming flow and the phenomenon of delayed shedding of the LEV.

\begin{figure}[!htbp]
    \centering
    \begin{subfigure}[b]{0.45\textwidth}
            \centering
            \includegraphics[width = 0.48\textwidth]{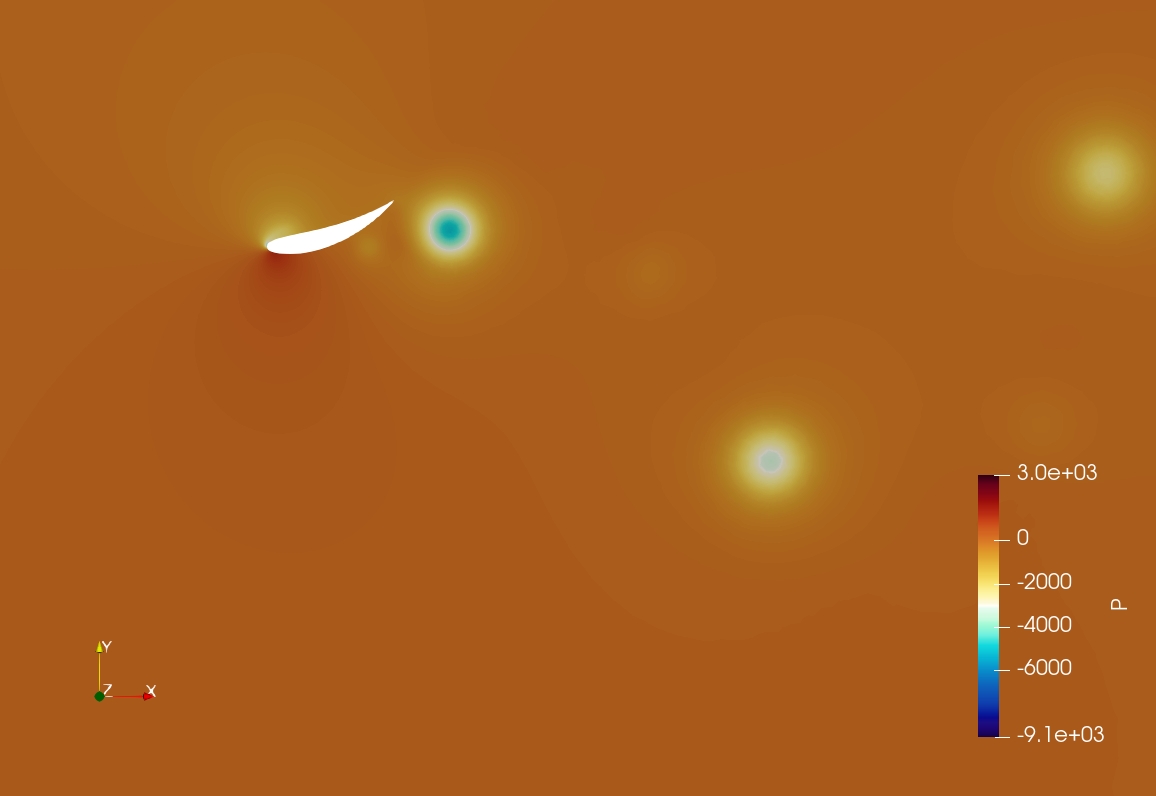}
            \hfill
            \includegraphics[width = 0.48\textwidth]{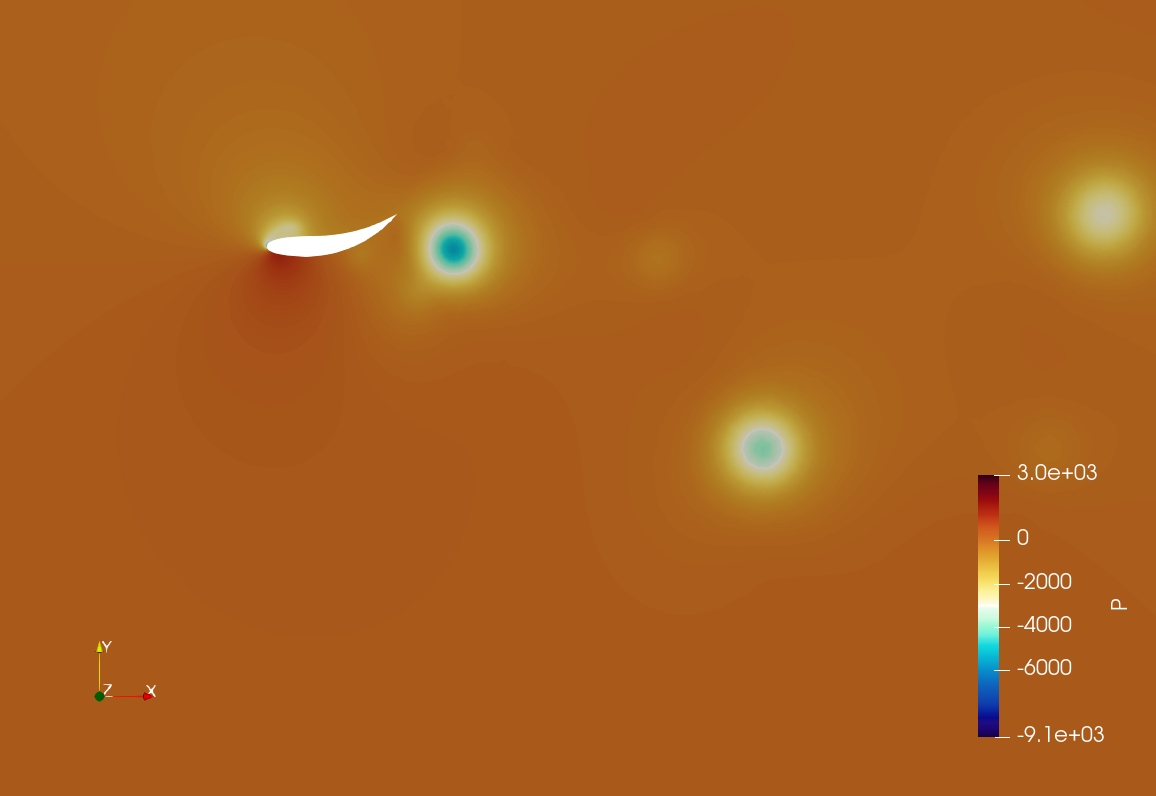}
            
            \includegraphics[width = 0.48\textwidth]{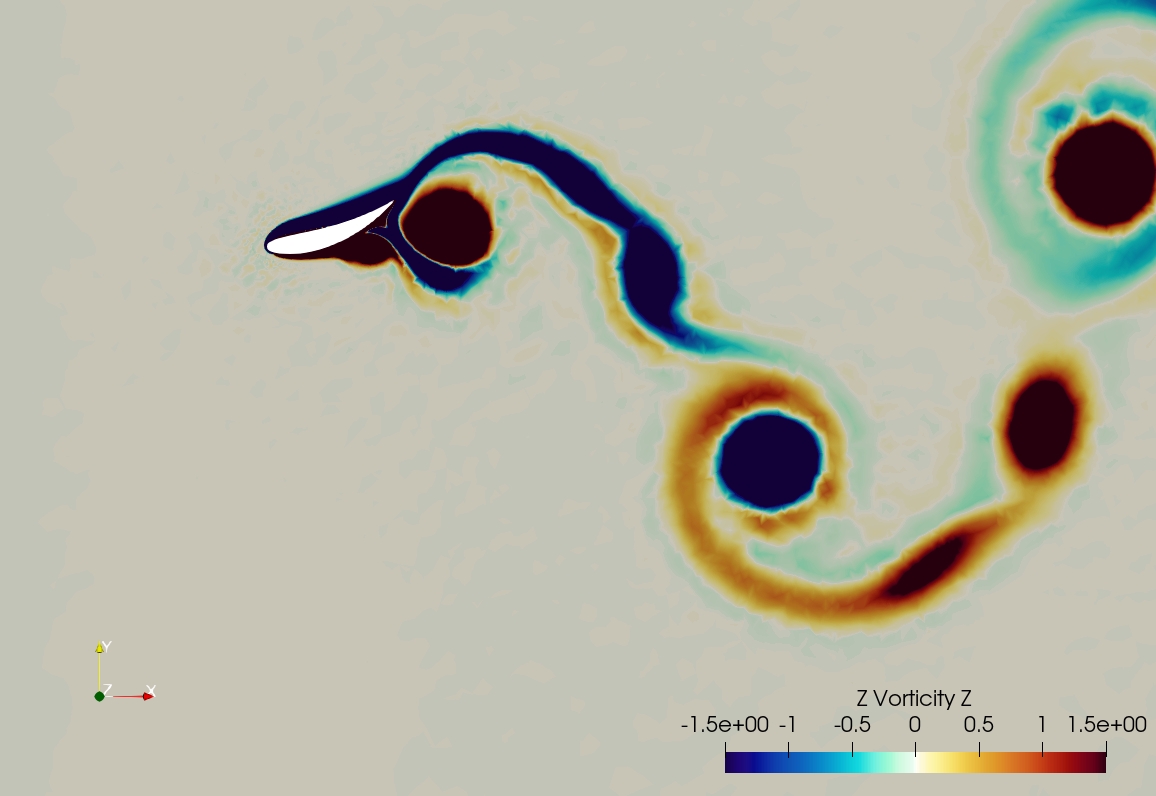}
            \hfill
            \includegraphics[width = 0.48\textwidth]{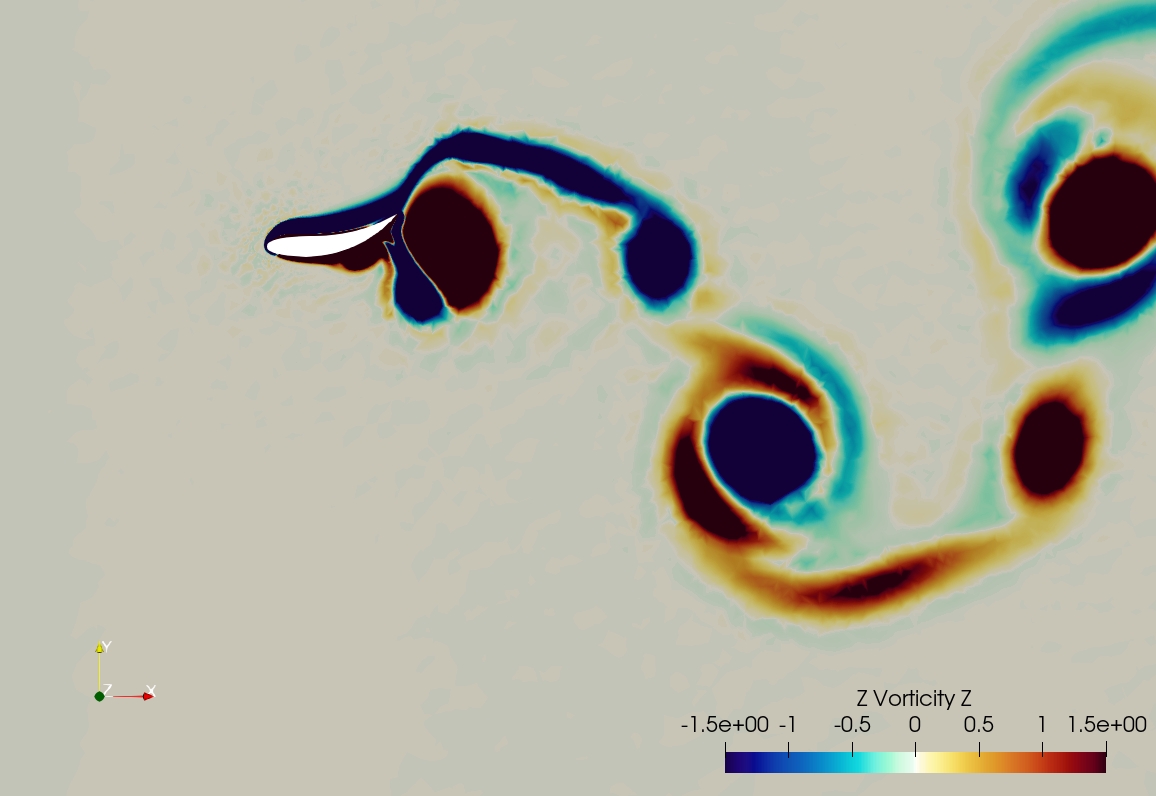}
        \caption{$t/T$ = 0.15}
    \end{subfigure}
    
    \begin{subfigure}[b]{0.45\textwidth}
            \centering
            \includegraphics[width = 0.48\textwidth]{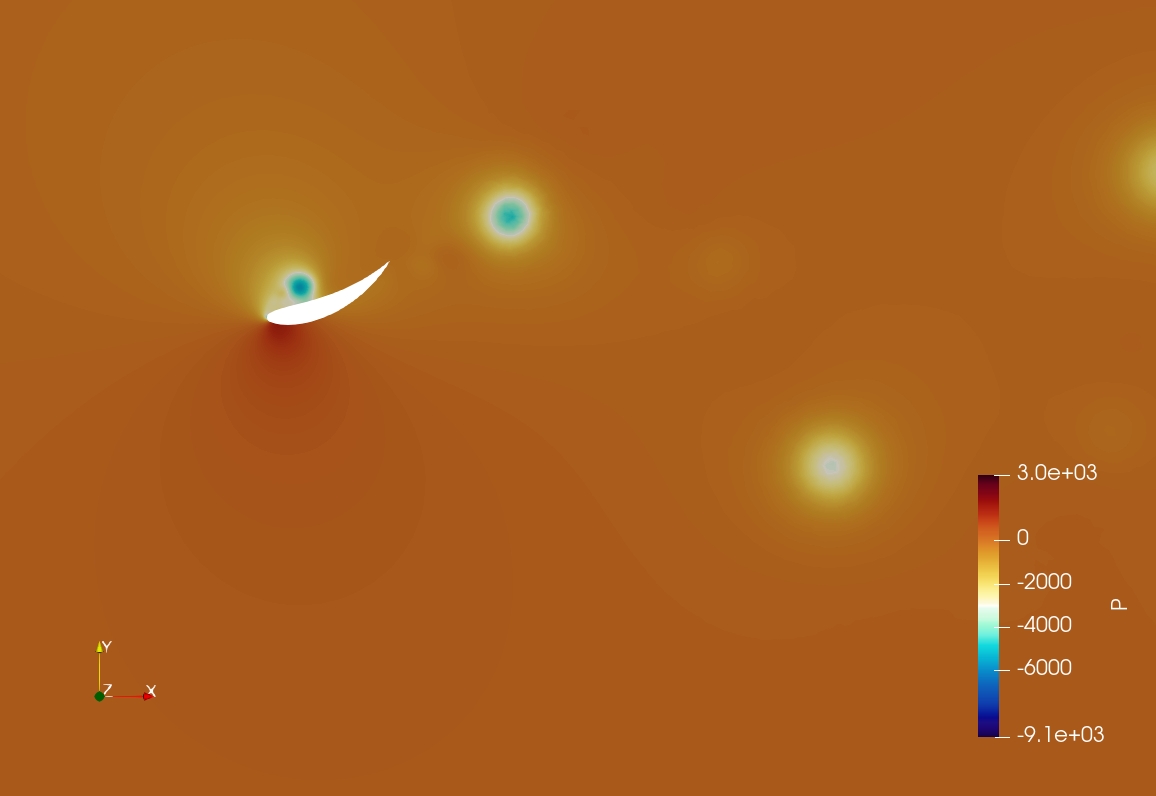}
            \hfill
            \includegraphics[width = 0.48\textwidth]{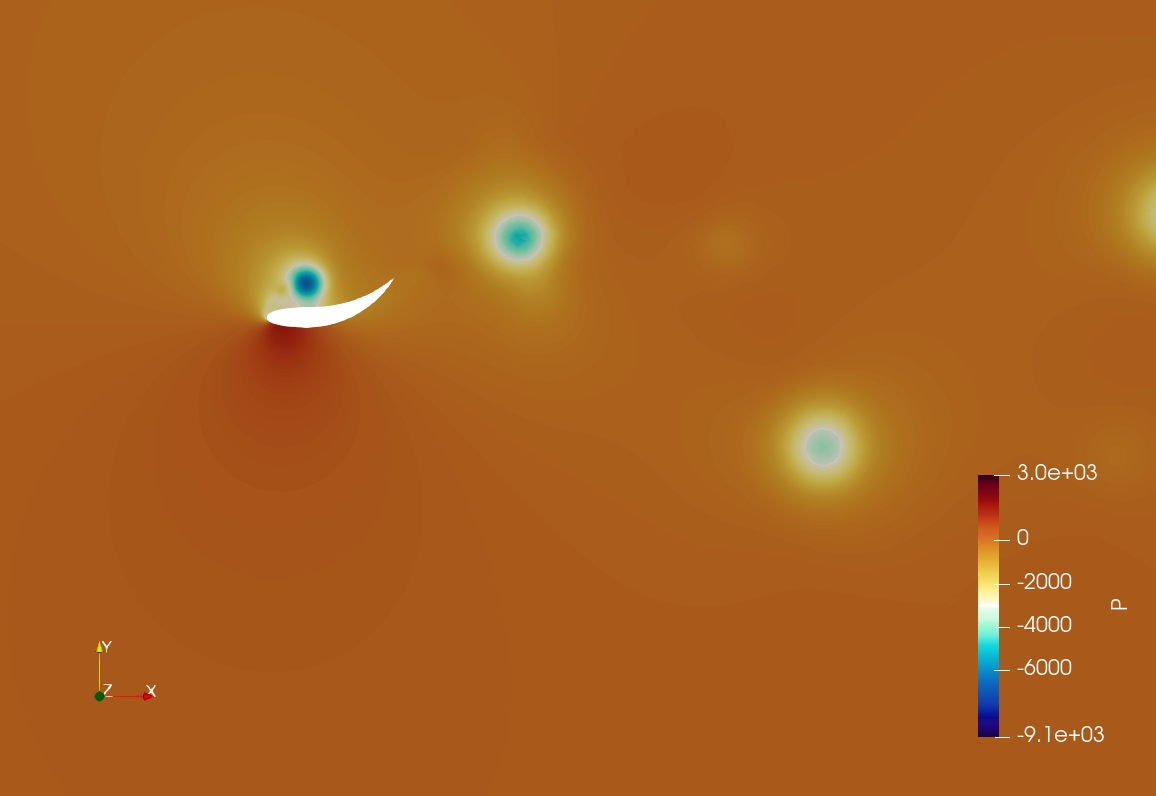}
            
            \includegraphics[width = 0.48\textwidth]{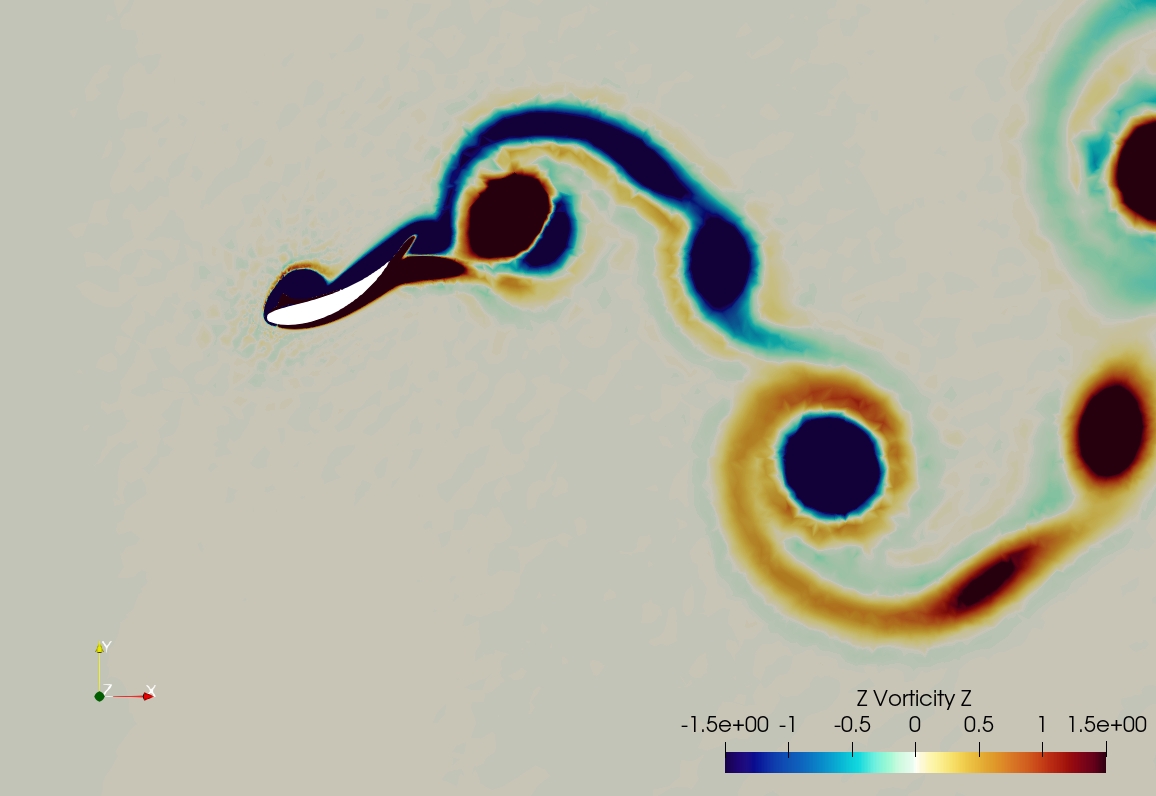}
            \hfill
            \includegraphics[width = 0.48\textwidth]{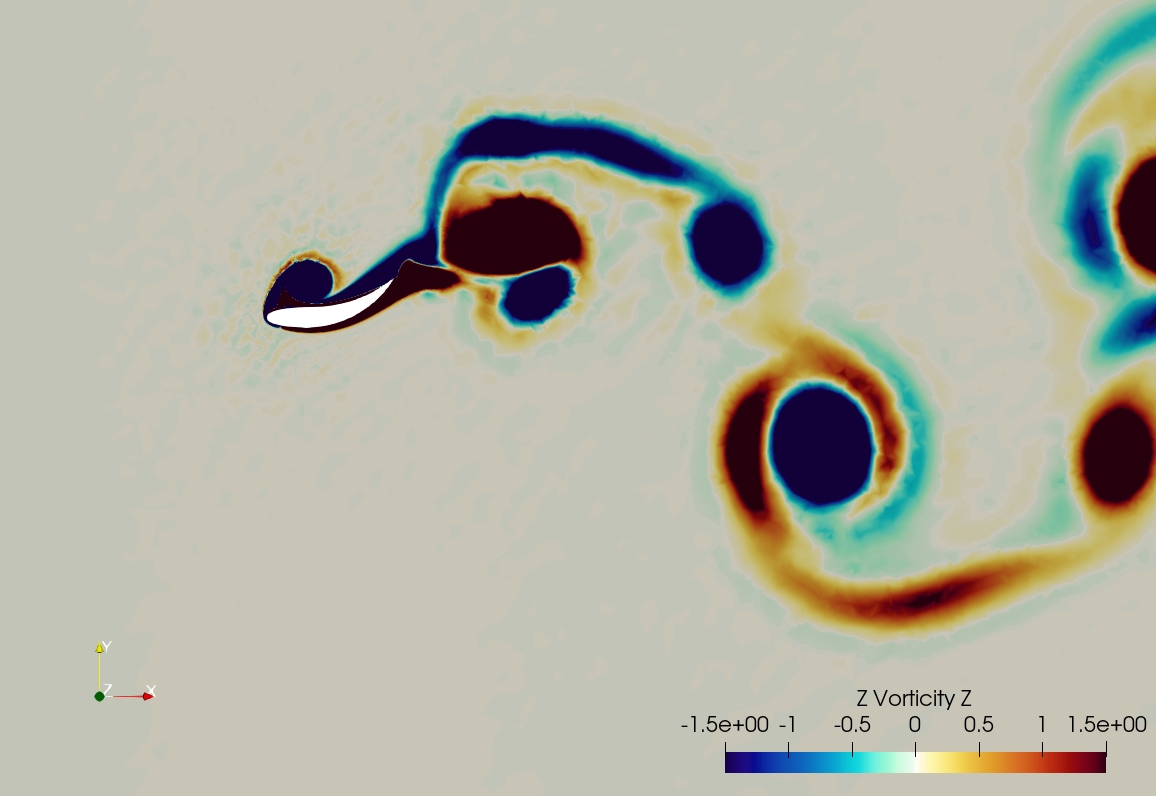}
        \caption{$t/T$ = 0.24}
    \end{subfigure}
    
    \begin{subfigure}[b]{0.45\textwidth}
            \centering
            \includegraphics[width = 0.48\textwidth]{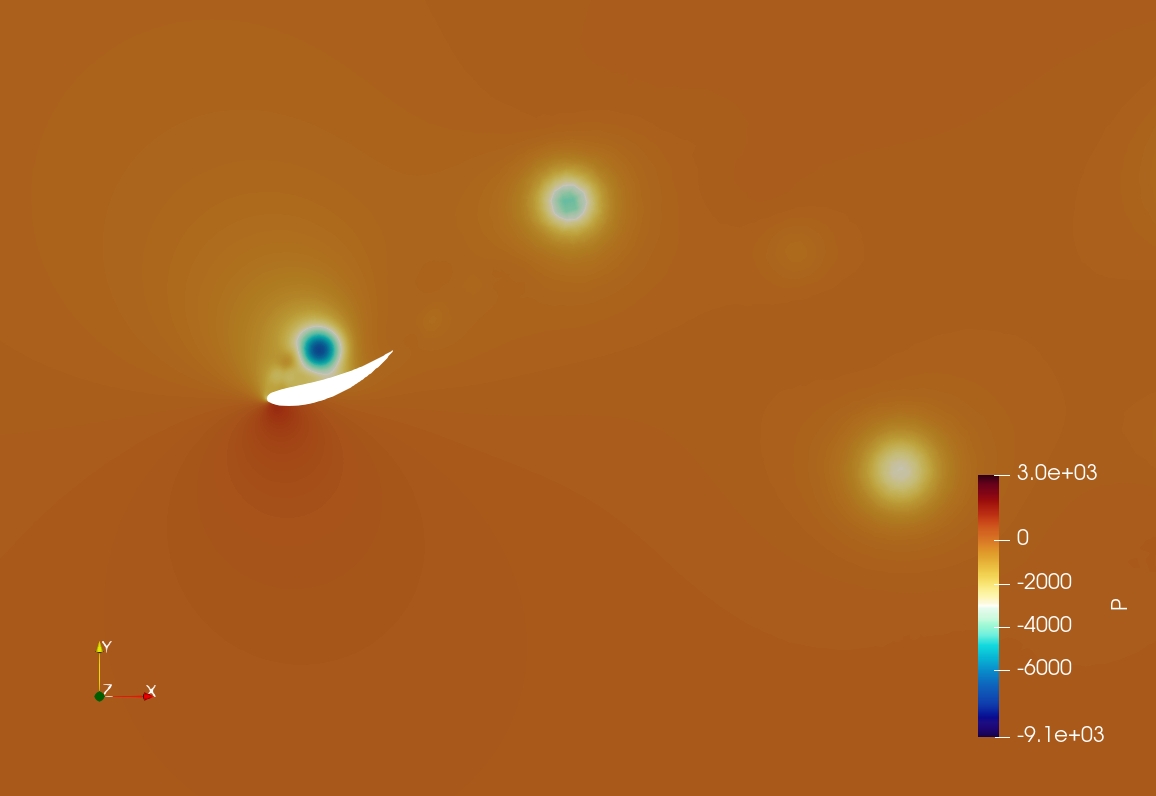}
            \hfill
            \includegraphics[width = 0.48\textwidth]{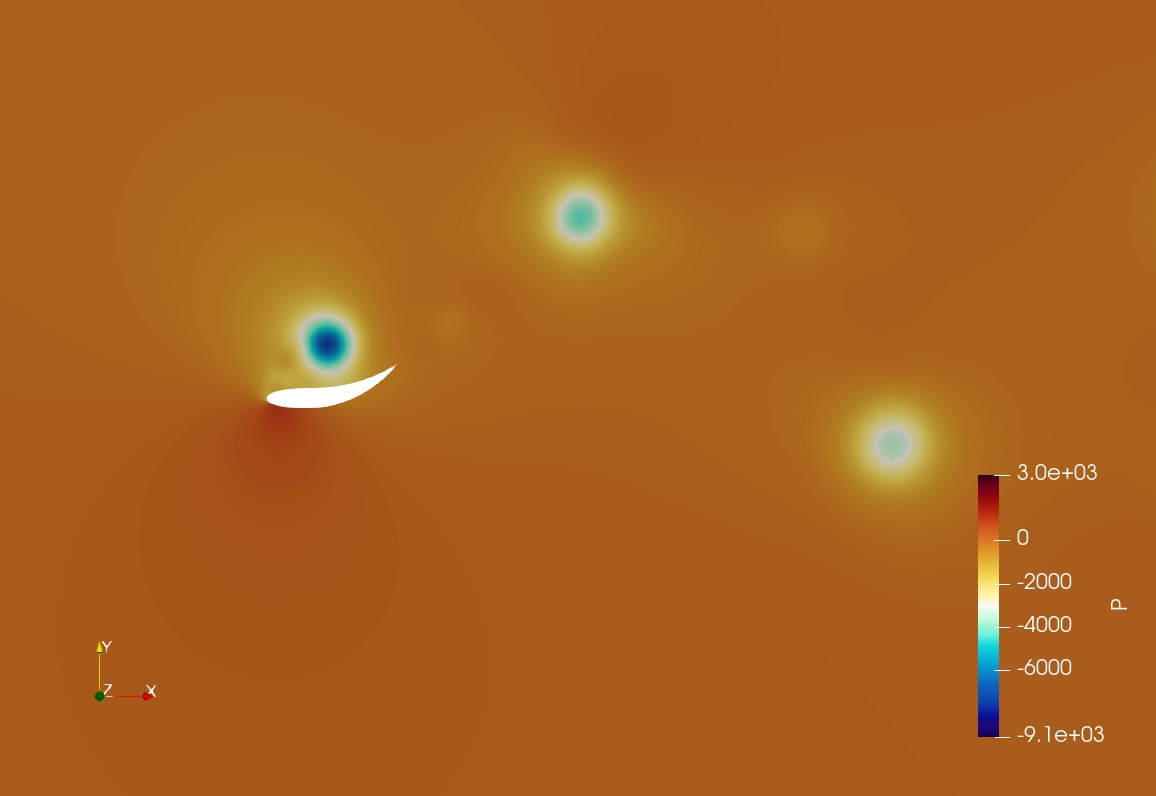}
            
            \includegraphics[width = 0.48\textwidth]{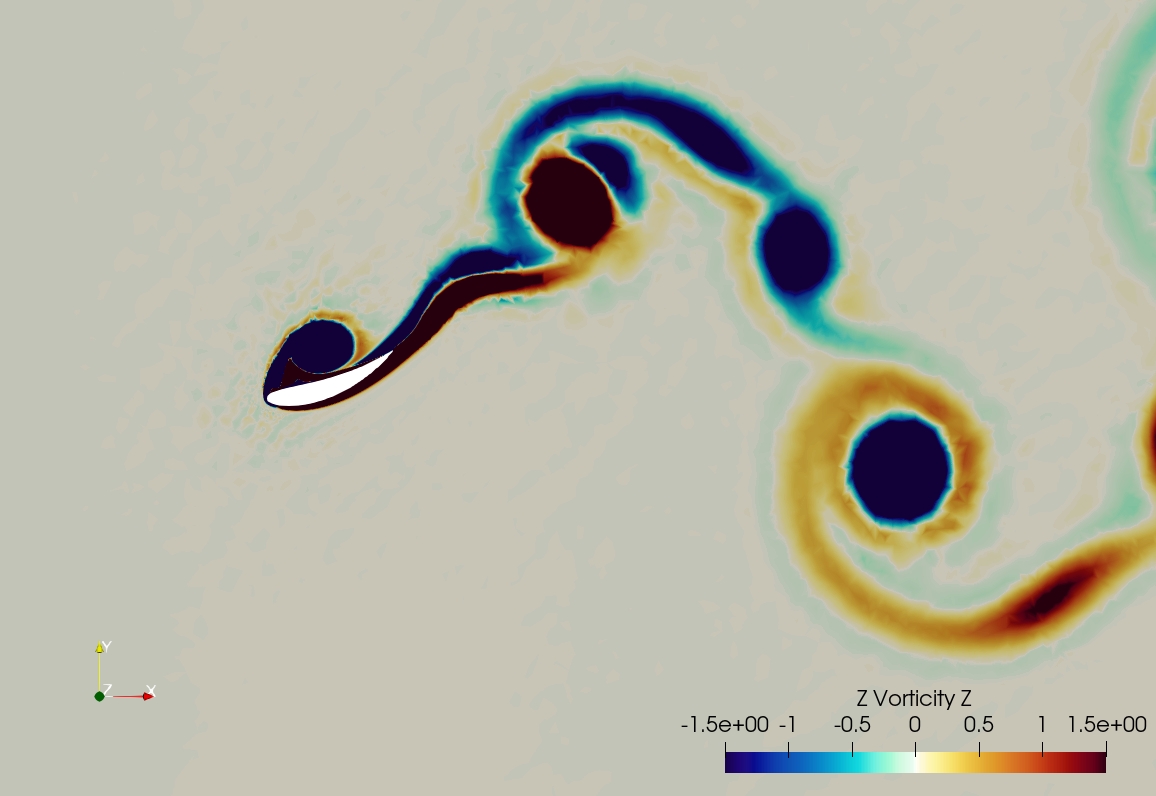}
            \hfill
            \includegraphics[width = 0.48\textwidth]{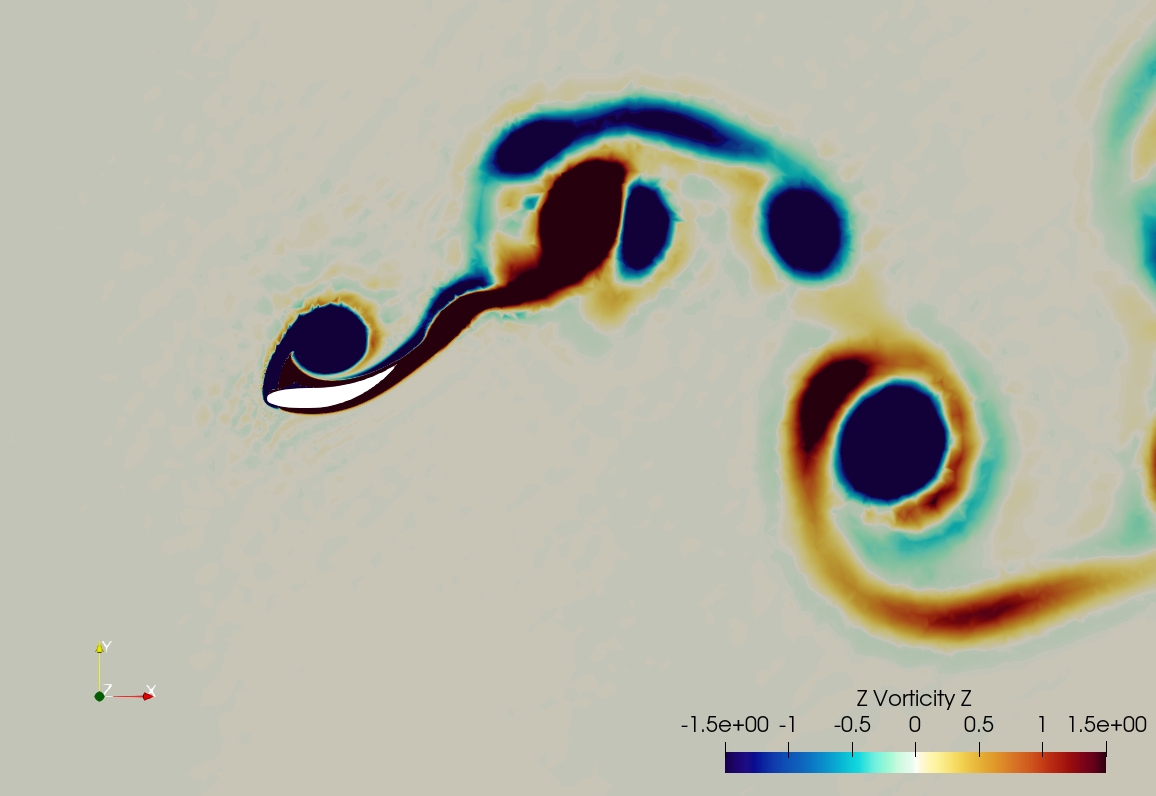}
        \caption{$t/T$ = 0.34}
    \end{subfigure}
    \caption{PRESSURE AND VORTICITY CONTOURS FOR THE MORPHING FOIL WITH MORPH AMPLITUDE $50^\circ$ AND MORPH POSITION OF 0\% (LEFT COLUMN) AND 30\% (RIGHT COLUMN) AT TIME INSTANCES: (a) $t/T = 0.15$, (b) $t/T = 0.24$ AND (c) $t/T = 0.34$}
    \label{Position comp}
\end{figure}

\begin{figure}[!htbp]
    \centering
    
        \includegraphics[trim = {10cm 0 10.5cm 1cm}, clip, width = 0.4\textwidth]{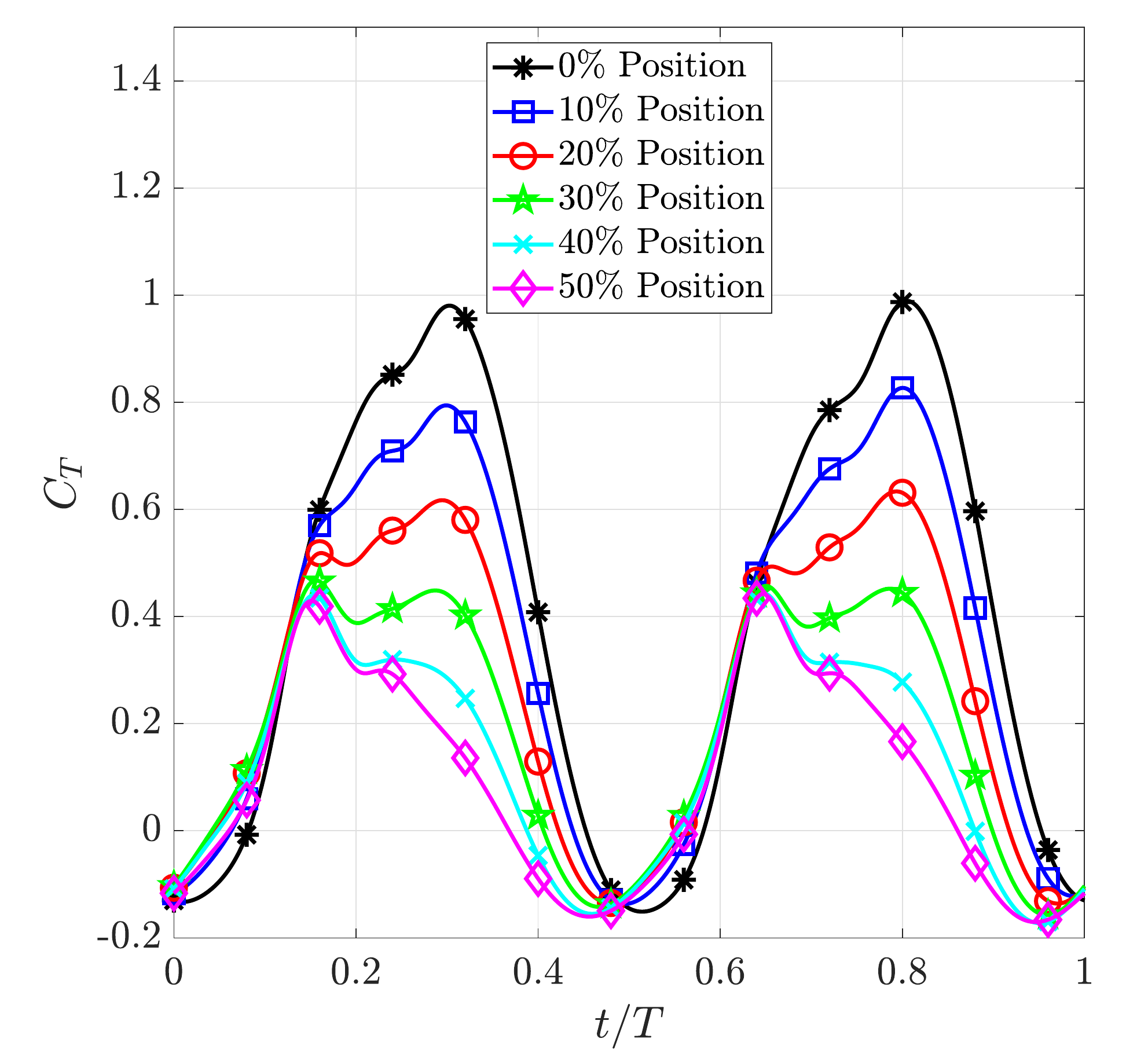}
    \caption{VARIATION OF $C_T$ OVER A TIME PERIOD FOR MORPH AMPLITUDE OF 50$^\circ$}
    \label{CT_amp_50}
\end{figure}

\underline{\textbf{Effect of Morph Position}}:
To further extend our discussion to the effect of morph position, we select the fixed morph amplitude of $50^{\circ}$ and compare the flow visualizations at the morph positions of 0\% and 30\%. The pressure and vorticity contours for the two  representative cases are shown in Fig. \ref{Position comp}. Due to the morphing motion of the complete foil for 0\% morph position, the surface of the foil near the leading edge moves with an absolute velocity opposite to the heave velocity during downstroke. 
This allows the foil surface to move closer to the developing LEV and prevent its early separation, resulting in higher thrust generation (Fig. \ref{Position comp}(c)). This is also reflected in the temporal variation of the thrust coefficient shown in Fig. \ref{CT_amp_50}, where $C_T$ continues to increase up to the quarter time period (mid of the downstroke) where morph is highest, and then drops subsequently as the foil reduces the trailing edge morph. However, at higher morph positions, the trailing edge is too far behind to meet the LEV at the point of separation. Since there is no morphing till 30\% of chord in the representative case considered, the leading 30\% of the foil is purely heaving. The higher absolute velocity solely due to heaving is encouraging separation, and simultaneously, the morphed trailing edge is too far away from the vortex. 

The morphed section of the foil contributes to the change in the projected area to the incoming flow, thus resulting in the favorable pressure difference across the morphed section which propels the foil. As discussed earlier, with increased morph position, the trailing edge of the foil (the section that is morphing) loses the influence of the LEV-ds(n) earlier leading to vortex separation and lower thrust values. Furthermore, the mean thrust coefficient \(C_{T,\mathrm{mean}}\) does not increase with amplitude at higher positions (Fig. \ref{CT_CL_Charts}(a)) as the morphed section of the foil is too far from the vortex to have any noticeable effect. 
Therefore, we observe that morphing facilitates the transit and influence of the LEV on the foil by keeping it attached although it delays the shedding of the vortex.

\section*{CONCLUSIONS AND FUTURE WORK}

An extensive numerical study is carried out to understand the enhancement in propulsive performance due to the inverted von-K$\mathrm{\acute{a}}$rm$\mathrm{\acute{a}}$n vortex street in a combined morphing and heaving foil.
The key findings from the present study are:
\begin{enumerate}
\item The coefficient of thrust ($C_T$) depends on both, the morph position and the morph amplitude. Higher amplitudes are generally observed to have a positive impact on increasing $C_T$, while higher positions of morphing negatively impact the thrust performance.

\item The lift coefficient ($C_L$) is dependent on the effective angle of attack $\alpha_e$ of the motion, and the variation of lift is resultant of the positive impact of morph position ($C_L$ increases with increasing morph position) and the negative impact of morph amplitude ($C_L$ decreases with increasing morph amplitude).


\item Higher amplitudes of morphing have larger projected area to the incoming flow, and the vortex remains closer to the surface of the foil for a longer duration. $C_{T,\mathrm{mean}}$ thus increases with the morph amplitude for lower morph positions.

\item An increase in morph position leads to decrease in $C_T$, because the LEV separates too early and the morphed trailing edge is not influenced by the suction pressure created by the vortex.

\end{enumerate}

The biological flapping motion observed in the wings of birds and fins of fishes can be realized by considering the pitching motion of the foil along with heaving and morphing, which forms a topic for future study. Effect of Reynolds number can also be investigated, extending the work to turbulent regime. 



%
% ...


\bibliographystyle{asmems4}
\bibliography{references}

\end{document}